\documentclass[paper,11pt]{article}
\pdfoutput=1

\usepackage{mathrsfs}
\usepackage{graphicx}
\usepackage{epstopdf}
\usepackage{color}
\usepackage[centertags]{amsmath}
\usepackage{amsfonts}
\usepackage{amssymb} 
\usepackage{amsthm}
\usepackage{slashed}
\usepackage{framed}
\usepackage{enumerate,enumitem}
\usepackage{amsbsy}
\usepackage{jheppub}

\def\({\left(}
\def\){\right)}
\def\[{\left[}
\def\]{\right]}

\def\one{{\rm 1\kern -.9mm l}}                             %

\def\beq{\begin{equation}}
\def\eeq{\end{equation}}
\def\beqa{\begin{eqnarray}}
\def\eeqa{\end{eqnarray}}
\newcommand{\eqa}{\begin{eqnarray}}
\newcommand{\ena}{\end{eqnarray}}

\newcommand{\bT}{\bar{T}}

\newcommand{\bbT}{\mathbb{T}}
\newcommand{\bfT}{{ \bf T }}

\newcommand\blank[1]{}

\newcommand{\tr}{{\rm Tr}}

\newcommand{\bP}{ {\bf P } }

\newcommand{\wT}{ \mathbb{T} }

\newcommand{\bbalpha}{\alpha\kern -4.95pt\alpha}

\newcommand\en{\end{equation}}
\newcommand\bea{\begin{eqnarray}}
\newcommand\eea{\end{eqnarray}}
\newcommand\nn{\nonumber}

\newcommand{\One}{{\hbox{{\rm 1{\hbox to 1.5pt{\hss\rm1}}}}}}
\renewcommand{\One}{{\mathbb 1}}
\renewcommand{\One}{{\rm 1\!\!1}}

\newcommand{\DD}{\triangle} 

\newcommand{\ba}{\begin{eqnarray}}
\newcommand{\ea}{\end{eqnarray}}

\newcommand{\be}{\begin{equation}}
\newcommand{\ee}{\end{equation}}

\renewcommand{\&}{& {\!\!\!\!\!\!}}

\setlist[itemize]{leftmargin=*}

\def\XXint#1#2#3{{\setbox0=\hbox{$#1{#2#3}{\int}$ }
\vcenter{\hbox{$#2#3$ }}\kern-.6\wd0}}

\def\tr{{\rm tr~}}

\newcommand{\sigmaBES}{\sigma }
\newcommand{\sigtw}{\mathcal{P} }

\newcommand{\hs}{\frac{\sqrt{3}}{2}}
\renewcommand{\d}{\partial}

\newcommand{\<}{{\langle}}
\renewcommand{\>}{{\rangle}}

\newcommand{\re}{\relax{\rm I\kern-.18em R}}

\renewcommand{\sp}{p\hspace{-.40em}/}

\renewcommand{\sp}{p\hspace{-.40em}/}

\newcommand{\bfo}{{\bar{4}}}

\def\su2{{SU(2)}}

\def\[{\left[}
\def\]{\right]}

\def\s{\sigma}

\def\({\left(}
\def\){\right)}
\def\[{\left[}
\def\]{\right]}

\def\<{\langle}
\def\>{\rangle}

\def\bT{{\bf T}}

\def\s*{\ *_{\!\!\!\!\!\!\!\!\!\,_{\,_\text{\scriptsize{sym}}}}}
\def\hs*{\ \hat{*}_{\!\!\!\!\!\!\!\!\!\,_{\,_\text{\scriptsize{sym}}}}}
\def\d{\partial}

\def\i2{\frac{i}{2}}

\def\bQ{{\bf Q}}
\def\bP{{\bf P}}

\def\wT{{\mathbb T}}

\def\spi{\relax{\rm \pi\kern-0.5em /}}
\def\sA{\relax{\rm A\kern-0.5em /}}
\def\sp{\relax{\rm p\kern-0.5em /}}
\def\sd{\relax{\rm \d\kern-0.5em /}}
\def\sk{\relax{\rm k\kern-0.5em /}}
\def\sn{\relax{\rm n\kern-0.5em /}}
\def\sP{\relax{\rm P\kern-0.7em /}}
\def\sBethe{\relax{\rm \Bethe\kern-0.5em /}}

\def\One{1\hskip-.16cm1}

\def\Om{\Phi}
\def\om{\phi}
\def\os{y}

\def\ch{\kappa}

\newcommand{\al}{\varepsilon}
\newcommand{\mQ}{{\mathcal{Q}}}
\newcommand{\balpha}{\mathit{s}}
\newcommand{\bu}{\mathit{s}}
\title{The full Quantum Spectral Curve for AdS$_4$/CFT$_3$}
\author[1]{Diego Bombardelli,}
\author[1]{Andrea Cavagli\`a,}
\author[2]{Davide Fioravanti,}
\author[3,4]{Nikolay Gromov}
\author[1]{and Roberto Tateo}

\affiliation[1]{Dipartimento\ di Fisica and INFN, Universit\`a di Torino, Via P.\ Giuria 1, 10125 Torino, Italy.}
\affiliation[2]{INFN and Dipartimento di Fisica e Astronomia, Universit\`a di Bologna, Via Irnerio 46, 40126 Bologna, Italy.}
\affiliation[3]{Mathematics Department, King's College London, The Strand, London WC2R 2LS, UK}
\affiliation[4]{St.Petersburg INP, Gatchina, 188 300, St.Petersburg,   Russia.}

\emailAdd{diegobombardelli@gmail.com}
\emailAdd{cavaglia@to.infn.it}
\emailAdd{fioravanti@bo.infn.it}
\emailAdd{nikgromov@gmail.com}
\emailAdd{tateo@to.infn.it}
 \abstract{
 The spectrum of planar $\mathcal{N}=6$ superconformal Chern-Simons theory, dual to type IIA superstring theory on $AdS_4 \times CP^3$, is accessible at finite coupling using integrability. 
 Starting from the results of [\href{http://journals.aps.org/prl/abstract/10.1103/PhysRevLett.113.021601}{arXiv:1403.1859}], we study in depth the basic integrability structure underlying the spectral problem, the Quantum Spectral Curve. The new results presented in this paper open the way to the quantitative study of the spectrum for arbitrary operators at finite coupling. 
 Besides, we show that the Quantum Spectral Curve is embedded into a novel kind of Q-system, which reflects the $OSp(4|6)$ symmetry of the theory and leads to exact Bethe Ansatz equations. The discovery of this algebraic structure, more intricate than the one appearing in the $AdS_5/CFT_4$ case, could be a first step towards the extension of the method to $AdS_3/CFT_2$.
}
\begin{document}
\maketitle
\section{Introduction}

The idea of a duality between gauge and string theory was put forward many years ago by 't Hooft \cite{tHooft:1973alw}, who noticed that the perturbative expansion in $SU(N_c)$ 
Yang-Mills theory in the large $N_c$ limit naturally organizes in terms of the topology of Feynman diagrams, mimicking the genus expansion of string theory.

The first concrete realization of the  duality  \cite{Maldacena:1997re,Gubser:1998bc,Witten:1998qj} conjectures the exact equivalence of  $\mathcal{N}=4$ super Yang-Mills (SYM) theory 
and type IIB string theory on $AdS_5\times S^5$. The precise identification of observables and parameters in the two theories relates the perturbative region of 
each model to the deep non-perturbative regime of the other. For this reason, the correspondence makes powerful predictions, but is also very difficult to test. 

 An important turning point in this field was the discovery of fingerprints of integrability, at both weak and strong coupling \cite{MZ,Bena:2003wd}, in the planar limit of this duality. 
 At least in this limit, it is hoped that the theory will be exactly solved adapting integrable model tools, and remarkable progress has been made on the study of various observables, including Wilson loops and correlation functions. 

In particular, the problem of computing the conformal spectrum of the theory was tackled by tailoring integrable QFT techniques to this new setting, in particular the Bethe Ansatz \cite{MZ, Beisert:2005fw, Beisert:2006ez}, the TBA, the Y and T-systems \cite{BFT,AF,GKKV,Extended,Balog:2011nm,Wronskian,FiNLIE}, leading to the discovery of the very effective Quantum Spectral Curve (QSC) formulation \cite{QSC, Gromov:2014caa}. The latter is a very satisfactory simplification and probably the most elementary formulation of the problem. 
 Thanks to the mathematical simplicity of the QSC, it appears that, in the near future,  the spectral problem may be completely solved also in a practical/computational sense. Already, the QSC method allows to compute the spectrum numerically with high precision \cite{Gromov:2015wca,Hegedus:2016eop} and to inspect analytically interesting regimes such as the BFKL limit \cite{QCDPomeron,Gromov:2015vua} or the weak coupling expansion \cite{Marboe:2014gma,MarboeTalk,Marboe:2017dmb}. It has also been generalized to so-called $\gamma$ deformations \cite{Thook} and to the quark-antiquark potential \cite{QSCCusp,QSCPotential}. 

Another remarkable example of AdS/CFT correspondence was introduced by Aharony, Bergman, Jafferis and Maldacena (ABJM) in \cite{Aharony:2008ug}. The gauge side of the duality corresponds to the $\mathcal{N}=6$ superconformal Chern-Simons theory with gauge group $U(N) \times U(N)$, with  opposite Chern-Simons levels, $k$ and $-k$, for the two $U(N)$ factors. We will be concerned with the planar limit, where $k, N \rightarrow \infty$ with the 't Hooft coupling $\lambda = \frac{k}{N}$ kept finite and the dual gravity theory becomes type IIA superstring theory on $AdS_4 \times  CP^3$. In this regime, integrability emerges, making the ABJM model the only known example of 3d quantum field theory which can be exactly solved \cite{Minahan:2008hf, Gaiotto:2008cg, Stefanski:2008ik, Arutyunov:2008if, Gromov:2008bz} (see also the review \cite{Klose:2010ki}). 
 
The spectral problem in ABJM theory was approached exploiting the experience gained in $AdS_5/CFT_4$. Anomalous dimensions of  single trace operators with asymptotically large quantum numbers are described at all loop by the so-called Asymptotic Bethe Ansatz equations, conjectured in \cite{Gromov:2008qe} and derived from the exact worldsheet S-matrix of \cite{Ahn:2008aa}. 
The exact result, including all finite-size corrections for short operators, is formally described by 
 an infinite set of TBA equations, proposed in \cite{Bombardelli:2009xz,Gromov:2009at}. 
 These equations were solved numerically for a particular operator in \cite{LevkovichMaslyuk:2011ty}. 
 However, solving excited states TBA equations with high precision is a challenging task already for very simple models \cite{DT,Bazhanov:1996aq, DoreyPerturbedCFT}. Besides, the form of the TBA equations depends on the state and  possibly also on the range of the coupling considered, so that they can be studied only on a case-by-case basis. 
 
 It is important to look for a simpler formulation which overcomes these problems. Starting from a precise knowledge of the analytic properties of the TBA
solutions \cite{ABJMdisco}, the basic equations characterizing the Quantum Spectral Curve of  the ABJM model were obtained in \cite{Cavaglia:2014exa}. 
 These results were used to compute the so-called slope function in a near-BPS finite coupling regime \cite{Gromov:2014eha} and to develop a generic algorithm for the weak coupling expansion in the $SL(2)$-like sector  \cite{Anselmetti:2015mda}. 
  
Although we stress that, as proved by the applications discussed above, the results of \cite{Cavaglia:2014exa} contain  all the analytic information necessary to solve the spectral problem, several important aspects of the full picture were still missing. 
 First of all, the concrete recipe to describe states within the QSC framework was discussed in \cite{Cavaglia:2014exa} only for the $SL(2)$-like sector. 
 Secondly, the set of equations obtained in  \cite{Cavaglia:2014exa}, 
 the $\bP\mu$/$\bP\nu$-system, can be  associated, in the classical limit, to degrees of freedom related to the $CP^3$ part of the whole  $AdS_4 \times  CP^3$ target space. A dual system of equations, only briefly mentioned in   \cite{Cavaglia:2014exa},  
may be instead associated to $AdS_4$ classical degrees of freedom. The interplay between the two systems is important for the development of the state-of-the-art solution algorithm at finite coupling \cite{Gromov:2015wca}, as well as at weak coupling for generic states \cite{Gromov:2015vua,MarboeTalk}. 
 Furthermore, the full algebraic structure was still not transparent, and for example 
  the link between the formulation of \cite{Cavaglia:2014exa} and the Asymptotic Bethe Ansatz of \cite{Gromov:2008qe} was difficult to see. In this paper we will fill these gaps and present the necessary elements for the quantitative solution of the spectral problem for an arbitrary operator at finite coupling. Besides, we reveal an interesting underlying representation theory structure, which could allow for generalizations and may in particular help in the solution of the spectral problem for $AdS_3/CFT_2$ dualities (see \cite{SfondriniReview} for a recent review).

To conclude this introduction, let us review an important fact. In contrast with $\mathcal{N}$=4 SYM, in ABJM theory integrability leaves unfixed the so-called interpolating function $h(\lambda)$ \cite{Gaiotto:2008cg, Grignani:2008is}, which parametrizes the dispersion relation of elementary spin chain/worldsheet excitations and enters as an effective coupling constant in the integrability-based approach, in particular in the QSC equations. An important conjecture for the exact form of this function, passing several tests at  weak and strong coupling \cite{Bianchi:2014ada}, was made in \cite{Gromov:2014eha} by a comparison with the structure of localization results. This conjecture was extendend in \cite{Cavaglia:2016ide} to encompass the  ABJ model \cite{Aharony:2008gk}, which is based on a more general gauge group $U(N)\times U(M)$ and possesses two 't Hooft couplings $\lambda_1$, $\lambda_2$ in the planar limit. According to the proposal of  \cite{Cavaglia:2016ide} (based on important observations of \cite{Bak:2008vd,Minahan:2009te,Minahan:2010nn,Bianchi:2016rub}), at the level of the spectrum the only difference between the ABJM and ABJ theories lies in the replacement of $h(\lambda)$ with an explicitly defined $h^{\text{ABJ}}(\lambda_1, \lambda_2)$ (see \cite{Cavaglia:2016ide}). In the following we will simply denote the ABJM/ABJ interpolating function as $h$. 

The contents of this paper are presented in detail below.

\noindent In {\bf Section \ref{sec:symmetries}}, we discuss the bosonic symmetry underlying the problem, namely $SO(3,2) \times SO(6)$, the isometry group of $AdS_4 \times CP^3$. We will introduce important vector and spinor notation used in the rest of the paper. Besides, we comment on the interesting fact that the isometry group of $CP^3$  effectively appears in the Quantum Spectral Curve as $SO(3,3)$, rather than $SO(6)$. 

\noindent In {\bf Section \ref{sec:PMuPNu}}, we review the results of \cite{Cavaglia:2014exa} and discuss how they reflect the $CP^3$ symmetry. We discuss a subtle modification of the analytic properties (initially overlooked in \cite{Cavaglia:2014exa}), which is needed for the study of certain non-symmetric sectors of the theory. The modified equations contain an extra nontrivial function of the coupling, which can be interpreted at weak coupling as the momentum of a single species of magnons. 

\noindent In {\bf Section \ref{sec:QOmega}}, we present an explicit construction of new variables, the functions $\bQ_I$, $\bQ_{\circ}$ and $\tau_i$, which satisfy a dual system of Riemann-Hilbert equations reflecting the symmetry of $AdS_4$. 

\noindent In {\bf Section \ref{sec:asymptotics}}, we treat in full generality the boundary conditions which need to be imposed on the solutions of the QSC at large value of the spectral parameter in order to describe a physical state. This is the place where the quantum numbers of the state make an appearance. We also discuss the correspondence between the functions $\bP$ and $\bQ$ and quasi-momenta of the spectral curve in the classical limit. 

\noindent In {\bf Section \ref{sec:gluing}}, based on results obtained in \cite{Gromov:2015vua,ContiNumerical}, we discuss a set of exact relations which are perhaps the most convenient way to repack the analytic properties discussed in Sections \ref{sec:PMuPNu}, \ref{sec:QOmega}. It is also shown how these equations encode the quantization of the spin. 

\noindent In {\bf Section \ref{sec:Qsystem}}, we embed the previous results into a larger set of functional relations which may be considered as (part of) a Q-system. Q-systems are familiar in the theory of integrable models \cite{Krichever:1996qd,Pronko:1999gh} and in the ODE/IM framework \cite{Dorey:2006an}: they are powerful sets of functional relations that, supplemented  by simple analytic requirements, become equivalent to exact Bethe equations. The structure of Q-systems is completely fixed by symmetry: for example, the QQ relations appearing in the $\mathcal{N}$=4 SYM case are the same as the ones for $SU(4|4)$ spin chains. For the  $OSp(4|6)$ superalgebra relevant to ABJM theory, however, this algebraic construction was not known in the literature. While we do not treat in full generality the representation theory aspects, we construct explicitly an enlarged set of Q functions, and prove that they satisfy exact Bethe equations reflecting the full supergroup structure. 
Generalizing arguments of \cite{Gromov:2014caa}, we will show that, in  the limit of large volume, some of these exact Bethe equations reduce to the Asymptotic Bethe Ansatz.   

\noindent The paper also contains five Appendices:

In {\bf Appendix \ref{app:derivations}}, we discuss the details of the derivation (already summarized in \cite{Cavaglia:2014exa}) of the QSC from the analytic properties of the T-system \cite{ABJMdisco}. In {\bf Appendix \ref{app:gamma}}, we list some useful algebraic identities used in the derivation of the Q-system  relations. In {\bf Appendix \ref{app:ABderive}}, we deduce some of the constraints on the asymptotics of $\bP$ and $\bQ$ functions. In {\bf Appendix \ref{sec:appWC}}, we discuss the weak coupling limit of the QSC and show the emergence of the 2-loop Bethe equations of \cite{MZ}. We exploit this link to prove the identification between the parameters entering the asymptotics of the QSC and the quantum numbers. Finally, in {\bf Appendix \ref{app:dictionary}} we review the dictionary between $OSp(4|6)$ quantum numbers and  number of Bethe roots appearing in various versions of the (Asymptotic) Bethe Ansatz, which could be useful for the reader wanting to apply the prescription of Section \ref{sec:asymptotics} to concrete states.

\section{Symmetries and conventions}\label{sec:symmetries}
ABJM theory is invariant under the supergroup $OSp(4|6)$, whose bosonic subgroups are associated to the isometries of $AdS_4$ and $CP^3$. 
 We will see that the Quantum Spectral Curve equations encode elegantly this symmetry structure. 
 Let us briefly introduce the main group-theoretic constructions related to  the bosonic symmetries.  
 
  \begin{itemize}
 \item $CP^3$: the isometry group of $CP^3$ is the orthogonal group $SO(6) \simeq SU(4)$. 
 The invariant $6 \times 6$ symmetric tensor naturally associated to this symmetry is the metric. This tensor enters the QSC equations\footnote{In \cite{Cavaglia:2014exa}, this tensor was denoted as $\chi_{AB}$.}, and will be denoted in this paper as $\eta_{AB}$. Peculiarly, we will see that it appears in the QSC with a $(+++---)$ signature. 
 The concrete form of $\eta_{AB}$ to be used in the rest of this paper is
\beq\label{eq:eta}
\eta_{AB} = \eta^{AB} = \left(
\begin{array}{cccccc}
 0 & 0 & 0 & 1 & 0& 0\\
 0 & 0 & -1 & 0 & 0 & 0 \\
 0 & -1 & 0 & 0 & 0 & 0 \\
 1 & 0 & 0 & 0 & 0 & 0 \\
 0 & 0 & 0 & 0 & 0 & 1 \\
  0 & 0 & 0 & 0 & 1 & 0
\end{array}
\right) ,
\eeq
where $\eta^{AB}$ is the inverse matrix, i.e. $\eta_{AB} \, \eta^{BC}= \delta_A^C$. This particular choice for $\eta_{AB}$ emerged naturally from the derivation of the QSC, summarized in Appendix \ref{app:derivations}. As explained there, the specific form of $\eta_{AB}$ in (\ref{eq:eta}) is partly conventional, but its signature cannot be modified without spoiling the reality properties of the system. The fact that the $CP^3$ symmetry appears effectively as $SO(3,3)$ can be understood heuristically considering the classical limit, where the basic variables of the QSC are related to the quasi-momenta  of the algebraic curve (see Section \ref{sec:classical}). 
The quasi-momenta describing a string moving in $CP^3$ are defined through the diagonalization of a $SO(6)$ block of the classical monodromy matrix. An $SO(2n)$ orthogonal matrix in general cannot be diagonalized with a real transformation, so that the signature of the metric is not preserved in the eigenvectors  basis; moreover, the signature changes precisely to the one typical of $SO(n, n)$.  

Let us introduce some conventions. We will use different index labels for objects with different symmetry properties. The indices $A, B, C = 1, \dots, 6$ will be assumed to carry the vector representation of $SO(3,3)$, and will always be lowered and raised with the metric $\eta_{AB}$ and its inverse $\eta^{AB}$, respectively. 
It will be useful to consider also spinor representations of $SO(3,3)$. The relevant $8 \times 8$ gamma matrices are defined by 
\beq
\left\{ \Gamma_{8\times 8}^A , \Gamma_{8 \times 8}^B \right\} = \eta^{AB} \, \text{Id}_{8 \times 8}. 
\eeq 
In even dimensions,  gamma matrices can always be written in a chiral form:
\beq
\Gamma^A = \left(\begin{array}{c  c } 0 & \sigma^A_{ab} \\
{( \bar{\sigma}^A ) }^{ab} \;\; & 0 
\end{array} \right),
\eeq
where the matrices $\sigma^A_{ab}$ and ${( \bar{\sigma}^A ) }^{ab}$ satisfy
\beq
\sigma^A_{ab} \, (\bar{\sigma}^{B} )^{bc} + \sigma^B_{ab} \, (\bar{\sigma}^{A} )^{bc} = \eta^{AB} \, \delta_a^c .
\eeq
While all our equations will be covariant, it is convenient to specify a concrete basis. The matrices $\sigma^A_{ab}$ and ${( \bar{\sigma}^A ) }^{ab}$ are defined in our conventions by
\beq
V_A \sigma^A_{ab} =  \left( \begin{array}{cccc} 0 & - V_1 & - V_2 & - V_5 \\  V_1 & 0 & - V_6 & - V_3 \\ V_2 & V_6 & 0 & - V_4 \\ V_5 & V_3 & V_4 & 0 \end{array} \right), \;\;\;\; V_A { ({ \bar{\sigma} }^A) }^{ab} = \left( \begin{array}{cccc} 0 & V_4 & -V_3 & V_6 \\ -V_4 & 0 & V_5 & -V_2 \\ V_3 & -V_5 & 0 & V_1 \\ -V_6 & V_2 & -V_1 & 0 \end{array} \right),
\eeq
for an arbitrary vector $(V_1,\dots, V_6)$.
Lower-case indices $a, b, c$ will always be taken to run over $1, \dots, 4$ and will be reserved for the spinor representations.
Note that there is a distinction between upper and lower spinor indices, as they belong to the chiral and anti-chiral spinor representations, respectively, which are equivalent to the representations $\bf{4}$ and $\bar{\bf{4}}$ of $SU(4) \simeq SO(6)$. 
Another natural tensor that will make an appearance in the equations is the anti-symmetrized product of gamma matrices, 
\beq\label{eq:M2}
( \sigma^{AB}  )_{a}^{ \;b} \equiv -\frac{1}{2} \left(( \sigma^{A}  )_{ac }  {( { \bar \sigma }^{B}  )}^{cb } - ( \sigma^{B}  )_{ac}  {( { \bar \sigma }^{A}  )}^{cb} \right) .
\eeq
\item $AdS_4$:
the isometry group of $AdS_4$ is $SO(3,2) \simeq Sp(4)$. We will denote the metric of this orthogonal group as $\rho_{IJ}$, and our concrete choice will be:
\beq\label{eq:rho}
\rho_{IJ} = \left(
\begin{array}{cccccc}
 0 & 0 & 0 & 1 & 0\\
 0 & 0 & -1 & 0 & 0  \\
 0 & -1 & 0 & 0 & 0 \\
 1 & 0 & 0 & 0 & 0  \\
 0 & 0 & 0 & 0 & \frac{1}{2} 
\end{array}
\right) ,  \;\;\;\; \rho^{IJ} \equiv ( \rho^{-1} )^{IJ} = \left(
\begin{array}{cccccc}
 0 & 0 & 0 & 1 & 0\\
 0 & 0 & -1 & 0 & 0  \\
 0 & -1 & 0 & 0 & 0 \\
 1 & 0 & 0 & 0 & 0  \\
 0 & 0 & 0 & 0 & 2 
\end{array}
\right).
\eeq
In the following, we shall always reserve the indices $I, J, K$, running over $1, \dots,5$, for the vector representation of $SO(3,2)$. 

Let us remind the reader of the isomorphism between $SO(3, 2)$ and $Sp(4)$, the group of linear maps preserving a $4 \times 4$ anti-symmetric two-form. One way to see this is to view $SO(3,2)$ as obtained from $SO(3,3)$ by reducing to the subspace orthogonal to a preferred vector $v$, with $v \cdot v = -1$. 

Then we see that an anti-symmetric two-form naturally emerges: $\ch_{ij} \equiv v_A \, (\sigma^A)_{ij} $. Let us denote a projection of the $\sigma$, $\bar{\sigma}$ matrices on the subspace orthogonal to $v$ as $\Sigma_I$, $\bar{\Sigma}_I$, respectively, with $I=1, \dots, 5$. By construction, they satisfy the intertwining relations $\bar{\Sigma}_{I}^{ij} = \ch^{i i_1} \, (\Sigma_{I} )_{i_1 i_2} \, \ch^{i_2 j}$, showing that there are in fact only five independent matrices $\Sigma_I$. The latter give a four dimensional representation of Clifford algebra:
\beq
\left\{ \Gamma_{4 \times 4} ^I , \Gamma_{4 \times 4}^J \right\} = \rho^{IJ} \, \text{Id}_{4 \times 4}, 
\eeq
with
\beq
 (\Gamma_{4 \times 4}^I )_i^j \equiv  (\Sigma^I)_{ik} \, \ch^{kj} =  \ch_{ij} (\bar{\Sigma}^I)^{jk} . 
\eeq 
In the following, we will use indices $i, j, k, l$, running over $1, \dots, 4$, to refer to the four-dimensional representation of $SO(3,2)$. Finally, one can introduce the anti-symmetric combinations 
 \beq\label{eq:MSigma}
 ( \bar{\Sigma}^{IJ}  )_{i}^{ \;j} \equiv -\frac{1}{2} \left(( \Sigma^{I}  )_{ik }  {( { \bar \Sigma }^{J}  )}^{kj } - ( \Sigma^{J}  )_{ik}  {( { \bar \Sigma }^{I}  )}^{kj} \right),
 \eeq
which play the role of generators of $SO(3,2)$. By construction, these generators leave invariant the two-form $\ch_{ij}$: therefore the spinor representation of $SO(3,2)$ is identified with the fundamental representation of $Sp(4)$. 

In our concrete case, we see that the metric (\ref{eq:rho}) is obtained from (\ref{eq:eta}) by restricting to the subspace orthogonal to $v=(0,0,0,0,-1,1)$. 
Our choice for the $\Sigma$ matrices will be
\beq\label{eq:Sigma}
\Sigma^I \equiv \left( \sigma^1, \sigma^2, \sigma^3 , \sigma^4,  \sigma^5 + \sigma^6  \right),\;\;\;\;\;\;\;
\bar{\Sigma}^I \equiv \left( \bar{\sigma}^1, \bar{\sigma}^2, \bar{\sigma}^3 , \bar{\sigma}^4,  \bar{\sigma}^5 + \bar{\sigma}^6  \right),
\eeq
and the two-form $\ch_{ij}$ reads
\beq
\ch_{ij} \equiv v_A \, (\sigma^A)_{ij} =  \left( \begin{array}{cccc} 0 &0 & 0 & 1 \\ 0 & 0 & -1 & 0 \\ 0 & 1 & 0 &0 \\ -1 & 0  & 0 & 0 \end{array} \right) .
\label{kijdef}
\eeq
\end{itemize}

\section{Formulation of the QSC from the TBA}\label{sec:PMuPNu}
In this Section, we recall the first version of the QSC equations proposed in \cite{Cavaglia:2014exa}. These equations were obtained through a reduction of the T-system, supplemented by analyticity properties extracted from the TBA \cite{Extended,Cavaglia:2014exa}, and ultimately take the form of a nonlinear Riemann-Hilbert problem defined on the complex domain of the spectral parameter $u$. In the $u$-plane, the Q functions have a characteristic pattern of branch points, whose positions depends on the coupling constant $h$ as specified below. These branch points will all be of square-root type. 
This peculiar kind of analytic structure for the Q functions, beside $AdS_5/CFT_4$, is also characteristic of some non-relativistic integrable systems such as the Hubbard model \cite{Cavaglia:2015nta}. The derivation of the QSC equations is discussed in Appendix \ref{app:derivations}. 
 
\subsection{Equations in vector form and analyticity conditions}\label{sec:PmuForm}
\begin{figure}
\begin{minipage}{0.45 \linewidth}
\centering
\includegraphics[scale=0.3]{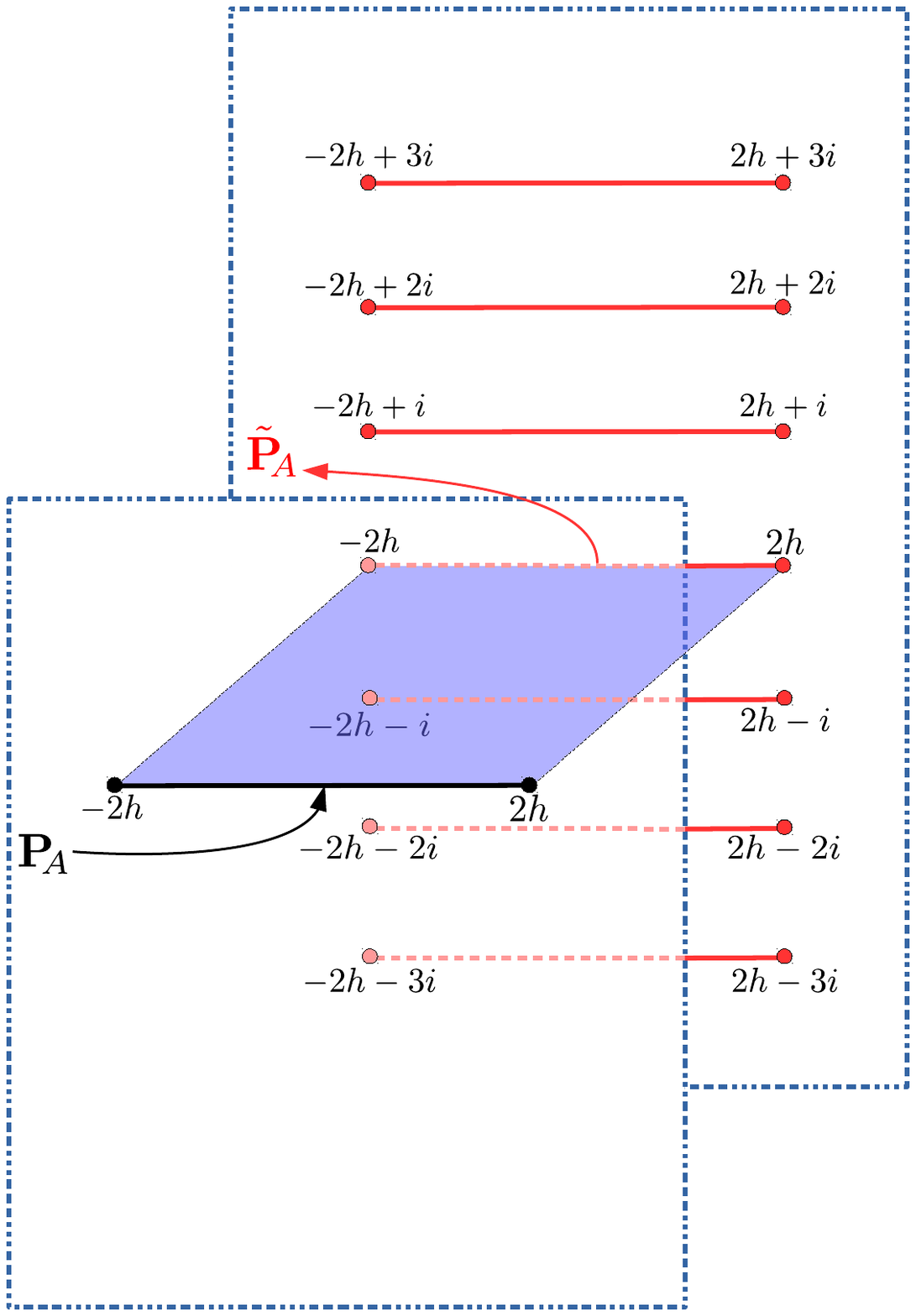}
\caption{Cut structure of the $\bP_A$ functions, with a single cut on the first sheet. We denote with $\widetilde \bP_A$ the analytic continuation to the next sheet, through the cut on the real axis. }
\label{fig:cutP}
\end{minipage}
\hspace{0.3cm}
\begin{minipage}{0.45 \linewidth}
\vspace{2cm}
\includegraphics[scale=0.25]{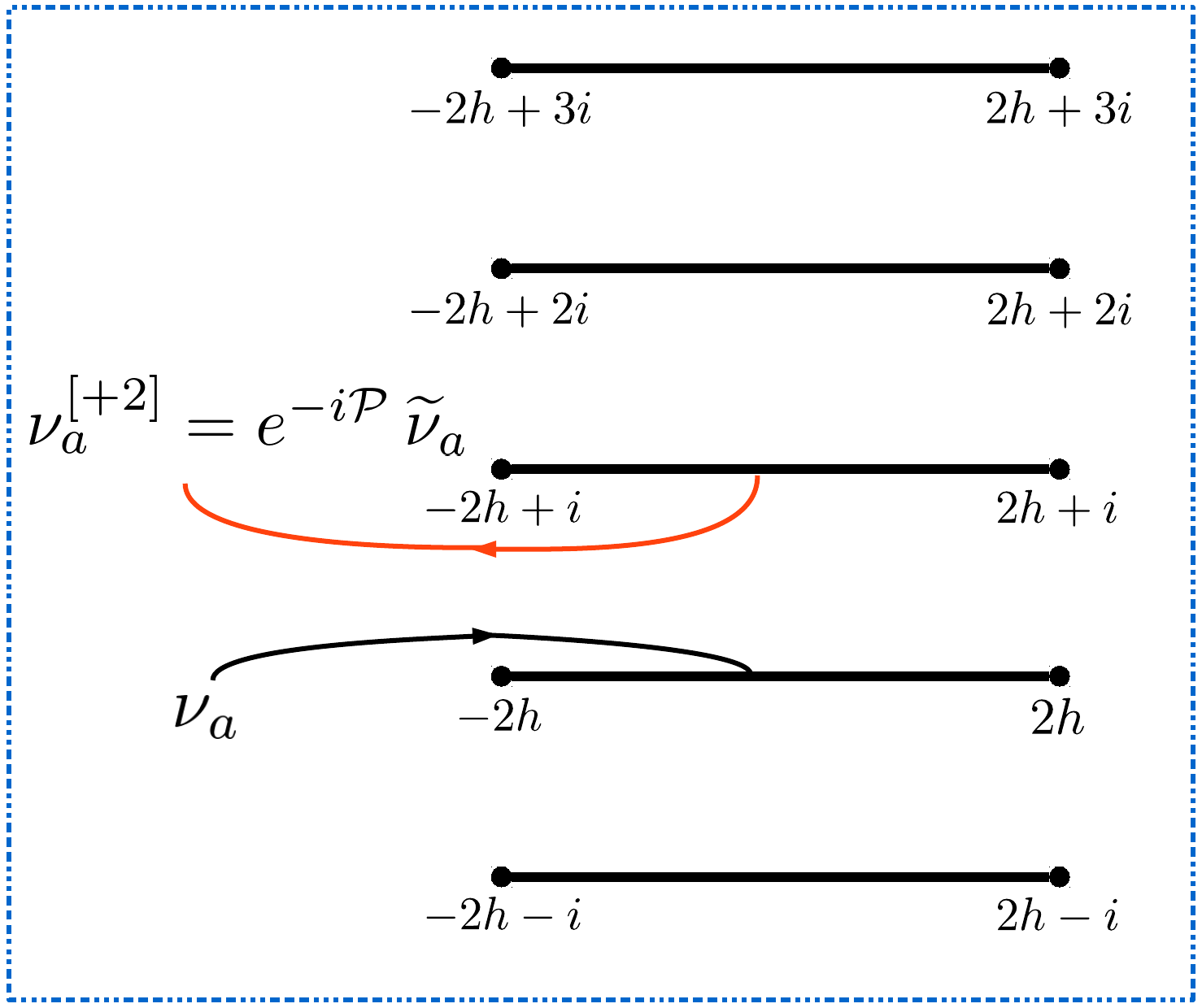}
\vspace{1.15cm}
\caption{The quasi-periodicity property of $\nu_a$ functions on a sheet with long cuts corresponds to $\nu_a(u+i)= e^{-i \sigtw} \, \widetilde\nu_a(u) $ on the defining sheet with short cuts.}
\label{fig:cutNu}
\end{minipage}
\end{figure}

 In the first version of the equations derived from TBA, the basic variables are: six functions $\left\{\bP_A(u)\right\}_{A=1}^6$, and a $6 \times 6$ anti-symmetric matrix $\left\{\mu_{A B}(u) = -\mu_{BA}(u) \right\}_{A,B=1}^6$. 
 They are constrained by the following quadratic conditions: 
\beq
 \bP_5 \bP_6 - \bP_2 \bP_3 + \bP_1 \bP_4 = 1 , 
\quad \quad \quad\mu_{AB} \, \eta^{BC} \, \mu_{CD} \,  = 0 , \label{eq:constraint} 
\eeq
 where $\eta^{AB}$ is defined in (\ref{eq:eta}). 
 All these functions live on an infinite-sheet cover of the $u$-plane, which, however, is built out of a simple set of rules. On what we will consider the first Riemann sheet, the functions $\bP_A(u)$ have a single branch cut, running from $-2 h$ to $+2 h$, see Figure \ref{fig:cutP}. 
 We assume that they have power-like asymptotics at large $u$, which means that they can be written as a Laurent series in the Zhukovsky variable $x(u)$:
 \beq
\bP_A(u)=(x(u))^{-M_A} \, \sum_{n=0}^{\infty} \frac{c_{A,n}}{x^n(u)},\quad \quad \quad x(u)=\frac{\left(u+\sqrt{u- 2h} \sqrt{u+2 h}\right)}{2h}.\label{eq:expP}
\eeq
The functions $\mu_{AB}(u)$ instead display an infinite ladder of branch cuts, at $u \in (-2 h \, , \, +2 \, h ) + i \mathbb{Z}$. They however have the following analyticity property (\emph{mirror periodicity}\footnote{This property means that $\mu_{AB}$ is $i$-periodic on the long-cuts section of the Riemann surface, known as the mirror sheet \cite{QSC}.}):
 \beq
\widetilde{\mu}_{AB}(u) = \mu_{AB}(u+i) , \label{eq:periomu}
 \eeq
where the symbol tilde is used throughout the paper to denote analytic continuation around any of the branch points at $\pm 2 h$ (see Figure \ref{fig:cutP}), while the shift on the rhs is evaluated avoiding all branch cuts. \\
Finally, the discontinuities of $\bP_A$ and $\mu_{AB}$ across the cut on the real $u$-axis are related by 
\beqa
\label{eq:Pmut}
{ \widetilde \bP}_A - \bP_A = \mu_{A B} \, \eta^{BC} \, \bP_C , \;\;\;\;\; {\widetilde \mu }_{AB} - \mu_{AB} = \bP_A\widetilde \bP_B -  \bP_B\widetilde \bP_A .
\eeqa
In addition, as common for the Q functions in integrable models, we should impose a regularity condition for the basic variables $\bP_A$ and $\mu_{AB}$. The precise statement of this condition, however, cannot be formulated in terms of the matrix entries $\mu_{AB}$, but of more fundamental building blocks which we introduce below. 
\subsection{Equations in spinor form}
As already discussed in \cite{Cavaglia:2014exa}, the matrix $\mu_{AB}$ can be decomposed in terms of $4+4$ functions $\nu_a$, $\nu^a$, as\footnote{ Notice that in \cite{Cavaglia:2014exa} a different notation was used and the functions $\nu^a$ were labeled as $\bar{\nu}$, the precise relation being $\left\{ \nu^1 , \nu^2, \nu^3, \nu^4 \right\}^{\text{here}} \, = \, \left\{ -\bar{  \nu }_4 , \bar{\nu}_3 , -\bar{\nu}_2 , \bar{\nu}_1 \right\}^{\text{\cite{Cavaglia:2014exa}} }$.
}
{ \small \beq
\mu_{AB} =
\left(
\begin{array}{cccccc}
 0 & \nu_1 \nu^{4} &- \nu_2 \nu^{3} &  -\nu^3 \nu_{3}-\nu^4 \nu_{4} & -\nu_1 \nu^3 & \nu^4 \nu_2 \\
 -\nu_1 \nu^{4} & 0 & -\nu^3 \nu_3 - \nu_1 \nu^1 & \nu_3 \nu^2  & \nu_1 \nu^2 & \nu^4 \nu_3 \\
 \nu_2 \nu^{3} &  \nu^3 \nu_3 + \nu_1 \nu^1 & 0 & -\nu_4 \nu^1  &\nu^3 \nu_4 & \nu_2 \nu^1 \\
 \nu^3 \nu_{3}  +  \nu^4 \nu_{4} & -\nu_3 \nu^2 & \nu_4 \nu^1 & 0  & -\nu^2 \nu_4 & \nu_3 \nu^1\\
 \nu_1 \nu^3 &  -\nu_1 \nu^2 &  -\nu^3 \nu_4 &  \nu^2 \nu_4 & 0 & -\nu^3 \nu_3 -\nu_2 \nu^2 \\
   -\nu^4 \nu_2 & -\nu^4 \nu_3  &  -\nu_2 \nu^1 & - \nu_3 \nu^1 & \nu_2 \nu^2 + \nu^3 \nu_3 & 0
\end{array}
\right) ,
\label{eq:parametrise}
\eeq
}
which, using the sigma matrices introduced in Section \ref{sec:symmetries}, can be compactly written as
\beq\label{eq:muparametrise}
\mu_{AB}  = \nu^a \; ( \sigma_{AB}  )_{a}^{ \; b} \;\nu_b .
\eeq
The constraint $(\mu \eta )^2 = 0$ is now equivalent to the condition
\beq
\nu^a \, \nu_a = 0.
\label{eq:constraint2}
\eeq
 Motivated by the weak coupling analysis of \cite{Cavaglia:2014exa,Anselmetti:2015mda}, we  will impose that the functions $\nu_a$, $\nu^a$ are analytic on any sheet of the Riemann surface, with the exception of the square-root branch points at $u \in i \mathbb{Z} \pm 2 h$, and that they remain bounded as these points are approached. Besides, for physical values of the charges we assume that $\nu_a(u)$, $\nu^a(u)$ exhibit power-like asymptotics for $u \rightarrow \infty$.  Under these conditions, the splitting (\ref{eq:muparametrise}) contains nontrivial analytic information,  and may be argued to be essentially unique\footnote{ It is unique apart for trivial rescalings $\nu_a \rightarrow \nu_a \, z$, $\nu^a\rightarrow \nu^a/z$, where $z$ is a constant independent of $u$. This freedom is however removed by the choice of the normalization of equations (\ref{eq:Pnu1}),(\ref{eq:Pnu2}) below. }. 
 The new functions $\nu_a$ and $\nu^a$ should therefore be regarded as more fundamental objects than $\mu_{AB}$. Indeed, at weak coupling, $\nu_1$ and $\nu^4$ are proportional to the Baxter polynomials containing the two types of momentum-carrying roots entering the 2-loop Bethe Ansatz of \cite{Minahan:2008hf}, see Appendix \ref{sec:appWC2}. 
 
The weak coupling analysis also reveals that the periodicity of $\mu_{AB}$ on the mirror sheet, equation (\ref{eq:periomu}), in general translates into quasi-periodicity for the basic functions $\nu_a$, $\nu^a$  (see Figure \ref{fig:cutNu}). In the subsector considered in \cite{Anselmetti:2015mda}, these functions could be either periodic or anti-periodic, and this is a general feature of a large sector of states discussed in Section \ref{sec:LR}. For a completely generic state, however, we have\footnote{Notice that $\sigtw$ has to be the same for all the components of $\nu_a$, due to the fact that in (\ref{eq:parametrise}) all combinations of $\nu_a \nu^b$ are present, for every $a$, $b$.}
\beqa\label{eq:perioanti}
{ \widetilde \nu }_a(u)  = e^{i\sigtw} \, \nu_a(u+i) , \;\;\;\;\; { \widetilde \nu }^a(u)  =  e^{-i\sigtw} \, \nu^a(u+i) ,\label{eq:phase}
\eeqa
where the phase $\sigtw$ depends on the state under consideration and may be, in general, a nontrivial function of the coupling constant $h$. We will make more comments on this quantity in Section \ref{subsec:P} below. 
 
 It is now convenient to pack the six $\bP$ functions into an anti-symmetric $4 \times 4$ tensor $\bP_{ab}$, defined as
\beqa\label{eq:Pabdef}
{ \bf P }_{ab} = { \bf P }_A \sigma^A_{ab} =  \left( \begin{array}{cccc} 0 & - \bP_1 & - \bP_2 & - \bP_5 \\  \bP_1 & 0 & - \bP_6 & - \bP_3 \\ \bP_2 & \bP_6 & 0 & - \bP_4 \\ \bP_5 & \bP_3 & \bP_4 & 0 \end{array} \right),
\eeqa
while the inverse matrix reads
\beqa
\bP^{ab} = { \bf P }_A { ({ \bar{\sigma} }^A) }^{ab} &=&  \left( \begin{array}{cccc} 0 & { \bf P}_4 & -\bP_3 & \bP_6 \\ -\bP_4 & 0 & \bP_5 & -\bP_2 \\ \bP_3 & -\bP_5 & 0 & \bP_1 \\ -\bP_6 & \bP_2 & -\bP_1 & 0 \end{array} \right).\label{eq:Pinv0main}
\eeqa
The constraint (\ref{eq:constraint}) can now be rewritten as the condition that $\bP_{ab}$ has unit Pfaffian:
\beq
\label{eq:constrPMu}
\text{Pf}( \bP_{ab} ) = 1 .
\eeq
Besides, it is possible to verify that the discontinuity equations (\ref{eq:Pmut}) can be split nicely as
\beqa
{ \widetilde {\bP} }_{ab} - { \bf P }_{ab} &=& \nu_a { \tilde \nu }_b - \nu_b { \tilde \nu }_a , \;\;\;\;\; { \widetilde {\bP} }^{ab} - { \bf P }^{ab} = -\nu^a { \tilde \nu }^b + \nu^b { \tilde \nu }^a ,
\label{eq:Pnu1}\\
{\tilde \nu }_a &=& -{ \bf P }_{ab} \; \nu^b , \;\;\;\;\; {\tilde \nu }^a = -{ \bf P }^{ab} \; \nu_b .
\label{eq:Pnu2}
\eeqa
As discussed in \cite{Cavaglia:2014exa}, in this form the equations are, from a purely algebraic point of view, exactly the same as the $\bP \mu$-system of $\mathcal{N}=4$ SYM \cite{QSC, Gromov:2014caa}, with the redefinitions
\beq\label{eq:AdS54}
\nu_a \rightarrow  ( \bP_a )^{\text{SYM}} , \;\;\;\;\; \nu^a \rightarrow (\bP^a )^{\text{SYM}} , \;\;\;\;\;  \bP_{ab} \rightarrow (\mu_{ab} )^{\text{SYM}}.
\eeq
 The analytic properties characterizing the $AdS_5/CFT_4$ case are however completely different: the map between the two models in (\ref{eq:AdS54}) requires to change all periodic functions into single-cut functions, and viceversa\footnote{The very existence of this relation is naturally quite surprising and, on the level of pure speculation, one may wonder if the two theories can somehow be connected through a continuous interpolation.}. 
 
  Equations (\ref{eq:constraint2}),(\ref{eq:constrPMu}),(\ref{eq:Pnu1}) and (\ref{eq:Pnu2}) should be supplemented with the requirement that all functions are bounded and free of singularities on every sheet of the Riemann surface, and with some information on their large-$u$ asymptotics, see Section \ref{sec:asyQtau}. 
 This set of conditions is in principle already constraining enough to determine the spectrum, but it is difficult if not impossible to solve in practice at finite coupling. For this purpose it is necessary to embed them in the wider set of equations derived in Sections \ref{sec:QOmega} and \ref{sec:gluing}. 

\subsection{Interpretation of the phase $\sigtw$ at weak coupling}
\label{subsec:P}
The phase $\mathcal{P}$ appearing in (\ref{eq:phase}) has an interesting  interpretation at weak coupling. Recall that the ABJM spin chain admits two types of momentum-carrying excitations \cite{Aharony:2008ug, Ahn:2008aa}, also known as A and B particles and corresponding to excitations of type $4$ and $\bar{4}$ in our notations. These pseudoparticles satisfy collectively the zero momentum condition:
\beq
\sum_{j=1}^{K_4} p_{4, j} + \sum_{j=1}^{K_{\bar{4}}} p_{\bar4, j} = 0 , \;\;\;\;\text{mod}(2 \pi).\label{eq:ZMC0}
\eeq
The total momentum of a single type of excitations is instead in general a nontrivial function of the coupling: it can be defined in the regime of validity of the Asymptotic Bethe Ansatz as
\beq
P^{({4})}_{\text{ABA}} = - P^{(\bar{{4}} )}_{\text{ABA}} = \sum_{j=1}^{K_4} p_{4, j} = - \sum_{j=1}^{K_{\bar{4}}} p_{\bar4, j} ,\;\;\;\;\text{mod}(2 \pi), 
\label{eq:Pmom}
\eeq
where  
\beq
p_{\bu , j} = -i \log( x_{\bu, j}^+/x_{\bu, j}^-), \;\;\;\;x_{\bu, j}^{\pm} = x( u_{\bu, j} \pm i/2 ), \;\; \bu= 4, \bar{4} ,
\eeq
and $\left\{u_{4, j} \right\}_{j=1}^{K_4}$, $\left\{u_{\bar{4}, j}\right\}_{j=1}^{K_{\bar{4}}}$ denote the momentum-carrying Bethe roots, see \cite{Gromov:2008qe}. 
We will show that the phase $\sigtw$ agrees with (\ref{eq:Pmom}) up to the first two orders at weak coupling, 
\beq\label{eq:idePmom}
\sigtw = P^{({4})}_{\text{ABA}} + \mathcal{O}(h^4) .
\eeq
Notice that this also implies that at leading order $\sigtw$ is quantized in units of the spin chain length $L$: $\sigtw + \mathcal{O}(h^2)  \in  2 \pi \mathbb{Z}/L $. 
This is a manifestation of the fact that at weak coupling A and B particles are decoupled on the spin chain and their momenta must be independently quantized. 

At order $\mathcal{O}(h^0)$, the identification (\ref{eq:idePmom})  can be proved to follow directly the analytic properties of the QSC. This is discussed in detail in Appendix \ref{sec:appWC2}, see equation (\ref{eq:proofP0}) there. 
Further, in Section \ref{sec:ABA}, we derive an explicit expression for $\sigtw$  for finite $h$ in the large volume limit -- equation (\ref{eq:PABA}) -- which extends (\ref{eq:idePmom}) up to the next order at weak coupling. 

For a generic short operator at finite coupling, the above mentioned large-volume result is not applicable, and therefore $\sigtw$ is in principle an undetermined, state-dependent function of the coupling. This could raise some questions on the completeness of the system of QSC equations. It is part of our proposal that $\sigtw$ should not be seen as an input, but is rather fully fixed, for every state, from the self-consistency of the QSC. In particular, we expect that this phase can be computed as an output, alongside the anomalous dimension, from the numerical solution of the QSC using the method of \cite{Gromov:2015vua}\footnote{We plan to return on this issue shortly \cite{ContiNumerical}.}. For instance, one method to reconstruct $\sigtw$ exactly in terms of quantities that are easily accessible for the numerical algorithm is presented in Appendix \ref{app:Prec}. It would be interesting to clarify whether this phase admits a  meaningful physical interpretation at finite $h$. 

\section{Construction of the $AdS_4$-related Q functions}\label{sec:QOmega}
As we will discuss in Section \ref{sec:classical}, the equations presented above are associated, in the classical limit, to the $CP^3$ degrees of freedom, and in particular the $\bP_A$ functions are quantum versions of the classical quasi-momenta living in this part of the target space.
 We shall now show how to construct an equivalent version of the QSC which is more appropriate to the description of $AdS_4$ degrees of freedom, and contains, in the classical limit, the four quasi-momenta parametrizing the motion of a  classical string solution in $AdS_4$. As in the case of $AdS_5/CFT_4$ considered in \cite{Gromov:2014caa}, this entails a swap between the \emph{physical} and the \emph{mirror} section of the Riemann surface. In addition, we will see that this alternative system naturally encodes the relevant symmetry group $SO(3,2)$, which was not explicitly visible in the previous formulation.  
 
\subsection{The $Q_{a|i}$ and $\bQ_{ij}$ functions}
It is convenient to introduce the standard notation for shifts of the rapidity variable $u$:
 \beq
F^{[\pm n]}\equiv F\left(u\pm \frac{in}{2}\right)\,;\quad F^{\pm}\equiv F\left(u\pm\frac{i}{2}\right)\,;\quad F^{\pm\pm}\equiv F(u\pm i),
\eeq
where we will always assume that shifts are performed on the section of the Riemann surface where all cuts are short. 

The first step of our construction is the definition of a $4 \times 4$ matrix $Q_{a|i}$, through the 4th order finite difference equation
\beq\label{eq:defQai}
Q_{a|i}^+ = \bP_{ab} \, (\bP^{bc} )^{[-2]} \, Q_{c|i}^{[-3]} .
\eeq 
Notice that exactly the same  equation is satisfied by $\nu_a^+$, as can be verified by combining (\ref{eq:perioanti}) and (\ref{eq:Pnu2}):
\beq\label{eq:4thdiff}
\nu_{a}^{[+2]} = \bP_{ab} \, (\bP^{bc} )^{[-2]} \, \nu_{c}^{[-2]}.
\eeq 
In particular, the index $i$ in (\ref{eq:defQai}) does not enter the matrix structure of the equation. We will take this index to run from $1$ to $4$, labelling a set of independent solutions of this fourth-order equation, distinguished by different asymptotic behaviours at large $u$ (see Section \ref{sec:asymptotics}). Despite the fact that they satisfy the same finite-difference relation, the analytic properties of $\nu_a$ and $Q_{a|i}$ will be different: we shall require that $Q_{a|i}(u)$ has no singularities in the whole region $\text{Im}(u)>0$. Notice that, because of the cut of $\bP_{ab}$ on the real axis, (\ref{eq:4thdiff}) implies that $Q_{a|i}$ has an infinite ladder of short branch cuts in the lower half plane, starting at $\text{Im}(u) = -1/2$. \\
It will be convenient to define $Q^{a}_{\,|i} \equiv (\bP^{ab} )^- \, (Q_{b|i} )^{[-2]}$, so that (\ref{eq:defQai}) can be split as
\beq\label{eq:eq1}
Q_{a|i}^+=\bP_{ab} \, (Q^b_{\,|i} )^-, \;\;\;\;\; (Q^{a}_{\,|i} )^+ = \bP^{ab} \, Q_{b|i}^-.
\eeq
Now, let us construct the tensor
\beq\label{eq:chidef}
  k_{ij} \equiv Q_{a|i}^+ \, (Q^{a}_{\,|j} )^+ = Q_{a|i}^+ \, \bP^{ab} \, Q_{b|j}^-.
\eeq
Using (\ref{eq:eq1}), it is simple to see that $k_{ij}$ is invariant under a shift $u \rightarrow u + 2i$, and,  since by construction it is  free of cuts in the upper half plane and has power-like asymptotics, it must be a constant matrix. 
In addition, notice that (\ref{eq:eq1}) implies more precisely that $k_{ij}^+ = - k_{ji}^-$, so that $k_{ij}$ is an anti-symmetric matrix, i.e. a symplectic form. This shows that the space of the $i$-indices should be thought as carrying the fundamental representation of $Sp(4) \simeq SO(3,2)$, the isometry group of $AdS_4$. 
It is very pleasing that this symmetry, while completely hidden at the level of the equations discussed in Section \ref{sec:PMuPNu}, naturally emerges from the construction. 

From (\ref{eq:chidef}) we see that the specific form of $k_{ij}$ can be adjusted by taking different linear combinations of the columns of the matrix $Q_{a|i}$ (we are allowed to do this since the defining relation (\ref{eq:defQai}) is linear). We use this freedom to impose that $k_{ij} = \ch_{ij}$ as defined in (\ref{kijdef}). Note in particular that\footnote{This concrete choice is purely conventional, however notice that a different value for the Pfaffian of $\ch_{ij}$ would affect some of the equations below.} $\text{Pf}( \ch_{ij} ) = -1$.
  
Using (\ref{eq:chidef}), we can relate $Q^a_{\,|i}$ to the inverse transposed matrix of $Q_{a|i}$:
\beqa\label{eq:Qtrans}
Q^{a}_{\, |i} = Q^{a|j} \, \ch_{ji} ,
\eeqa
where $Q^{a|i} \equiv ( Q^{-T} )^{a|i}$, such that $Q_{a|j} \, Q^{a|j} = \delta_i^j$, $Q_{a|i} \, Q^{b|i} = \delta_a^b$. 
Another simple consequence of (\ref{eq:defQai}) is that the determinant $\text{det}(Q_{a|i})$ is invariant under shifts of $+2 i$; by the same arguments as above, it also must be a constant independent of $u$. Considering the Pfaffian of equation (\ref{eq:chidef}) and using the property $\text{Pf}( A^t B A ) = \text{det}(A) \, \text{Pf}( B )$, we see that
\beq
\text{det}\left( Q_{a|i} \right) = \text{det}\left( Q^{a}_{\,|i} \right)= \text{Pf}\( \ch_{ij} \) = -1 .
\eeq
We proceed now to construct an object whose indices live in the product of two $Sp(4)$ representations, as
\beq
\label{Qijdef}
\bQ_{ij} =  (Q^{a}_{\,|i})^+ \, Q_{a|j}^-  =  ( Q^{a}_{\,|i} )^+ \, \bP_{ab} \, (Q^b_{\,|j} )^+ .
\eeq
Let us discuss the algebraic properties of this tensor. First, from (\ref{Qijdef}), we see immediately that
\beq
\bQ_{ij} = -\bQ_{ji}, \;\;\;\;\; \text{Pf}(\bQ_{ij} ) = -1.\label{eq:PfQ}
\eeq
Being a $4\times 4$ anti-symmetric matrix, $\bQ_{ij}$ has six independent components. It will be convenient to decompose it into $\bf{5}$+$\bf{1}$-dimensional irreducible representations of $SO(3,2)$ using the invariant tensor $\ch$: the trivial representation is given by the trace
\beqa
\bQ_{\circ} = \bQ_{ij} \, \ch^{ij} = Q_{a|i}^- \, (Q^{a|i} )^+ ,
\eeqa
while the five dimensional vector representation is the traceless part:
\beqa
\bQ_{ij}^{\bf{5}} = \bQ_{ij} + \frac{1}{4} \, \ch_{ij} \, \bQ_{\circ} .\label{eq:Q5vecmain}
\eeqa
 The inverse matrix $\bQ^{ij}$, satisfying $\bQ_{ij} \, \bQ^{jk} = \delta_j^k$, can be computed as 
\beqa
\bQ^{ij} &=& 
\ch^{i i_1} \, \ch^{j i_2} \, ( Q_{a|i_1} )^+ \, \bP^{ab} \, (Q_{b|i_2} )^+ \\
&=& - ( Q^{a|i} )^- \, \bP_{ab} \, (Q^{b|j} )^-,
\eeqa
and it is simple to show (see Appendix \ref{app:QQinv}) that the following identity holds
\beqa
\bQ^{ij} =\ch^{i i_1} \, \ch^{j i_2} \, \bQ_{i_1 i_2} - \frac{\ch^{ij}}{2} \, \bQ_{\circ} .\label{eq:QQhatmain}\label{eq:QQinvmain}
\eeqa
Finally, the following relations constitute a natural  counterpart of (\ref{eq:eq1}) involving the $Sp(4)$-invariant indices:
\beqa\label{eq:translQ}
Q_{a|i}^+ = -Q_{a|j}^- \, \bQ^{jk} \, \ch_{ki} , \;\;\;\;\; (Q^{a}_{\,|i})^+ = -(Q^{a}_{\,|j})^- \, \ch^{jk} \, \bQ_{ki}.
\eeqa
Shortly, we will show that the elements $\bQ_{ij}$ have very simple analytic properties: starting from the upper half plane, they can be analytically continued to a Riemann section with the only branch cuts being the semi-infinite segments $(-\infty, -2h)$ and $(2h, \infty)$. 

\subsection{The $\tau_i$ functions}
We now construct a new set of four functions, denoted as $\tau_i$ and defined as
\beq\label{eq:TAUdef}
\tau_i = \nu^a \, Q_{a|i}^- . 
\eeq
Manifestly, these quantities exhibit an infinite series of short branch cuts. Applying (\ref{eq:eq1}) and (\ref{eq:perioanti}), we see that, under a shift $u \rightarrow u + i$, they transform as
\beq\label{eq:halfperiodtau}
\tau^{[+2]}_i = Q_{a|i}^{[+]} \, (\nu^a)^{[+2]} = \bP_{ab} \, (Q^{b}_{\,|i})^{-} \, ( - e^{i \sigtw} \, \bP^{ac} \, \nu_c )= e^{i \sigtw} \, \nu_a \, ( Q^{a}_{\,|i} )^-,
\eeq
and shifting this expression once more we find that $\tau_i$ are $2 i$-periodic on the Riemann section with short cuts:
\beq\label{eq:periotau}
\tau^{[+4]}_i = \tau_i .
\eeq
The $\tau_i$ functions may be seen as counterpart of the $\nu_a$ functions. Their analytic properties are very similar, with a characteristic swap of short and long cuts. However, notice that, while the functions $\nu_a$ and $\nu^a$  are distinct objects, carrying different irreps of $SO(3,3)$, there are only four independent functions $\tau_i$, corresponding to the spinor representation of $SO(3,2)$.

\subsection{The $\bQ\tau$-system} 
The functions $\bQ_{ij}(u)$ introduced above have, by their very definition, no singularities in the upper half plane, with two branch points at $u=\pm 2 h$ and an infinite ladder of short cuts further down in the lower half plane. 

Let us study the analytic continuation of $\bQ_{ij}$ and $\tau_i$ through the branch cut on the real axis. 
Combining (\ref{eq:periotau}) and (\ref{eq:halfperiodtau}), we have
\beqa\label{eq:taudef}
\tau_i =  e^{i\sigtw} \, \nu_a^{[+2]} \, ( Q^{a}_{\,|i} )^+ = \widetilde{\nu}_a \, ( Q^{a}_{\,|i} )^+ , 
\eeqa
and, since $Q^a_{\,|i}$ has no cuts in the upper half plane, we find
\beqa\label{eq:tautilde}
\widetilde{\tau}_i = \nu_a \, ( Q^{a}_{\,|i} )^+ = - \nu_a \, ( Q^{a}_{\,|j} )^- \, \ch^{jk} \, \bQ_{ki},
\eeqa
where we used (\ref{eq:translQ}) in the last step. By comparison with (\ref{eq:taudef}), we see that (\ref{eq:tautilde}) can be rewritten as
$ \widetilde{\tau}_i = - \bQ_{ij} \, \tau^ j $, 
where we have defined
\beqa\label{eq:tauUP}
 \tau^i \equiv e^{-i \sigtw} \, \ch^{ij} \, {\tau}_j^{[+2]}. 
 \eeqa 
Let us now consider the discontinuity of $\bQ_{ij}$: we find
\beqa\label{eq:Qijtilde}
 \widetilde{\bQ}_{ij} - \bQ_{ij} &=& ( Q^{a}_{\,|i} )^+ \, \left( \widetilde{\bP}_{ab}- \bP_{ab} \right)\, (Q^b_{\,|j} )^+ \nn\\
  &=&\left( (Q^{a}_{\,|i} )^+ \, \nu_a \right) \, \left( \widetilde{\nu}_b \, (Q^b_{\,|j} )^+ \right)- \left( (Q^{a}_{\,|i} )^+ \, \widetilde{\nu}_a \right) \, \left( {\nu}_b \, (Q^b_{\,|j} )^+ \right)\nn\\
  &=&  \widetilde{\tau}_i \, \tau_j- \widetilde{\tau}_j \, \tau_i .
\eeqa
All in all, we see that the discontinuities (\ref{eq:tautilde}) and (\ref{eq:Qijtilde}) take the form
\beqa\label{eq:QtilTautil}
&& \widetilde{\bQ}_{ij} - \bQ_{ij} =  \widetilde{\tau}_i \, \tau_j- \widetilde{\tau}_j \, \tau_i , \;\;\;\;\;\;\; 
\widetilde{\tau}_i = - \bQ_{ij} \, \tau^j . 
\eeqa
The second relation in (\ref{eq:QtilTautil}) shows how the phase $\sigtw$ appears in the $\bQ\tau$-system, through (\ref{eq:tauUP}).
 Finally, contracting (\ref{eq:TAUdef}) and (\ref{eq:halfperiodtau}) with $\ch^{ij}$, we find the constraint 
\beqa\label{eq:constrQTau}
\tau_i \, \tau^i = e^{-i \sigtw} \, \tau_i \, \ch^{ij} \, \tau_j^{[+2]} = -\nu_a \, \nu^a = 0.
\eeqa  
Equations (\ref{eq:QtilTautil}), with the constraints (\ref{eq:constrQTau}), (\ref{eq:PfQ}) may be considered as a counterpart of the $\bP\nu$-system (\ref{eq:constraint2}),(\ref{eq:constrPMu})-(\ref{eq:Pnu2}). 
While the equations take a very similar form, they are not identical from the algebraic point of view, due to the fact that the functions $\tau_i$ and $\tau^i$ are simply related, for a generic state, by a shift in the spectral parameter, as expressed by (\ref{eq:tauUP}). This distinction reflects the representation theory, as there is only one four-dimensional representation of $Sp(4)$. The difference can be fully appreciated by projecting the $\bQ \tau$ equations on irreducible representations; this is discussed below in Section \ref{sec:vectorQ}.
\begin{figure}[t!]
\begin{minipage}{0.5 \linewidth}
\centering
\centering
\includegraphics[scale=0.3]{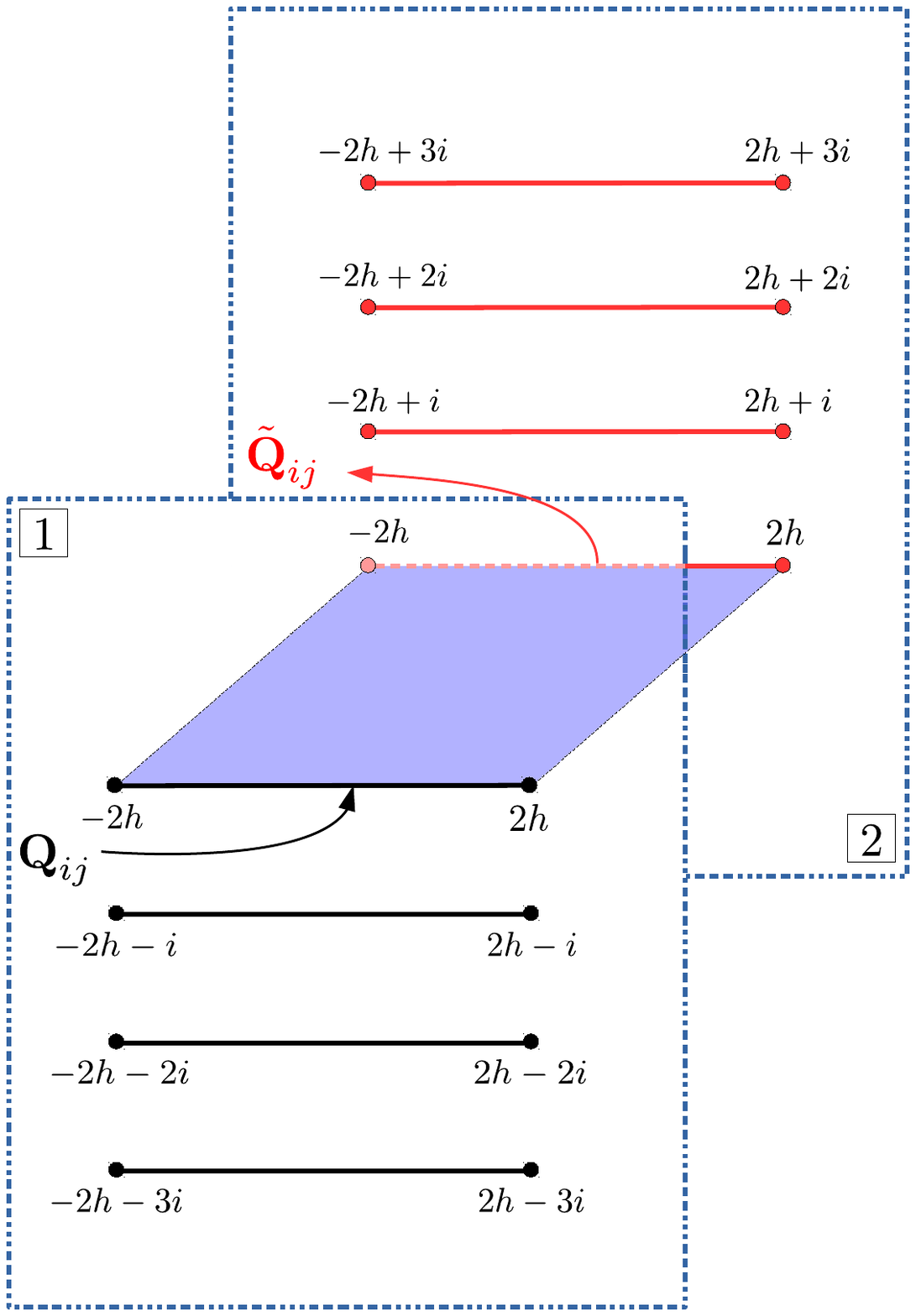}
\caption{Cut structure of the $\bQ$ functions in the physical Riemann section. On the first (second) sheet, $\bQ$ is analytic in the upper (lower) half plane.}
\label{fig:cutQ}
\end{minipage}
\hspace{0.1cm}
\begin{minipage}{0.5 \linewidth}
\includegraphics[scale=0.3]{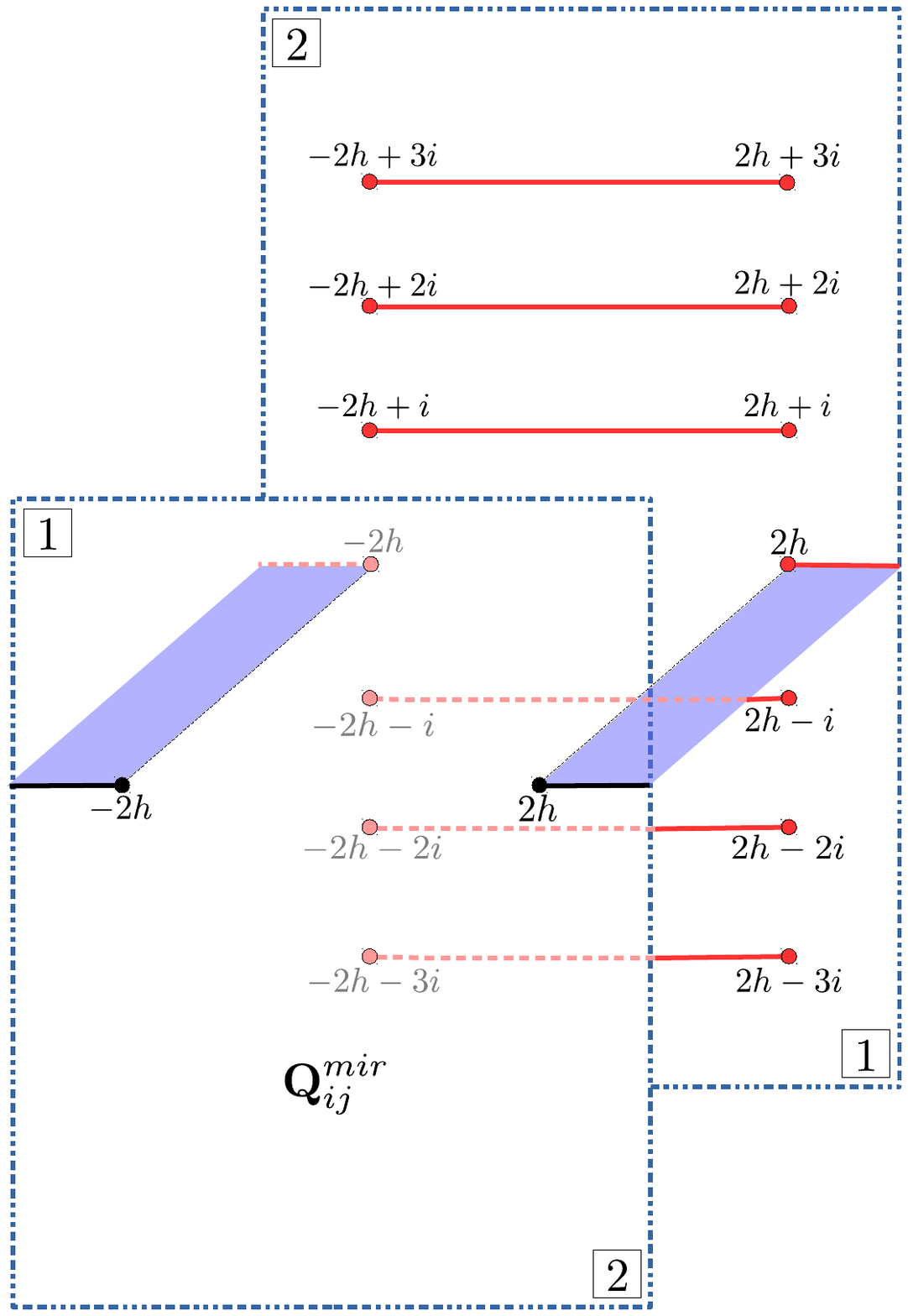}
\caption{Gluing the two analyticity regions from the sheets $1$ and $2$ of Figure \ref{fig:cutQ}, one defines the \emph{mirror} sheet, with a single long cut.}
\label{fig:cutQijmir}
\end{minipage}
\end{figure}

\subsubsection{$\bQ_{ij}$ on the mirror sheet}
Let us now prove that, when analytically continued from the upper to the lower half plane passing through the cut $(-2 h, 2 h)$, the matrix $\bQ_{ij}$ is analytic in the whole lower half plane (see Figure \ref{fig:cutQ}). Therefore, on an appropriate Riemann section, it has only a pair of long cuts stretching from $\pm 2 h$ to infinity (see Figure \ref{fig:cutQijmir}). This is a very strong analogy with the $AdS_5/CFT_4$ case considered in \cite{QSC}. 
 
We start by observing that, using (\ref{eq:constrQTau}) and the second equation in (\ref{eq:QtilTautil}), the discontinuity relation (\ref{eq:Qijtilde}) can be put in the form
\beq
\widetilde{\bQ}_{ij} = \bQ_{mn} \, \( \delta_i^m - \tau_i \, \tau^m  \) \, \( \delta_j^n - \tau_j \, \tau^n \) \equiv \bQ_{mn} \, f_i^m \, f_j^n ,
\label{eq:fdef}
\eeq
where we have defined a $2i$-periodic matrix function $f_i^j \equiv \delta_i^j - \tau_i \, \tau^j$. This relation can be recast as
\beq\label{eq:QTilLHPA}
\widetilde{\bQ}_{ij} = \( Q_{b|m}^- \, \bP^{ab} \, Q_{a|n}^- \) \,  \, f_i^m \, f_j^n  = \bP^{ab} \,  (Q_{b|i}^{\text{LHPA}} )^- \, (Q_{a|j}^{\text{LHPA}} )^- , 
\eeq
where
\beq\label{eq:QLHPA}
Q_{a|i}^{\text{LHPA}} \equiv Q_{a|j} \, ( f^j_i )^{+} = Q_{a|i} - Q_{a|j} \, (\tau^j )^{+} \, \tau_i^{+}  = Q_{a|i} + \nu_a^+ \, (\nu^b)^+  \, Q_{b|i} .
\eeq 
We will now show that $Q_{a|i}^{\text{LHPA}}$ has no branch cuts in the lower half plane (hence the superscript LHPA -- Lower Half Plane Analytic). Therefore, the representation (\ref{eq:QTilLHPA}) manifestly shows that the same is true for $\widetilde{\bQ}_{ij}$,  implying that $\bQ$ has a single  long cut on the mirror Riemann sheet.
 
To prove that $Q_{a|i}^{\text{LHPA}}$ has no cuts in the lower half plane,  we can exploit the fact that, due to the periodicity of $f_i^j(u)$,  it satisfies the same fourth order difference equation (\ref{eq:defQai}) fulfilled by $Q_{a|i}$. 
 Therefore, it is sufficient to check that it has no cut on the lines $\text{Im}(u) = -1/2$, $-3/2$: the difference equation (\ref{eq:defQai}) will then automatically imply that it is analytic everywhere in the lower half plane. This leaves us with just two conditions to check. The first discontinuity to study is 
\beq\label{eq:target1}
\DD( \, (Q_{a|i}^{\text{LHPA}})^- \,) = \DD( \, Q_{a|i}^- - Q_{a|j}^- \, \tau^j \, \tau_i \,),
\eeq
where we are using the notation $\DD( \mathcal{G} ) = \mathcal{G} - \widetilde{ \mathcal{G}}$. From the first relation in (\ref{eq:translQ}), we find
\beqa
\DD( Q_{a|i}^- ) &\equiv& Q_{a|i}^- - \widetilde  Q_{a|i}^- = - Q_{a|k}^+ \,  \ch^{kl} \, \( \bQ_{li} - \widetilde \bQ_{li} \)\\
 &=& - Q_{a|k}^+ \,  \ch^{kl} \, \( \tau_l \, \widetilde \tau_i - \tau_i \, \widetilde \tau_l \),\label{eq:passage2}
\eeqa
where we used (\ref{eq:QtilTautil}) in the last step. 
We may now to use the following identities, found by inverting (\ref{eq:taudef}),(\ref{eq:tautilde}): 
\beq\label{eq:nutau}
\nu_a= - Q_{a|i}^+ \, \ch^{ij} \, \widetilde \tau_j, \;\;\;\;\;\; 
\nu_a= -Q_{a|i}^- \tau^i ,
\eeq
 to transform (\ref{eq:passage2}) into
\beq
 \DD(Q_{a|i}^- ) = \widetilde \nu_a \, \widetilde \tau_i -\nu_a \, \tau_i = -\DD ( \nu_a \, \tau_i )  = \DD( Q_{a|j}^- \, \tau^j \, \tau_i ).
\eeq
The last equality shows the vanishing of the discontinuity  (\ref{eq:target1}). 
A completely analogous calculation would show that
\beq
\DD\[  (Q^a_{\,|j})^{-} \, (f^j_i )^{[-2]}  \] = 0 \label{eq:cutQai2},
\eeq
therefore also the next discontinuity is trivial
\beq
\DD\[ \, (Q_{a|i}^{\text{LHPA}} )^{[-3]} \, \] = \bP_{ab}^{[-2]} \; \DD\[  (Q^a_{\,|j})^{-} \; (f^j_i )^{[-2]} \] = 0 , 
\eeq
which concludes the proof.

\subsubsection{Vector form of the $\bQ\tau$-system}	
\label{sec:vectorQ}
We may rewrite the discontinuity equations (\ref{eq:QtilTautil}) in an alternative form, more similar to the $\bP \mu$-system. To do this, let us rearrange the components of $\bQ_{ij}^{\bf{5}}$ into a five-vector:
 \beq\label{eq:defQI}
 \bQ_I(u) \equiv -\frac{1}{2} \,\( \bQ_{ij}^{\bf{5}}(u) \, \bar{\Sigma}^{ij}_I\) , \;\;\; (I=1, \dots, 5 ) ,
 \eeq
 or equivalently
 \beq
 \bQ_{ij}^{\bf{5}}(u) = (\Sigma_I )_{ij} \, \rho^{IJ} \, \bQ_J(u) , \label{eq:Qtovec}
 \eeq
 where we are using the matrices $\Sigma^I$ and the metric $\rho^{IJ}$ defined in Section \ref{sec:symmetries}. In components, this definition reads 
 \beqa
\bQ_I &=&-\( \bQ_{12} , \bQ_{13} ,  \bQ_{24} , \bQ_{34} , \frac{1}{2} \, (\bQ_{14} + \bQ_{23}  )   \) ,\\
 \bQ_{ij}^{\bf5} &=&  \left( 
 \begin{array}{ccccc}
 0 & -\bQ_1 & -\bQ_2 &  -\bQ_5 \\
 \bQ_1 & 0 &  -\bQ_5 & -\bQ_3 \\
 \bQ_2 &  \bQ_5 & 0 & -\bQ_4 \\
  \bQ_5 & \bQ_3 & \bQ_4 & 0
\end{array} 
\right).
 \eeqa
 It is also convenient to define 
\beqa
\omega_{IJ}(u) &\equiv& \tau^k(u) \, ( \Sigma_{IJ} )_k^{\, i} \, \tau_i(u) , \;\;\;\;\;\;
\psi_I(u) \equiv  {\tau}^m(u) \, \ch_{mi}  \, \bar{\Sigma}_I^{ij} \, \tau_j(u) ,
\eeqa
or explicitly:
{\small
\beq
\omega_{IJ} = \left(
\begin{array}{ccccc}
 0 & \tau_1 \tau^{4} &- \tau_2 \tau^{3} &  -\tau^3 \tau_{3}-\tau^4 \tau_{4} & \frac{1}{2}( \tau_2 \tau^4-\tau_1 \tau^3) \\
 -\tau_1 \tau^{4} & 0 & -\tau^3 \tau_3 - \tau_1 \tau^1 & \tau_3 \tau^2  & \frac{1}{2}(\tau_1 \tau^2 + \tau_3 \tau^4) \\
 \tau_2 \tau^{3} &  \tau^3 \tau_3 + \tau_1 \tau^1 & 0 & -\tau_4 \tau^1  &\frac{1}{2}(\tau_2 \tau^1 + \tau_4 \tau^3)\\
 \tau^3 \tau_{3}  +  \tau^4 \tau_{4} & -\tau_3 \tau^2 & \tau_4 \tau^1 & 0  & \frac{1}{2}(\tau^1 \tau_3-\tau^2 \tau_4)\\
\frac{1}{2}(\tau_1 \tau^3 - \tau_2 \tau^4)  &  -\frac{1}{2}(\tau_1 \tau^2 + \tau_3 \tau^4) &  -\frac{1}{2}(\tau_2 \tau^1 + \tau_4 \tau^3)& \frac{1}{2}(\tau^2 \tau_4 - \tau^1 \tau_3) & 0 
\end{array}
\right),
\eeq}
\beq
\psi_I= \( -\tau_1 \tau^3 - \tau_2 \tau^4 ,\, \tau_1 \tau^2 - \tau_3 \tau^4  ,\, - \tau_2 \tau^1 + \tau_4 \tau^3 , \,-\tau^2 \tau_4 - \tau^1 \tau_3 ,\, \tau_2 \tau^2 + \tau_3 \tau^3  \).
\eeq
From (\ref{eq:periotau}),(\ref{eq:tauUP}), it is simple to prove  that the components of $\omega_{IJ}(u)$ are $i$-periodic functions, while the components of $\psi_I$ are anti-periodic under the same shift: 
\beq
\omega_{IJ}^{[+2]} = \omega_{IJ} , \;\;\;\;\; \psi_I^{[+2]} = -\psi_I .
\eeq
In terms of these new variables, the nonlinear constraints (\ref{eq:PfQ}),(\ref{eq:constrQTau}) take the form
\beq\label{eq:constrQvec}
\frac{ \bQ_{\circ}^2 }{16} -1  = \bQ_5^2 - \bQ_2 \, \bQ_3 + \bQ_1 \, \bQ_4  \; , \;\;\;\;\;\;\;\; \omega_{IJ} \, \rho^{JK} \, \omega_{KL} = -\frac{1}{2}\, \psi_I \, \psi_L  \; , \;\;\;\;\;\;\;\; \psi_I \, \rho^{IJ} \, \psi_J = 0 \; ,
\eeq
while the discontinuity equations (\ref{eq:QtilTautil}) can be rewritten as
\begin{align*}\label{eq:Qomega}
\widetilde{\bQ}_I - \bQ_I &= - \omega_{IJ} \; \rho^{JK} \,  \bQ_K  + \frac{1}{4} \, \psi_I \, \bQ_{\circ} ,& \widetilde{\omega}_{IJ}-\omega_{IJ} &= \bQ_I \, \widetilde{ \bQ}_{J} -  \bQ_{J} \, \widetilde{ \bQ}_{I}  , \\
\widetilde{\bQ}_{\circ} - {\bQ}_{\circ} &= 2\, \psi_J \, \rho^{JK} \, \bQ_K, & \widetilde{\psi}_I - \psi_I  &= 
\frac{1}{2} \, \( \bQ_I \, \widetilde{ \bQ}_{\circ} -  \bQ_{\circ} \, \widetilde{ \bQ}_{I}  \)
.
\end{align*}
 
\subsection{Reduction to $4 \leftrightarrow \bar{4}$ symmetric states}\label{sec:LR}
In this Section we consider the reduction of the QSC equations to a large subsector characterized by perfect symmetry between the contributions of A- and B-type excitations. In terms of the ABA, this subsector is characterized by the equality of the sets of momentum-carrying Bethe roots, $\left\{u_{4, k} \right\}_{k=1}^{K_4} = \left\{u_{\bar{4}, k} \right\}_{k=1}^{K_{\bar{4}} } $. 
As discussed in Appendix \ref{app:derivations}, this case is selected by the conditions: 
\beq\label{eq:nusym}
 \bP_5=\bP_6,\quad \nu^a = \ch^{ab} \nu_b .
\eeq
In this case we have the relation $\bP^{ab} = \ch^{al} \,\bP_{lm} \, \ch^{mb} $ and we see that necessarily, $e^{i\sigtw}$ is either $1$ or $-1$. By studying the large-$u$ asymptotics of equation (\ref{eq:defQai}), we find that, in this case, the elements of the matrices $Q_{a|i}$, $Q^{a|i}$ may be chosen as related by the symmetry:
\beq\label{eq:Qsym}
Q^{a}_{\,|i} = -e^{i \sigtw} \, \ch^{ab} \, Q_{b|j} \, \mathbb{K}^{j}_i , 
\eeq
with 
\beq
\mathbb{K}^{i}_j = \left(
\begin{array}{cccc}
 1 & 0 & 0 & 0 \\
 0 & -1& 0 & 0 \\
 0 & 0 &  -1& 0 \\
 0 & 0 & 0 & 1
\end{array}\right).
\eeq
This means also that
\beq
Q_{a|i} \,\ch^{ab} \, Q_{b|k} \, \hat{\ch}^{kl} = \delta_i^{\,l} ,
\eeq
where $\hat{\ch}^{ki} \equiv- e^{i \sigtw} \,  (\kappa \, \mathbb{K} )^{ki}=- e^{i \sigtw} \, (\mathbb{K} \, \ch)^{ki}$. 
The symmetry imposes the following condition:
\beq
\bQ_{ij} = - \mathbb{K}_i^{k_1} \, \bQ_{k_1 k_2} \, \mathbb{K}^{k_2}_j  -\frac{\ch_{ij}}{2} \, {\bQ}_{\circ}, \;\;\;\;\; 
\eeq
which implies
\beq
\bQ^{\bf{5}}_{ij} = - \mathbb{K}_i^{k_1} \, \bQ^{\bf{5}}_{k_1 k_2} \, \mathbb{K}^{k_2}_j  .
\eeq
Taking (\ref{eq:nusym}),(\ref{eq:Qsym}) into account in (\ref{eq:halfperiodtau}), we see that in this subsector the periodicity of $\tau_i$ is  enhanced to 
\beq
\tau_i^{[+2]} =  \tau_k \, \mathbb{K}^k_i ,
\eeq
which means that $\tau_1$ and $\tau_4$ are $i$-periodic, while $\tau_2$, $\tau_3$ are $i$-anti-periodic. Since we expect all these functions to have power-like asymptotics for physical operators, we see, from the condition of anti-periodicity, that
\beq
\lim_{u \rightarrow \pm \infty} \tau_2 = \lim_{u \rightarrow \pm \infty} \tau_3 = 0 .
\eeq
This resut will be important in the following.  
Finally, in terms of the variables of Section \ref{sec:vectorQ}, the reduction to the symmetric subsector can be obtained setting $\bQ_5=\psi_5=\omega_{5I}=\omega_{I 5} =0$. 

\section{Asymptotics and global charges}\label{sec:asymptotics}
\label{sec:asyQtau}
\subsection{Large-$u$ behaviour and quantum numbers}
The Riemann-Hilbert type equations  described in Sections \ref{sec:PMuPNu} and \ref{sec:QOmega} have to be supplemented with appropriate constraints on the large-$u$ behaviour of the functions entering the QSC. We will assume, in analogy with \cite{Gromov:2014caa}, that all the functions we have described scale as powers of $u$ for large values of the spectral parameter, in particular 
\beq \label{eq:Pasy}
\bP_A(u) \sim \mathcal{A}_A \, u^{-M_A} . 
\eeq
An important observation is that, since the $\bP$ functions have a single short  cut on the first Riemann sheet, they must have  trivial monodromy around infinity, which forces $M_A \in \mathbb{Z}$. For the spectrum problem, we found 
that these parameters should be paired up as\footnote{This two-by-two pairing of the charges is equivalent to requiring that all terms in the equation (\ref{eq:constraint}) are of the same order at large-$u$. We suspect that relaxing this condition, without modifying the power-like character of the asymptotics, may lead only to trivial or singular solutions of the QSC equations. 
} $M_1=-M_4$, $M_2=-M_3$, $M_5=-M_6$. The three independent integer parameters contained in the asymptotics (\ref{eq:Pasy}) can be identified with the three $SO(6)$ R-charges $J_1, J_2, J_3$, corresponding to three angular momenta parametrizing the motion of the string in $CP^3$:
\beq
M_A = \(J_2+1  , \,J_1, \, -J_1, \,-J_2-1, \, -J_3, \, J_3\).
\label{MAdef}
\eeq
The $AdS_4$ charges $\Delta$ and $S$, corresponding to the conformal dimension and spin of the gauge theory operator, respectively, enter the QSC through the asymptotics of the $\nu_a$ functions. Equivalently, they can be read off the coefficients $\mathcal{A}_A$ in (\ref{eq:Pasy}), which satisfy the constraints 
\beq\label{eq:AAasy}
\mathcal{A}_B \, \mathcal{A}^B = 2 \, \frac{\prod_{I=1}^5 \left(M_B - \hat{M}_I \right) }{ \prod_{C \neq B }^6 (M_B-M_C ) }, \;\;\;\; (B=1, \dots, 6),
\eeq
(with no summation implied on the index $B$), where the 5-vector $\hat{M}$ is defined as
\beq
\hat{M}_I =  \left(\Delta + S+1 \,,\, \Delta - S \, ,\, -\Delta + S \, ,\, -\Delta -S-1 \, , \, 0\right).
\label{MIdef}
\eeq
The above identifications (\ref{MAdef}),(\ref{MIdef}) between parameters and quantum numbers will be deduced in Appendix \ref{sec:appWC} considering the weak coupling limit of the QSC equations. 
 Notice that the charges $(\Delta, S, J_1, J_2, J_3)$ used above are defined relatively to the Dynkin diagram of Figure \ref{fig:dynkinetap1}. We remind the reader that, for supersymmetric algebras, the definition of the  charges 
depends on a choice of grading of the Dynkin diagram; if a different grading were chosen, relations (\ref{MAdef}) and (\ref{MIdef}) would be slightly different. 
However, we stress that the parameters $M_A$ and $\hat M_I$ appearing in the  asymptotics of the QSC are invariant under these changes, and unambiguously associated to a given multiplet (see \cite{Gromov:2014caa} for a detailed discussion). Concretely, we may read the charges from the Asymptotic Bethe Ansatz description of the state:  
\beqa
&J_1=L-K_1, \quad J_2= L-K_4 -K_{\bar{4}}+K_3 , \quad J_3= K_{4} - K_{\bar{4}},& \label{eq:chargesexcitations1}\\
&\Delta-S=L+K_2-K_1 + \gamma, \quad \Delta+S=L+K_3 - K_2 + \gamma ,&\label{eq:chargesexcitations2}
\eeqa
where $L$ is the length parameter and $K_i$ denotes the number of Bethe roots of type $i$ in the so-called $\eta=+1$ version of the ABA \cite{Gromov:2008qe}, while $\gamma$ is the anomalous dimension. For more details and a dictionary between different forms of the ABA, see Appendix \ref{app:dictionary}. 

The large-$u$ asymptotics of the matrix $Q_{a|i}(u)$ may be determined by studying (\ref{eq:defQai}). There are four possible asymptotic behaviours where $Q_{a|i}$ scales as a power of $u$, parametrized in terms of the charges $M_A$, $\hat{M}_I$ entering the equation through (\ref{eq:Pasy}),(\ref{eq:AAasy}). 
 By choosing a suitable linear combination of solutions, we shall impose that different columns of $Q_{a|i}$ have distinct leading asymptotics, ordered in such a way that $|Q_{a|i}| > |Q_{a|j}|$ for $i < j$ for large $u$. To describe the possible scaling behaviours, it is convenient to introduce:
\beqa\label{eq:NaNi}
\mathcal{N}_a &=& \( \frac{1}{2} (-M_1-M_2 -M_5), \frac{1}{2} (-M_1+M_2+M_5), \frac{1}{2}(M_1-M_2 + M_5), \frac{1}{2} ( M_1+M_2 - M_5 ) \), \nn\\
\mathcal{N}^a &=& \( \frac{1}{2} (M_1+M_2 + M_5), \frac{1}{2} ( M_1-M_2-M_5), \frac{1}{2}(- M_1+ M_2 -  M_5), \frac{1}{2} ( - M_1- M_2 + M_5 ) \), \nn\\
\hat{\mathcal{N}}_i &=& \( \frac{1}{2} (\hat M_1 + \hat M_2 ), \frac{1}{2} (\hat M_1-\hat M_2 ), \frac{1}{2}(\hat M_2 - \hat M_1 ), \frac{1}{2} (-\hat M_1-\hat M_2) \).
\eeqa
With these definitions, we have
\beq
\bP_{ab}(u) \sim u^{ \mathcal{N}_a + \mathcal{N}_b}, \;\;\;\;\; Q_{a|i}(u) \sim u^{\mathcal{N}_a + \hat{\mathcal{N}}_i } , \;\;\;\;\;\; Q^{a}_{\,|i}(u) \sim u^{\mathcal{N}^a + \hat{\mathcal{N}}_i } ,
\label{eq:Qaiasym}
\eeq
while $\nu_a$ and $\nu^a$ have the same leading asymptotic behaviour as $Q_{a|1}$, $Q^{a}_{\,|1}$, namely:
\beq
\nu_a(u)\sim u^{\mathcal{N}_a + \hat{\mathcal{N}}_1},\quad \nu^a(u)\sim u^{\mathcal{N}^a + \hat{\mathcal{N}}_1}.
\eeq
The asymptotics of $\bQ_{ij}$ can be computed from the definition (\ref{Qijdef}), and turn out to be, for the vector components,
\beqa
\bQ_I(u)
&\simeq& \( \mathcal{B}_1 
\, u^{ \hat{M}_1 -1 }, \mathcal{B}_2 
\, u^{\hat{M}_2 -1} , \mathcal{B}_3 
\, u^{ -\hat{M}_2 -1}, \mathcal{B}_4 
\, u^{ -\hat{M}_1 -1 } ,   \frac{\mathcal{B}_5}{u}  \) ,
\eeqa
where the coefficients $\mathcal{B}_I$ are constrained by consistency conditions similar to (\ref{eq:AAasy}):
\beqa
\mathcal{B}_I \, \mathcal{B}^I &=& \frac{1}{2} \, \frac{\prod_{A=1}^6 \left( \hat{M}_I-M_A \right) }{ \prod_{J \neq I }^5 ( \hat{M}_I- \hat{M}_J ) } , \;\;\;\; (I=1, \dots, 5) ,\label{eq:BIBI}\\
 \mathcal{B}_5 &=& \frac{i}{2} \frac{M_1 \, M_2 \, M_5 }{\hat M_1 \, \hat M_2 },\label{eq:B5}
\eeqa
(with no summation on the index $I$ in (\ref{eq:BIBI})).
The trace part satisfies
\beq
\bQ_{\circ}(u) = 4 + \frac{2 \, \mathcal{C}}{u^2} + \mathcal{O}\left(\frac{1}{u^3}\right),
\label{eq:Q0asy}
\eeq
where the constant $\mathcal{C}$ coincides with the value of the $OSp(4|6)$ Casimir:
\beq
\mathcal{C} = \frac{1}{4} \left( \hat{M}_1^2 + \hat{M}_2^2 - M_1^2 - M_2^2 - M_5^2 \right).\label{eq:Casimir}
\eeq
A derivation of the constraints (\ref{eq:AAasy}),(\ref{eq:BIBI}-\ref{eq:Casimir}) is discussed in Appendix \ref{app:ABderive}. 
Finally, let us comment on the asymptotics of the four functions $\tau_i(u)$. Since the latter are $2i$-periodic, and by construction grow less than exponentially for large $u$, they must approach a vector of constants at infinity. There is a certain amount of freedom in normalizing these constants, but we expect that for any physical state the components of $\tau_i$ with $i=2,3$ always vanish at large $u$:
\beq
\lim_{u \rightarrow \pm \infty } \tau_2(u) = \lim_{u \rightarrow \pm \infty } \tau_3(u) = 0. \label{eq:tau2tau3Van}
\eeq
In Section \ref{sec:LR} we established (\ref{eq:tau2tau3Van}) for the class of $4 \leftrightarrow {\bar{4}} $-symmetric operators. While we do not have a fully rigorous argument, we postulate that (\ref{eq:tau2tau3Van}) is true in general even for nonsymmetric states. As we discuss in Section \ref{sec:gluing}, the asymptotics (\ref{eq:tau2tau3Van}) implies the quantization of the spin and is the main ingredient for deriving the so-called gluing conditions, a powerful set of constraints encoding the main analytic properties of the system.   

\subsection{Classical limit}
\label{sec:classical}
The algebraic curve describing IIA string solutions on $AdS_4\times CP^3$ in the classical limit where $\Delta , S ,J_i = \mathcal{O}(h)$, $ h \sim \sqrt{\lambda/2} \rightarrow \infty$ was proposed in \cite{Gromov:2008bz}. In particular, a monodromy matrix was built on the basis of the Lax connection found in \cite{Stefanski:2008ik, Arutyunov:2008if} and its eigenvalues $\lambda_a \equiv e^{i q_a }$ were shown to define a ten-sheeted Riemann surface covering the domain of the relevant strong coupling spectral parameter, the Zhukovsky variable $x$. It is convenient to consider the logarithm of the eigenvalues, the so-called quasi-momenta, naturally grouped as $\{q_3,q_4,q_5,-q_3,-q_4,-q_5\}$ and $\{q_1,q_2,-q_1,-q_2\}$, corresponding respectively to the $SO(6)$ invariant $CP^3$ and the $Sp(4)$ invariant $AdS_4$ sectors of the monodromy matrix. The quasi-momenta are connected by logarithmic cuts\footnote{These cuts exist only in the classical limit and of course they should not be confused with the square-root branch cuts at $u = \pm 2 h + i \mathbb{Z}$ considered in the rest of the paper for the QSC.}, which may be viewed as condensates of Bethe roots. Classical string solutions can be studied by listing algebraic curves satisfying appropriate analytic properties (see \cite{Gromov:2008bz} for full details), and in particular the charges can be read off the asymptotics of the curve at large values of the spectral parameter:
\beq\label{eq:algebraic0}
\left( \begin{array}{c}  q_1(x) \\ q_2(x) \\ q_3(x) \\ q_4(x) \\ q_5(x) \end{array} \right)\sim 
\frac{1}{h \, x}\left( \begin{array}{c} \Delta + S \\ \Delta - S \\ J_1 \\  J_{2} \\  J_{3} \end{array} \right),\;\;\; x \sim \infty,
\eeq
where the quasi-momenta are ordered as in \cite{Gromov:2008bz}. 
In the classical limit, we expect that some of the $\bP$ and $\bQ$ functions of the QSC are related to the quasi-momenta as follows:
 \beqa\label{eq:classid1}
&&\bP_{1}(u) \sim e^{-h\int^{u/h} { \bf q }_4(z) dz}, \;\;\;\; \bP_{4}(u)  \sim e^{+h\int^{u/h} { \bf q}_4(z) dz} , \label{eq:classidP4}\\ 
 &&\bP_{2}(u) \sim e^{-h\int^{u/h} { \bf q}_3(z) dz}, \;\;\;\; \bP_{3}(u)  \sim e^{+h\int^{u/h} { \bf q}_3(z) dz} , \label{eq:classidP3}\\ 
 &&\bP_{5}(u) \sim e^{+h\int^{u/h} { \bf q}_5(z) dz}, \;\;\;\; \bP_{6}(u)  \sim e^{-h\int^{u/h} {\bf q}_5(z) dz} , \label{eq:classidP6}\\ 
 &&\bQ_{1}(u) \sim e^{+h\int^{u/h} {\bf q}_1(z) dz}, \;\;\;\; \bQ_{4}(u)  \sim e^{-h\int^{u/h} { \bf q}_1(z) dz} , \label{eq:classid14}\\ 
&&\bQ_{2}(u) \sim e^{+h\int^{u/h} {\bf q}_2(z) dz}, \;\;\;\; \bQ_{3}(u)  \sim e^{-h\int^{u/h} {\bf q}_2(z) dz}  ,
\label{eq:classid23}
\eeqa
where we use the notation ${ \bf q }_i( z ) \equiv q_i( z/2 + \sqrt{z-2}\sqrt{z+2}/2)$ for the quasi-momenta parametrized in terms of the rescaled spectral parameter $z = u/h$, which is the natural variable at strong coupling. 
 Using (\ref{eq:algebraic0}), one can verify that  (\ref{eq:classidP4})-(\ref{eq:classid23} are nicely consistent with  our asymptotics (\ref{MAdef})-(\ref{MIdef})\footnote{Indeed this expected semi-classical relation was an important guiding principle in guessing the way quantum numbers appear in the QSC. However,  since the charges are large in the classical limit, this reasoning only fixes the powers in the QSC asymptotics up to finite, state-independent shifts. }.
 
 Some of the limits (\ref{eq:classid23}), particularly the ones for $\bP_1$, $\bP_2$, $\bQ_1$, $\bQ_2$, $\bP_5$, $\bP_6$, can be derived from the large volume solution of the QSC, see the Section \ref{sec:ABA} below. In the rest of this Section, we discuss other consistency checks of the semi-classical identifications, as this will illustrate interesting analogies between classical and quantum curve (for a similar treatment, see Section 6 in \cite{Gromov:2014caa}). 

 One of the important features of the classical curve is the inversion symmetry \cite{Gromov:2008bz}:
\beq\label{eq:invsym}
\left( \begin{array}{c}  q_1(1/x) \\ q_2(1/x) \\ q_3(1/x) \\ q_4(1/x) \\ q_5(1/x) \end{array} \right)= 
\left( \begin{array}{c} -q_2(x) \\ -q_1(x) \\ 2\pi m-q_4(x) \\ 2\pi m-q_3(x) \\ q_5(x)\end{array} \right), \quad m \in \mathbb{Z}
\eeq
which is inherited by the transformation property of the monodromy matrix under the $\mathbb{Z}_4$ automorphism of $OSp(4|6)$ \cite{Stefanski:2008ik, Arutyunov:2008if}. Let us discuss how this property is related to the  Riemann-Hilbert type equations (\ref{eq:Pnu1}),(\ref{eq:QtilTautil}) valid for $\bP$ and $\bQ$ at finite coupling. 
 
 Consider first the case of $\bP$ functions. Their values on the second sheet is parametrized in terms of the matrix $\mu_{AB}$ which is $i$-periodic on the mirror section. In terms of the natural variable $z = u/h$, this periodicity becomes $i/h \rightarrow 0$ at strong coupling. Therefore, assuming that $\mu_{AB}$ admits a smooth classical limit, it must freeze to a constant value independent of $z$ \cite{Gromov:2014caa}, which can be normalized to be of order $\mathcal{O}(1)$. From two of the QSC equations (\ref{eq:Pmut}), we then find
\beq
\widetilde \bP_1 \sim \bP_3,\quad \widetilde \bP_2\sim \bP_4 ,
\label{eq:ancontP}
\eeq
where we have dropped all terms containing $\bP_1$ and $\bP_2$ on the rhs, since we see from (\ref{eq:classidP4}),(\ref{eq:classidP3}) that they are exponentially suppressed as $h \rightarrow \infty$. 
 On the other hand, analytically continuing to the second sheet the semi-classical expressions for $\bP_1$ and $\bP_2$, and using the inversion symmetry (\ref{eq:invsym}), one finds (see \cite{Gromov:2014caa} for details)
\beq
\widetilde \bP_1 \sim e^{+h\int^{u/h} { \bf q}_3(z) dz},\quad \widetilde \bP_2\sim e^{+h\int^{u/h} { \bf q}_4(z) dz} . \label{eq:P1class}
\eeq
 The comparison between (\ref{eq:P1class}) and (\ref{eq:ancontP})  motivates the semi-classical identification  for $\bP_3$ and $\bP_4$. 
 
This analysis cannot be straightforwardly repeated for the $\bQ$ functions, since the functions $\tau_i$ are periodic only on the short-cuts section, which becomes analytically disconnected from the $z$-plane at strong coupling. 
However, the inversion symmetry has a quantum analogue in the gluing conditions discussed in Section \ref{sec:gluing}, which connect $\widetilde \bQ_{ij}$ and the complex conjugate functions $\overline \bQ_{ij}$. From the analytic continuation of (\ref{eq:classid1})-(\ref{eq:classid23}), combined with the inversion symmetry, we may infer that in the classical limit
 \beq
 \label{eq:gluingclass}
 { \widetilde \bQ}_{3} \propto \overline{ \bQ }_{1} , \quad { \widetilde \bQ}_{4} \propto \overline{ \bQ }_{2}.
 \eeq
This is indeed consistent with the results of Section \ref{sec:gluing}. 
 
As a last comment, notice that there is no classical analogue for two of the components of the matrix $\bQ_{ij}$, namely the functions $\bQ_5$ and $\bQ_{\circ}$, which enter the basic Riemann-Hilbert constraints at finite coupling, but appear to completely decouple from the dynamics in the classical limit. This is a peculiar feature, as compared with the case of $AdS_5/CFT_4$, and it would be important to find a proper interpretation. One may also speculate that there is a connection with the fact that  part of the classical string solutions in ABJM theory are not captured by the classical spectral curve \cite{Sorokin:2011mj}. 

\subsection{Unitarity conditions} 
 The structure of the QSC also appears to automatically implement the unitarity bounds satisfied by the charges of a physical state. The discussion here will be very similar to the argument of Section C.2 of \cite{Gromov:2014caa}, so we will only sketch the main points. 
From the perspective of the QSC, the unitarity bounds arise from the requirement that the powers appearing in the asymptotics of $\bP$ and $\bQ$ functions are all distinct. This condition is very natural, since otherwise expressions like (\ref{eq:AAasy}) and (\ref{eq:BIBI}) for the coefficients $\mathcal{A}_A$, $\mathcal{B}_I$ would become singular. A further condition appears to be needed,  namely that, for all consistent solutions of the QSC, the powers entering the asymptotics of $\bQ$ functions are greater than the ones entering the asymptotics of $\bP$ functions: precisely, $|M_A | < |\hat{M}_I |$, $I \neq 5$. While it is more difficult to motivate this bound from first principles, it can be verified that it holds at weak coupling or in the large volume limit. 
Assuming a (purely conventional) ordering of magnitude for the components of $\bP_A$ and $\bQ_I$, we can therefore argue that all non-singular solutions of the QSC can be found restricting our attention to
\beq
 \hat{M}_1 > \hat{M}_2 > M_2 > M_1 > |M_5| .
 \label{eq:constr}
\eeq
With the identification (\ref{MAdef}),(\ref{MIdef}), we find that these conditions coincide with the unitarity bounds
\beqa
 J_2 \geq |J_3| , \;\;\; J_1 \geq 2 + J_2 , \;\;\; S \geq 0 , \;\;\; \Delta > S + J_1,
\eeqa 
or equivalently, in terms of excitation numbers (see \cite{Minahan:2009te}\footnote{Notice that, in \cite{Minahan:2009te}, the bounds are written in terms of the excitation numbers referring to a different version of the Bethe Ansatz, associated to the distinguished grading of the Dynkin diagram. The rules to convert between different conventions are reported in Appendix \ref{app:dictionary}. }): 
 \beqa
&&L + K_3 - 2 K_4 \geq 0 , \;\; L + K_3 - 2 K_{\bar{4}} \geq 0, \;\;\;  K_4 + K_{\bar{4}} - K_3 \geq 2 + K_1 , \\
&& K_3 + K_1 \geq 2 \, K_2 , \;\;\;\; K_2 + \gamma > 0 .
\eeqa
As a final comment, notice that, in principle, some of the inequalities (\ref{eq:constr}) could be saturated exactly in the weak coupling limit, where $\gamma \rightarrow 0$. Since the parameters $M_A$, as well as $\hat{M}_2 - \hat{M}_1$ (see Section \ref{sec:gluing}) are quantized, this is possible only for the condition $\hat M_2 > M_2$. The saturation of this bound for $\gamma \rightarrow 0$ is equivalent to the multiplet shortening condition:
 \beq\label{eq:shorte}
\Delta^{(0)}-S-J_1=0 ,
\eeq
where $\Delta^{(0)}$ is the classical conformal dimension, or equivalently $K_2=0$ in terms of excitation numbers. 
The states satisfying (\ref{eq:shorte}) have a peculiar characteristic in the QSC, namely they are the ones for which one of the $\bP$ functions vanishes at weak coupling. This is shown by the fact that for these operators $\mathcal{A}_2 \, \mathcal{A}_3 \rightarrow 0$ as $\hat M_2 - M_2\rightarrow 0$ in (\ref{eq:AAasy}). 
\section{Gluing conditions and spin quantization}
\label{sec:gluing}
 We shall now derive an exact relation (valid for real values of the charges) connecting the values of $\bQ_{ij}$ on the second sheet to the values of the complex conjugate function $\overline{ \bQ}_{ij}$. 
A similar result was first found in the $AdS_5/CFT_4$ context and exploited to solve the QSC in various regimes \cite{Gromov:2015wca, Gromov:2015vua}. In particular the equations presented below\footnote{The results presented in this Section were also obtained independently by Riccardo Conti using a slightly different argument \cite{ContiPrivate}.} may be used to solve the QSC numerically at finite coupling \cite{ContiNumerical}. For the derivation, we need an important technical assumption: we require that the matrix elements $Q_{a|i}$ can be expanded at large-$u$ as
\beq\label{eq:asyQ}
Q_{a|i}(u) \sim u^{\mathcal{N}_{a} + \hat{\mathcal{N}}_i } \, \sum_{m=0}^{\infty} \frac{B_{(a|i), m} }{u^m} , \;\;\;\; u\rightarrow + \infty .
\eeq
In words, (\ref{eq:asyQ}) means that there is no mixing among the powers occurring in the asymptotics of different columns of $Q_{a|i}$. This condition was dubbed ``pure asymptotics'' in \cite{Gromov:2015vua}, and can always be enforced using the freedom to take linear combinations of the columns of $Q_{a|i}$. We also assume that, for real values of the charges and the coupling, $\bP_{ab}$ can be chosen to be real\footnote{Throughout this section, reality and complex conjugation will be defined on the Riemann section with short cuts. Concretely, the reality of $\bP_A$ means that all coefficients $c_{A, n}$ in  (\ref{eq:expP}) are real.}. Under these conditions, the conjugate matrix elements $\overline{Q}_{a|i}$ satisfy the same difference equation (\ref{eq:defQai}) as $Q_{a|i}$. This implies that the two matrices are related through
  \beq\label{eq:defOme}
 \overline{Q}_{a|i}(u) = Q_{a|j}(u) \; ( \Omega^j_i (u) )^+,
  \eeq
  where $\Omega^j_i(u) $ 
  is a $2i$-periodic function of $u$: $\Omega^i_j(u+2 i) = \Omega^i_j(u)$. The condition of pure asymptotics (\ref{eq:asyQ}) implies that, as $u\rightarrow \infty$, the matrix $\Omega^i_j$ becomes diagonal.  Now, we recall the discontinuity relation (\ref{eq:fdef}):
\beq
{\widetilde\bQ}_{ij}(u) =  f_i^l(u) \, {\bQ}_{lk}(u) \, f^k_j(u),
\eeq  
where $  f_i^j(u) = \delta_i^j - \tau_i(u) \, \tau^j(u) $, which, combined with (\ref{eq:defOme}), gives
 \beq
\widetilde{\bQ}_{ij} = \mathcal{L}_i^l \; \ch_{lk} \;  \overline{\bQ}^{km} \; \ch_{mn} \; \mathcal{L}^n_j ,
\label{eq:QtQb}
 \eeq
 with 
\beq
\mathcal{L}^i_l(u)  = ( f(u)\, {\Omega}^{-1}(u) )_i^j .\label{eq:defL}
\eeq 
The crucial observation is now that $\mathcal{L}^i_j(u)$ must be a constant independent of $u$. In fact, the definition (\ref{eq:defL}) can be rewritten as 
$$\mathcal{L}^j_i = f_i^k \, Q_{a|k}^- \, ( \overline{Q}^{a|j} )^- = ( Q_{a|i}^{\text{LHPA}} )^- \,  ( \overline{Q}^{a|j} )^-,\nn
$$
and the last equality shows manifestly that $\mathcal{L}^j_i$ has no cuts  in the upper half plane, since this property is true for both $Q_{a|i}^{\text{LHPA}}$ and $\overline{Q}^{a|j}$. Because of its $2i$-periodicity, $\mathcal{L}^i_j$ is then entire in $u$, and, since it does not grow exponentially, it must be a constant. 

To determine the form of $\mathcal{L}_i^j$, we can study its definition at large $u$, where $\Omega^i_j$ becomes diagonal and many of the matrix elements of $f^i_j$ vanish due to the fact that $\tau_2, \, \tau_3 \rightarrow 0$. The structure is further specified by several consistency conditions. For instance, since $\mathcal{L}$ does not depend on $u$, we should certainly impose the equality of the following limits:
 \beqa\label{eq:condFF}
\mathcal{L}_i^j=\lim_{u\rightarrow +\infty}	\( f(u)\,\Omega^{-1}(u) \)_i^j = \lim_{u\rightarrow -\infty}	\(f(u)\, \Omega^{-1}(u) \)_i^j . 	\eeqa
To exploit this constraint, notice that the constant limits of $\Omega$ at $\pm \infty$ are related as follows:
\beq
 \lim_{u\rightarrow -\infty} \Omega^i_i(u) = \(  \lim_{u\rightarrow +\infty} \Omega^i_i(u) \)\;  e^{-2 \pi i (\mathcal{N}_{a} + \hat{\mathcal{N}}_i ) } .\label{eq:Oplusminus}
\eeq
This condition can be obtained studying the definition (\ref{eq:defOme}) as $u \rightarrow \pm \infty$, using the fact that the asymptotic behaviour of $Q_{a|i}(u)$ ($\overline{Q}_{a|i}(u)$, respectively) as $u \rightarrow - \infty$ must be connected to the one for $u \rightarrow + \infty$ through analytic continuation along a large semicircle in the upper (lower) half plane, where this function is free of singularities. 
 Considering relation (\ref{eq:condFF}) for $j=2,3$, and using (\ref{eq:Oplusminus}), we find 
 \beq
 e^{2 \pi i  (\mathcal{N}_{a} + \hat{\mathcal{N}}_i ) }= 1 ,
 \eeq
 for $i=2,3$, $\forall a$. This equation implies that $\hat{M}_2 - \hat{M}_1  = 2 S + 1 \in \mathbb{Z}$, namely the spin is integer or half-integer. The other conditions in (\ref{eq:condFF}) constrain the asymptotics of the non-zero components of $\tau$. Denoting $t_{i, \pm} \equiv \lim_{u\rightarrow \pm \infty} \tau_i$, we have in particular
\beq
 t_{1, \pm} \, t_{4, \pm} = \pm i\, e^{i \sigtw} \, \tan( \pi \, \hat{M}_1 ) .\label{eq:t1t4}
\eeq
Finally, evaluating $\mathcal{L}$ at large $u$ and using (\ref{eq:t1t4}), relation (\ref{eq:QtQb}) leads to the gluing conditions:
\begin{align}
\label{eq:gluing1}
\hspace{-0.7cm}\widetilde{\bQ}_{1} &= -\frac{e^{ i \pi  \hat M_1}}{\os_1\,\os_2\,\cos( \pi  \hat M_1 ) } \, \overline{\bQ}_{1} + \delta_1 \, \overline{\bQ}_{3} , \quad 
& \widetilde{\bQ}_{3} &= -\frac{e^{-i \pi \hat M_1}}{\os_2 \, \os_4\,\cos( \pi \hat M_1 ) } \, \overline{\bQ}_{3} + \frac{\os_3}{\os_2} \, \delta_2 \, \overline{\bQ}_{1} , \\
\hspace{-0.7cm}\widetilde{\bQ}_{2} &= - \frac{e^{i \pi \hat M_1}}{\os_1 \, \os_3 \,\cos( \pi \hat M_1 ) } \, \overline{\bQ}_{2} +  \frac{ \os_2 }{ \os_3 } \, \delta_1 \, \overline{\bQ}_{4} ,  \quad 
& \widetilde{\bQ}_{4} &= -\frac{e^{-i \pi \hat M_1}}{\os_4\,\os_3\,\cos( \pi \hat M_1 ) } \, \overline{\bQ}_{4} +\delta_2 \,  \overline{\bQ}_{2} ,  \label{eq:gluing4}\\
\hspace{-0.7cm}\widetilde{\bQ}_{\circ} &= \overline{\bQ}_{\circ} ,\quad &\widetilde{\bQ}_{5} &=  - \overline{\bQ}_{5} ,
\label{eq:gluing5}
\end{align}
where we are using the vector notation defined in Section \ref{sec:vectorQ}, $\delta_1 = e^{-i \sigtw} t_{1,+}^2/( \os_1 \os_2 ) $, $\delta_2 =-e^{-i \sigtw} t_{4,+}^2/( \os_3 \os_4 )$, and $\os_i \equiv \lim_{u \rightarrow + \infty} \Omega^i_i(u)$. For completeness we point out that the constants $y_i$, $\delta_i$ may in general depend on the coupling and on various normalization choices. For the implementation of the numerical method, it is only needed to know explicitly the value of $y_i$. These constants, which satisfy the consistency conditions $y_1=1/y_4=1/(y_1^* )$, $y_2=1/y_3=1/(y_2^*)$, are simply related\footnote{For real values of the coupling it is always possible to choose a normalization where $B_{(a|i), 0} \in \mathbb{R}$, so that $y_i=1$. } to the choice of normalization of the $Q_{a|i}(u)$ functions, and can be determined as:
\beq
y_i = (B_{(a|i), 0} )^*/B_{(a|i), 0} , \quad \forall a.
\eeq

The relations (\ref{eq:gluing1})-(\ref{eq:gluing5}) are similar to the ones obtained in \cite{Gromov:2015wca, Gromov:2015vua}, but slightly more complicated. Indeed, in the $AdS_5/CFT_4$ context a single $\overline{\bQ}$ function appears on the rhs of the gluing conditions, which are an almost direct lift of the inversion symmetry connecting pairs of quasi-momenta in the classical limit. In the present case, the quantum version is a bit more intricate. In particular, the explicit parametric dependence of the gluing conditions on the charge $\hat{M}_1$ needs to be taken into account in order to develop a numerical algorithm \cite{ContiNumerical}. 
As a last comment, we observe that the quantization of the spin is a direct consequence of the choice of vanishing asymptotics for two of the components of $\tau$. As shown in \cite{Gromov:2015wca}, it should be possible to relax this condition and consider continuous values of $S$ by admitting exponentially growing asymptotics in $\tau_2$ and $\tau_3$. 

\section{The Q-system}\label{sec:Qsystem}
In this Section we show how to embed the previous results into a larger set of functional equations reflecting the $OSp(4|6)$ symmetry.
 It is important to mention that, while the form of Q-systems associated to $GL(M|N)$-type superalgebras is known (see e.g. \cite{Kazakov:2007fy,Wronskian,Thook}), there appears to be no comprehensive understanding of this mathematical structure for orthosymplectic superalgebras. Here we take a bottom-up approach to the problem and try to construct the Q-system starting from the Q functions already introduced\footnote{Starting from these functions, we will define a Q-system where the Q functions are free of cuts in the upper half plane. An analogous construction, analytic in the lower half plane, could be performed starting from the Q functions $\bP_A$, $\widetilde \bQ_I$, and $Q_{a|i}^{\text{LHPA}}$ defined in (\ref{eq:QLHPA}). Notice that the two systems are connected through the $\nu$ or $\tau$ functions, which therefore play the role of a symmetry transformation of the Q-system (for an interesting discussion see \cite{Gromov:2014caa}). }: $\bP_A$, $\bQ_I$, $Q_{a|i}$, $Q^a_{\, |i}$, together  with the relations linking them, equations (\ref{eq:eq1}),(\ref{eq:chidef}),(\ref{Qijdef}). 
 We will explicitly define new Q functions and prove the validity of a set of functional relations which is rich enough to contain various forms of exact Bethe Ansatz equations (equivalent to the absence of poles for the Q functions) related to the $OSp(4|6)$ symmetry. 
 
\begin{figure}[t!]
\begin{minipage}{0.47\linewidth}
\hspace{-0.8cm}
\includegraphics[scale=0.24]{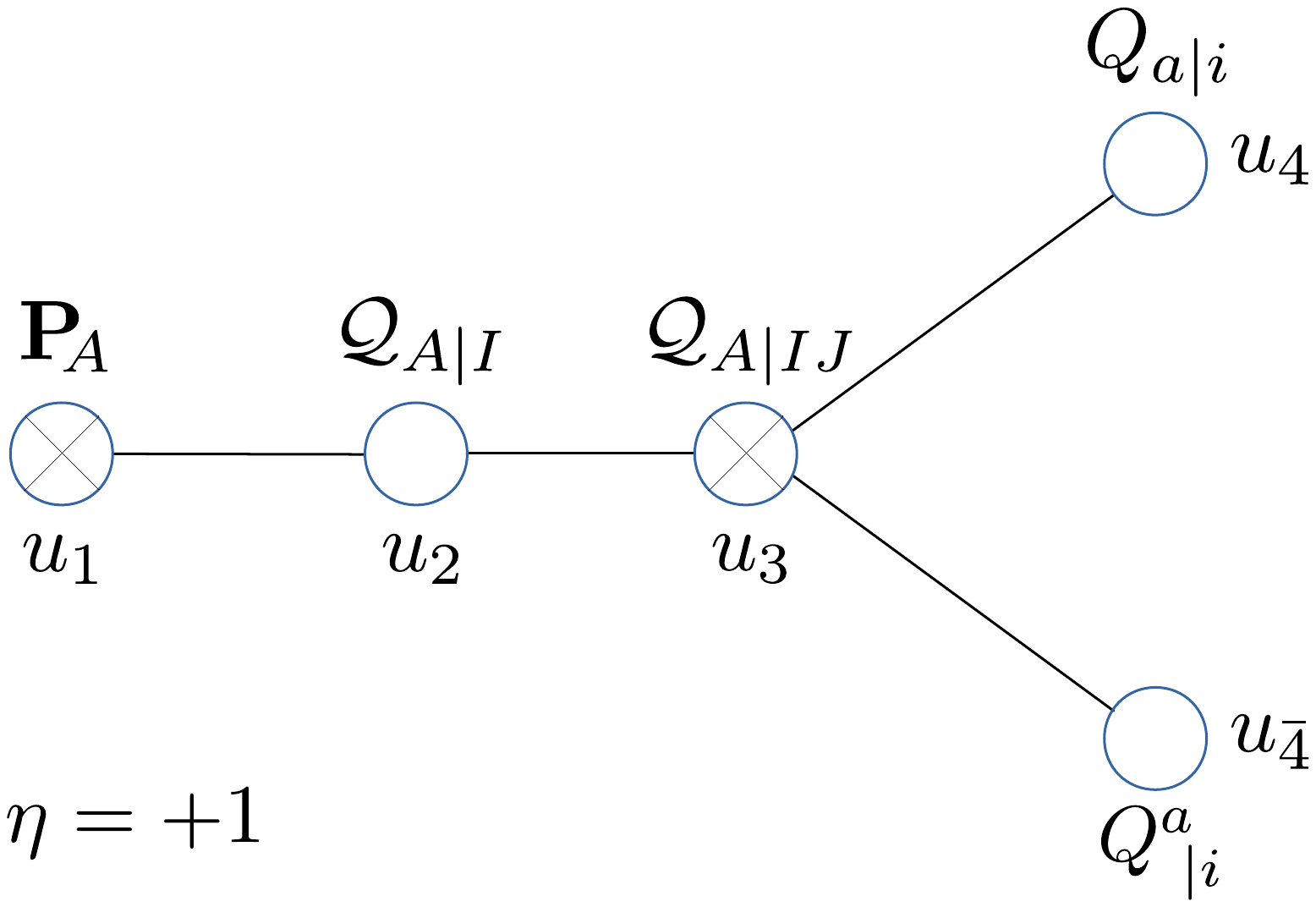}
\caption{ Chain of Q functions corresponding to the $\eta=+1$ grading of the Bethe Ansatz.}
\label{fig:dynkinetap1}
\end{minipage}
\hspace{0.5cm}
\begin{minipage}{0.47\linewidth}
\hspace{-0.6cm}
\includegraphics[scale=0.24]{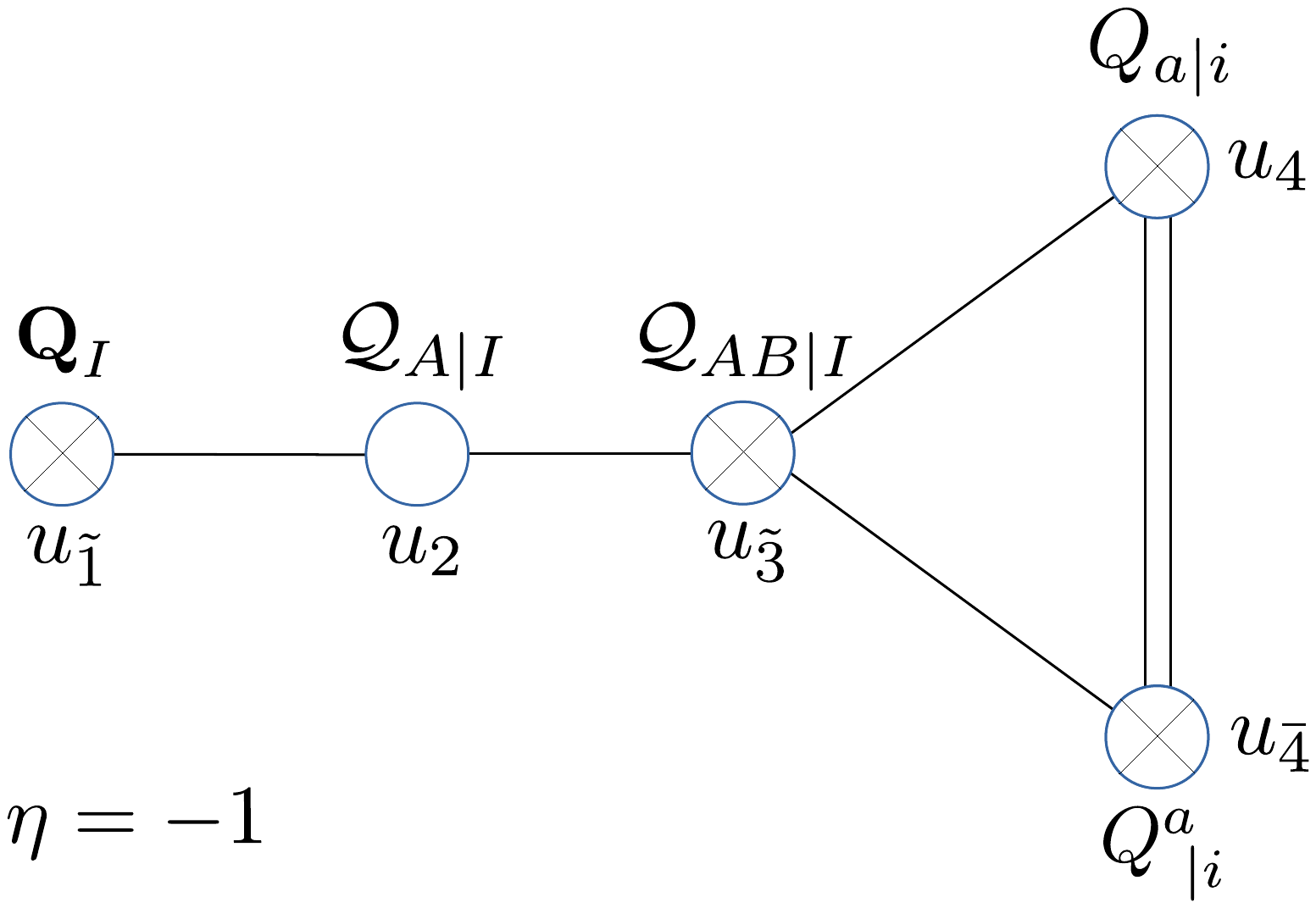}
\caption{Chain of Q functions corresponding to the $\eta=-1$ grading of the Bethe Ansatz. }
\label{fig:dynkinetam1}
\end{minipage}
\end{figure}

 Before starting the construction, let us describe some of its main characteristics. 
 Various types of Q functions will be assigned to particular nodes of the Dynkin diagram. We will almost exclusively consider the two versions of the diagram shown in Figures \ref{fig:dynkinetap1}, \ref{fig:dynkinetam1}, which are the ones associated to the two known forms of Asymptotic Bethe Ansatz. 
 The Q functions will have the general index structure\footnote{ Notice that also the Q functions $\bP_A$ and $\bQ_I$ fit this pattern and we could identify them with $\bP_A \equiv Q_{A| \emptyset}$, $\bQ_I \equiv Q_{\emptyset|I}$, where $\emptyset$ denotes the trivial representation.} $Q_{\bullet | \ast}$, where $\bullet$ and $\ast$ are (vector or spinor)  multi-indices carrying representations of $SO(3,3)$ and $SO(3,2)$, respectively, see Section \ref{sec:symmetries} for notations.   
 Various arguments, and in particular the weak coupling analysis, suggest that Q functions of types $\bP_A$ and $\bQ_I$ carry Bethe roots associated to the first node of the two diagrams, while the Q functions $Q_{a|i}$, $Q^a_{\, |i}$ should be linked to the nodes corresponding to the spinorial representations, see Figures \ref{fig:dynkinetap1}, \ref{fig:dynkinetam1}. The main task of this Section is to complete the picture by constructing Q functions and functional relations associated to the remaining nodes.  In analogy with the Q-system of \cite{Gromov:2014caa}, and in contrast to the case of standard Lie algebras, for every node of the diagram one may define equations of two basic types -- fermionic or bosonic. This feature of supersymmetric Q-systems is known to be related to the existence of different gradings of the Dynkin diagram. Choosing different chains of Q functions, we will recover different sets of exact Bethe equations. Finally, as a non-trivial check of the construction, we will recover the two forms of the ABA equations in the large volume limit. 
\subsection{Construction of the Q-system}\label{sec:constrQsyst}
\subsubsection*{First step: identifying $\mQ_{A|I}$}
We start the construction by some guesswork. From the form of the Bethe Ansatz, and taking inspiration from \cite{Gromov:2014caa}, it is natural to expect that one of the functional relations should read:
\begin{align}
\label{eq:QQF1}
&F_1:& \quad \mQ_{A|I}^+ -  \mQ_{A|I}^- & = \bP_{A} \, { \bQ }_{I} .&
\end{align}
We have marked this equation with the symbol $F_1$ to point out that it is a fermionic-type Q-system relation, based at the first node of the Dynkin diagram. This equation might be taken as a non-local definition of the $6 \times 5 $ matrix\footnote{Notice that we are denoting Q functions carrying capital indices such as $A\in \left\{1,\dots,6\right\}$ or $I\in \left\{1,\dots,5\right\}$ with the calligraphic font $\mQ$ in order to avoid possible confusion with $Q_{a|i}$ when the indices take some concrete value. So, for example, notice that $\mQ_{1|2} \neq Q_{1|2}$! } $\mQ_{A|I}$. However, this new type of Q functions can also be expressed as an explicit, local combination of the building blocks $Q_{a|i}$, $Q^a_{ \, |i}$, through the following quadratic combinations:
\beq
Q_{ab|ij} = Q_{a|i} \, Q_{b|j} - Q_{a|j} \, Q_{b|i} = \text{det} \left( \begin{array}{cc} Q_{a|i} & Q_{a|j}  \\ Q_{b|i} & Q_{b|j}  \end{array}\right),
\label{eq:defQabij}
\eeq
namely, the $2 \times 2$ minors of the $4 \times 4$ matrix $\left\{Q_{a|i}\right\}$. 
Notice that $Q_{ab|ij}$ is antisymmetric in both $(ab)$ and $(ij)$, and  therefore has $6 \times 6$ independent components. To match the $6 \times 5$ components of $\mQ_{A|I}$ we need of course to project the $(ij)$ indices on the vector representation. The correct identification, which will be important for the derivation of the rest of the Q-system, is simply:
\beq
 \mQ_{A|I} \equiv - \frac{1}{4} \, Q_{ab|ij} \, \bar{\sigma}_A^{ab} \, \bar{\Sigma}_I^{ij}. 
 \label{eq:defQAI}
\eeq
We will show below that this definition implies the validity of (\ref{eq:QQF1}). 

One could also consider the complementary projection on the singlet representation for the $(ij)$ indices, and define:
\beq
\mQ_{A| \circ} =  -\frac{1}{4} Q_{ab|ij} \, \ch^{ij} \, \bar{\sigma}_A^{ab}  .
\eeq
However, it turns out that all Q functions carrying the singlet representation of $SO(3,2)$, such as $\mQ_{A| \circ} $ and $\bQ_{\circ}$, drop out of the functional relations needed for the derivation of exact Bethe equations. It would be interesting to understand from the algebraic point of view whether they should be considered as part of the Q-system.

\subsubsection{Q-system relations for the nodes $1,2,3$}

To prove the validity of (\ref{eq:QQF1}), we start by rewriting the constraint $\text{Pf}( \bP_{ab} )=1$ as:
\beq
\bP_{ab} \, \bP_{cd} - \bP_{cb} \, \bP_{ad} - \bP_{ac} \, \bP_{bd} = \epsilon_{abcd} ,
\eeq
where $\epsilon_{abcd}$ denotes the completely antisymmetric Levi-Civita tensor. 
 Using this identity, it is immediate to prove that\footnote{ We are using the standard notation $[ \,, \,]$ for the antisymmetrization of indices, e.g. $H_{\left[i \,, j \right] }\equiv H_{ij}-H_{ji}$.}
 \beqa
 Q_{a|\,\left[i \right.}^+ \, Q_{b| \; \left.j\right]}^+ &=& \bP_{a a_1} \, \bP_{b b_1} \(Q^{a_1}_{\,| \, \left[i\right.}  \, Q^{b_1}_{\,|j \left.\right]} \)^-  \\
 &=& \frac{1}{2} \; \epsilon_{a a_1 b b_1} \, \(Q^{a_1}_{\,| \, \left[\right. i}  \, Q^{b_1}_{\,|j\left.\right]} \)^- + \frac{1}{2} \, \bP_{ab} \; \(  \bP_{a_1 b_1} (Q^{a_1}_{\,|\left[\right.i} )^- \, (Q^{b_1}_{\,|j\left.\right]} )^-  \) , \nn
 \eeqa
and, inserting (\ref{eq:QQinvmain}), we obtain
\beq\label{eq:alreadyQAI}
Q_{ab|ij}^+ + \frac{1}{2} \, \epsilon_{abcd} \, (Q^{cd}_{\,|ij} )^- = -\bP_{ab} \, \( \bQ_{ij} + \frac{\ch_{ij}}{2} \bQ_{\circ} \) .
\eeq
Projecting on vector indices as in (\ref{eq:defQAI}) and taking into account simple algebraic identities (see (\ref{eq:lowup3APP})), (\ref{eq:alreadyQAI}) yields precisely the fermionic equation (\ref{eq:QQF1}):
\begin{align}
\label{eq:F1}
&F_1:& \quad \mQ_{A|I}^+ -  \mQ_{A|I}^- & = \bP_{A} \, { \bQ }_{I} .&
\end{align} 
For completeness, we report also the identity obtained by tracing over $(ij)$:
\beq\label{eq:Fcirc}
\mQ_{A|\circ}^+ +  \mQ_{A|\circ}^-  =   \frac{1}{2} \, \bP_{A} \, { \bQ }_{\circ} .
\eeq
 As anticipated, (\ref{eq:Fcirc}) is apparently decoupled from the rest of the Q-system and will not play a role in the following considerations. Bosonic-type Q-system relations for the first node can be introduced straightforwardly. They take the standard form:
\begin{align}\label{eq:B1}
&B_1:& \bP_{A}^+ \, \bP_B^- - \bP_{A}^- \, \bP_B^+ &= \mQ_{AB|\emptyset} , &\\
&B_{1\ast}:& \bQ_{I}^+ \, \bQ_J^- - \bQ_{I}^- \, \bQ_{J}^- &= \mQ_{\emptyset|IJ} ,&
\label{eq:B1*}
\end{align}
which can be interpreted as definitions of the new two-index objects $Q_{AB|\emptyset}$ and $Q_{\emptyset|IJ}$. These Q functions do not sit on the diagrams in Figures \ref{fig:dynkinetap1}, \ref{fig:dynkinetam1}, but appear in other choices of gradings, such as the distinguished one (see discussion below). 

The construction of functional relations for the second and third nodes is  standard and follows the usual fusion rules, cf \cite{Gromov:2014caa}. 
In particular, associated to the third node we define the Q functions
\beqa
\mQ_{A|IJ} &\equiv&  \bQ_I \, \mQ_{A|J}^- - \bQ_J \, \mQ_{A|I}^- =  \bQ_I \, \mQ_{A|J}^+ - \bQ_J \, \mQ_{A|I}^+ , \label{eq:defQAIJ1}\\
\mQ_{AB|I} &\equiv& \bP_A \, \mQ_{B|I}^- - \bP_B \, \mQ_{A|I}^- = \bP_A \, \mQ_{B|I}^+ - \bP_B \, \mQ_{A|I}^+, \label{eq:defQAIJ2}
\eeqa
which satisfy bosonic-type relations for the second node:
\begin{align}\label{eq:B2}
&B_2:& \mQ_{A|IJ} \, \bP_A &= \mQ_{A|I}^+ \, \mQ_{A|J}^- -  \mQ_{A|J}^+ \, \mQ_{A|I}^- ,& \\
&B_{2\ast}:& \mQ_{AB|I} \, \bQ_I &= \mQ_{A|I}^+ \, \mQ_{B|I}^- -  \mQ_{B|I}^+ \, \mQ_{A|I}^-.&
\label{eq:B2*}
\end{align}
Using equation $F_1$ (\ref{eq:F1}), we can also straightforwardly establish the following fermionic-type functional relations for the second node:
\begin{align}\label{eq:F2}
&F_2:& \mQ_{A|I} \, \mQ_{AB} &= \mQ_{AB|I}^+ \, \bP_A^- - \bP_A^+ \, \mQ_{AB|I}^-, \\ 
&F_{2\ast}:& \mQ_{A|I} \, \mQ_{IJ} &= \mQ_{A|IJ}^+ \, \bQ_I^- - \bQ_I^+ \, \mQ_{A|IJ}^-.&
\label{eq:F2*}
\end{align}
Now let us derive the relations centered around the third node. Using (\ref{eq:defQAIJ1})-(\ref{eq:defQAIJ2}), it is simple to obtain the bosonic-type equations
\begin{align}\label{eq:B3}
&B_3:& \mQ_{AB|IJ} \, \mQ_{AB|\emptyset} &= \mQ_{AB|I}^+ \, \mQ_{AB|J}^- - \mQ_{AB|I}^- \, \mQ_{AB|J}^+ , &\\ 
&B_{3\ast}:& \mQ_{AB|IJ} \, \mQ_{\emptyset |IJ} &= \mQ_{A|IJ}^+ \, \mQ_{B|IJ}^- - \mQ_{A|IJ}^- \, \mQ_{B|IJ}^+ ,&
\label{eq:B3*}
\end{align}
while the definitions (\ref{eq:defQAIJ1}),(\ref{eq:defQAIJ2}) and relation (\ref{eq:F1}), imply the validity of the fermionic identity
\begin{align}\label{eq:F3}
&F_3:&  \mQ_{A|IJ} \, \mQ_{AB|I} &= \mQ_{AB|IJ}^+ \, \mQ_{A|I}^- -  \mQ_{AB|IJ}^- \, \mQ_{A|I}^+ ,&
\end{align}
where
\beq
\label{eq:QABIJ}
\mQ_{AB|IJ} \equiv \mQ_{A|I} \, \mQ_{B|J} - \mQ_{B|I} \, \mQ_{A|J}.
\eeq
As we may expect from the Dynkin diagram, the newly defined object in (\ref{eq:QABIJ}) represents the fusion of the spinorial Q functions $Q_{a|i}$ and $Q^a_{\;|i}$.  Indeed, let us prove that it can be rewritten as:
\beq
\label{eq:QABrew}
\mQ_{AB|IJ} = (\sigma_{AB} )^b_a \; Q^{a}_{\,|i} \, Q_{b|j} \, {\Sigma}_{IJ}^{ij} ,
\eeq
where $ {\Sigma}_{IJ}^{ij} \equiv \frac{1}{2} \; (\bar{\Sigma}_I \, \ch \, \bar{\Sigma}_J - \bar{\Sigma}_J \, \ch \, \bar{\Sigma}_I   )^{ij} $. 
This equation will be crucial for the derivation of closed sets of exact Bethe equations. To derive (\ref{eq:QABrew}), start from the definition of $\mQ_{A|I}$ in (\ref{eq:defQAI}) and rewrite (\ref{eq:QABIJ}) as
\beq
\mQ_{AB|IJ}  = \frac{1}{4} \( Q^a_{\,|i} \, Q^b_{\,|j} \, Q_{c|k} \, Q_{d|l} \) \, \( ({\sigma}_{ A} )_{ab} \; ( \bar{\sigma}_{B } )^{cd}- ({\sigma}_{B} )_{ab} \; ( \bar{\sigma}_{A} )^{cd} \) \; \bar{\Sigma}_I^{ij} \; \bar{\Sigma}_J^{kl}.\label{eq:com2}
\eeq
 Using formula (\ref{eq:commu}) for the commutator of sigma matrices appearing in (\ref{eq:com2}), we find 
\beqa\label{eq:blank}
\mQ_{AB|IJ} &=& \( Q^a_{\,|i} \,  Q_{c|k} \; ( \sigma_{AB} )_a^{\, c} \) \, \bar{\Sigma}_I^{ij} \, \( Q^{b}_{\,|j} \, Q_{b|l} \) \, \bar{\Sigma}_J^{kl}  = \( Q^a_{\,|i} \,  Q_{c|k} \; ( \sigma_{AB} )_a^{\, c} \) \, \bar{\Sigma}_I^{ij} \, \ch_{jl} \, \bar{\Sigma}_J^{lk} \nn\\
&=& \( Q^a_{\,|i} \,  Q_{c|k} \; ( \sigma_{AB} )_a^{\, c} \) \, \bar{\Sigma}_{IJ}^{ik},
\eeqa
where, in the last step, we have used the  anti-symmetry in $(IJ)$ of the whole expression by definition of $\mQ_{AB|IJ}$. 

\subsubsection{Q-system relations for the nodes $4$ and $\bar{4}$}
Let us now derive the functional relations centered at the spinor nodes. The two bosonic Q-system equations (centered at nodes $4$ and $\bar4$, respectively) are:
\begin{align}
&B_4:  &( \bar{\sigma}_A )^{ab} \, \( Q_{a|i}^+ \; Q_{b|j}^- \) \, ( \Sigma_{IJ} )^{ij} &= \mQ_{A|IJ} , &\label{eq:idB4}\\
&B_{\bar4}:&  ( {\sigma}_A )_{ab} \, \( (Q^{a}_{\,|i} )^+ \; (Q^{b}_{\,|j} )^- \) \, ( \Sigma_{IJ} )^{ij} &=  \mQ_{A|IJ} ,&
\label{eq:Bbar4}
\end{align}
while the fermionic-type relations, which cross the two spinor nodes, read
\begin{align}\label{eq:F4}
&F_4:&  ( \sigma_{AB} )_{a}^{\,b} \; \( (Q^{a}_{\,|i} )^+ \; Q_{b|j}^- \) \; ( \bar{\Sigma}_I )^{ij} &= \mQ_{AB|I} ,&\\
&F_{\bar4}:&   ( \sigma_{AB} )_{a}^{\,b} \; \( (Q^{a}_{\,|i} )^- \; Q_{b|j}^+ \) \; ( \bar{\Sigma}_I )^{ij} &= \mQ_{AB|I} .&
\label{eq:Fbar4}
\end{align}
To prove (\ref{eq:idB4}), start from the combination
\beq
 \( Q_{a|i}^+ \; Q_{b|j}^- - Q_{b|i}^+ \; Q_{a|j}^- \) \, ( \Sigma_{IJ} )^{ij}.
\eeq
Using (\ref{eq:translQ}), (\ref{eq:QQinvmain}), (\ref{eq:Q5vecmain}), we can eliminate all positive shifts through
\beqa
Q_{a|i}^+ &=& 
\frac{1}{4} \,Q_{a|i}^- \, \bQ_{\circ} +  Q_{a|m}^- \, \( \kappa^{ml} \, {\bQ}_I \; \Sigma^I_{li} \) , 
\eeqa
and we find\footnote{Notice that the terms proportional to $\bQ_{\circ}$ cancel out of the equation due to the symmetry $(\Sigma_{IJ} )^{ij}= (\Sigma_{IJ} )^{ji}$, see Appendix \ref{app:gamma}.}: 
\beqa\label{eq:passage}
 \( Q_{a|i}^+ \; Q_{b|j}^- - Q_{b|i}^+ \; Q_{a|j}^- \) \, ( \Sigma_{IJ} )^{ij} &=&  Q_{ab|mj}^- \; \( \kappa^{ml} \, {\bQ}_K \; \Sigma^K_{li} \) \, (\Sigma_{IJ} )^{ij}  \\
&=& \frac{1}{2} \; Q_{ab|mj}^- \; \( \bQ_I \, (\bar{\Sigma}_J )^{mj} -  \bQ_J \, (\bar{\Sigma}_I )^{mj} \) \nn\\
&=&  Q_{ab|I}^- \;\bQ_J -  Q_{ab|J}^- \;\bQ_I ,
\eeqa
where we have used identity (\ref{eq:propTr}) to simplify the product of $\Sigma$ matrices in (\ref{eq:passage}).  Contracting with $(\bar{\sigma}_A)^{ab}$ and comparing with (\ref{eq:defQAIJ1}) yields (\ref{eq:idB4}). Similarly, to prove (\ref{eq:F4}), we consider
\beq
 \( ( Q^{a}_{\,|i} )^+ \; Q_{b|j}^- - (Q^{a}_{\,|j} )^+ \; Q_{b|i}^- \) \, ( \sigma_{AB} )_{a}^{\,b},
\eeq
 and replace all Q functions with positive shifts using $(Q^a_{\,|i} )^+ = \bP^{a a_1} \, Q_{a_1|i}^- $: 
\beqa
 \(  (Q^{a}_{\,|i} )^+ \; Q_{b|j}^- - (Q^{a}_{\,|j} )^+ \; Q_{b|i}^- \) \, ( \sigma_{AB} )_{a}^{\,b} &=& -Q^-_{a_1 b | ij} \, \bP^{a_1 a} \, ( \sigma_{AB})_a^{\,b} \\
 &=& \frac{1}{2} \; Q^-_{a_1 b | ij} \, \( \bar{\sigma}_C \, \sigma_A \, \bar{\sigma}_B-  \bar{\sigma}_C \, \sigma_B \, \bar{\sigma}_A \)^{a_1 b} \; \bP^C \nn\\
 &=& - \bP_A \; Q_{B|ij}^- + \bP_B \; Q_{A|ij}^- = - Q_{AB|ij} ,\nn
\eeqa
where we have used (\ref{eq:Pinv0main}) in the second equality and identity (\ref{eq:sigss}) in the third. 
Finally, projecting on the vector component out of the antisymmetric indices $(ij)$, we get (\ref{eq:F4}).

\subsection{Exact Bethe equations}\label{sec:exactBA}
Let us now show how to obtain exact Bethe equations for the zeros of Q functions. We will obtain equations formally identical to the various versions of 2-loop Bethe Ansatz proposed in \cite{Minahan:2008hf}, based on the underlying $OSp(4|6)$ symmetry, with the important difference that, at finite coupling, Q functions are nontrivial functions of the spectral parameter living on infinitely many sheets (and, in general, with infinitely many zeros). In the weak coupling limit, the branch cuts shrink to zero size and are usually replaced by poles. However, for particular choices of indices the Q functions reduce to polynomials at weak coupling, and the exact equations discussed here reduce to the 2-loop Bethe Ansatz of \cite{Minahan:2008hf}. This is discussed in detail in Appendix \ref{sec:appWC}. 

To derive a version of the Bethe Ansatz related to the $\eta=1$ grading of the Dynkin diagram, we need to consider a chain of functional relations made of equations of type $F_1$ (\ref{eq:F1}), $B_{2}$  (\ref{eq:B2}) and $F_3$  (\ref{eq:F3}) for the first, second and third  nodes respectively, and $B_4$  (\ref{eq:idB4}) and $B_{\bar4}$  (\ref{eq:Bbar4}) for the nodes at the bifurcation.
 For concreteness, let us make a specific choice of indices, and consider the following sequence of Q-system relations
\begin{align}
&F_1:&& \mQ_{2|2}^+ - \mQ_{2|2}^- = \bP_{2} \, \bQ_{2}, \label{eq:f1} \\
&B_{2}:&& \mQ_{2|1}^+ \,  \mQ_{2|2}^- -  \mQ_{2|2}^+ \,  \mQ_{2|1}^- = \mQ_{2|12} \, \bP_{2},\label{eq:f2}\\
&F_3:&& ( Q_{1|1} \, Q^4_{\,|1} )^+ \; \mQ_{2|2}^- - (Q_{1|1} \, Q^4_{\,|1})^- \; \mQ_{2|2}^+  = \mQ_{12|2} \, \mQ_{2|12} ,\label{eq:f3}\\
&B_4:&& ( Q_{1|1}  )^+ \, Q_{3|1}^- - (Q_{3|1}  )^+ \, Q_{1|1}^- = \mQ_{2|12} , \label{eq:b4}\\
&B_{\bar4}:&& ( Q^4_{\,|1}  )^+ \, ( Q^2_{\,|1} )^- - (Q^2_{\,|1}  )^+ \, (Q^4_{\,|1} )^- = \mQ_{2|12}, \label{eq:b4b}
\end{align}
where we used (\ref{eq:QABrew}) to evaluate
\beq
\label{eq:Q1212}
\mQ_{12|12} =  Q_{1|1}\,Q^4_{\,|1} .
\eeq
Relations (\ref{eq:f1})-(\ref{eq:b4b}), supplemented with the requirement that no Q functions have poles, imply a set of exact BA equations for the zeros of the Q functions
\beq
\bP_2 , \;\;\; \mQ_{2|2}, \;\;\; \mQ_{2|12} , \;\;\;   Q_{1|1}, \;\;\; Q^4_{\,|1} . \label{eq:chain1}
\eeq
Let us denote the zeros of these functions as $\left\{ u_{\bu, k } \right\}$, with $\bu = 1,2, 3, 4, \bar{4}$, respectively (where the index $k$ runs over different zeros of a given Q function). 

Taking the ratio of (\ref{eq:b4}) evaluated at points $u_{4, k} + i/2$ and $u_{4, k} -i/2$, where $u_{4, k}$ is a generic zero of $Q_{1|1}$, gives the massive node Bethe equation
\beq
-1 =  \left. \frac{Q_{1|1}^{++} }{Q_{1|1}^{--} } \, \frac{ \mQ_{2|12}^- }{\mQ_{2|12}^+ } \right|_{u_{4,k}} \;\;,\;\;\;\;\text{ with } Q_{1|1}(u_{4,k} )= 0 ,\label{eq:4eta1}
\eeq
and similarly from (\ref{eq:b4b}) one gets
\beq
-1 = \left. \frac{Q^4|_{\,|1}^{++} }{Q^4|_{\,|1}^{--} } \, \frac{ \mQ_{2|12}^- }{\mQ_{2|12}^+ } \right|_{u_{4,k}}  \;\;,\;\;\;\;\text{ with }Q^4_{\,|1}(u_{\bar{4},k} ) = 0.\label{eq:bar4eta1}
\eeq
Auxiliary equations for the fermionic nodes are obtained simply by evaluating (\ref{eq:f1}) and (\ref{eq:f3}) at the respective zeros  $u_{1, k}$ and $u_{3, k}$ of their rhs:
\beqa
1 &=& \left. \frac{ \mQ_{2|2}^- }{\mQ_{2|2}^+ } \right|_{u_{1,k}}  \;\;\;\;\;\;\;\;\;\;\;\;\;\;\;\;\;\;\;,\;\;\;\;\text{ with }\bP_{2}(u_{1,k} ) = 0,\\
1 &=& \left. \frac{Q_{1|1}^{+} }{Q_{1|1}^{-} } \frac{Q^4|_{\,|1}^{+} }{Q^4|_{\,|1}^{-} } \, \frac{ \mQ_{2|2}^- }{\mQ_{2|2}^+ } \right|_{u_{3,k}}  \;\;,\;\;\;\;\text{ with }\mQ_{2|12}(u_{3,k} ) = 0,\label{eq:f3su2}
\eeqa
while the Bethe equation for the second node is obtained by taking the ratio of (\ref{eq:f2}) computed at $u_{2,k}+i/2$ and $u_{2,k}-i/2$:
\beq
-1 = \left. \frac{ \mQ_{2|2}^{--} }{\mQ_{2|2}^{++} } \, \frac{ \mQ_{2|12}^+ }{\mQ_{2|12}^- } \, \frac{ \bP_{2}^+ }{\bP_{2}^- } \right|_{u_{2,k}}  \;\;,\;\;\;\;\text{ with }\mQ_{2|2}(u_{2,k} ) = 0.
\label{eq:2eta1}
\eeq
 In Section \ref{sec:ABA}, we will show that  in the large volume limit these equations reduce to the $\eta=1$ form of the ABA \cite{Gromov:2008qe}. We can describe an alternative grading by using relation $B_{2\ast}$ (\ref{eq:B2*}) instead of $B_2$ for the second node and the fermionic-type equations (\ref{eq:F4}),(\ref{eq:Fbar4}) for the nodes $4$ and $\bar{4}$.
Consider for example the chain of Q functions
\beq
\bQ_2 , \;\;\; \mQ_{2|2}, \;\;\; \mQ_{12|2} , \;\;\;  Q_{1|1}, \;\;\; Q^4_{\,|1} , \label{eq:chain2}
\eeq
connected by the Q-system relations
\begin{align}
&F_1: &&\mQ_{2|2}^+ -\mQ_{2|2}^- = \bP_{2} \, \bQ_{2} , \\
&B_{2\ast}: && \mQ_{1|2}^+ \,  \mQ_{2|2}^- -  \mQ_{2|2}^+ \,  \mQ_{1|2}^- = \mQ_{12|2} \, \bQ_{2},\\
&F_3: &&( Q_{1|1} \, Q^4_{\,|1} )^+ \; \mQ_{2|2}^- -  (Q_{1|1} \, Q^4_{\,|1})^- \; \mQ_{2|2}^+ = \mQ_{12|2} \, \mQ_{2|12} , \\
&F_4: && (Q^4_{\,|1}  )^+ \, Q_{1|3}^- - (Q^4_{\,|3}  )^+ \, Q_{1|1}^- = \mQ_{12|2} , \label{eq:f4a}\\
&F_{\bar4}: && (Q^4_{\,|1}  )^- \, Q_{1|3}^+ - (Q^4_{\,|3}  )^- \, Q_{1|1}^+ =  \mQ_{12|2} \label{eq:f4b}.
\end{align}
Using the pole-free condition, they straightforwardly lead to exact BA equations corresponding to the Dynkin diagram of Figure \ref{fig:dynkinetam1}:
\begin{align}
1 = &\left. \frac{Q^4|_{\,|1}^{++} }{Q^4|_{\,|1}^{--} } \, \frac{ \mQ_{12|2}^- }{\mQ_{12|2}^+ } \right|_{u_{4,k}}, & &\text{with }Q_{1|1}(u_{4,k} ) = 0, \label{eq:f4sl2}\\
1 = & \left. \frac{Q_{1|1}^{++} }{Q_{1|1}^{--} } \, \frac{ \mQ_{12|2}^- }{\mQ_{12|2}^+ } \right|_{u_{\bar{4},k}}, & &\text{with }Q^4_{\,|1}(u_{\bar{4},k} ) = 0 ,\label{eq:fbar4sl2}\\
1= & \left. \frac{Q_{1|1}^{+} }{Q_{1|1}^{-} } \frac{Q^4|_{\,|1}^{+} }{Q^4|_{\,|1}^{-} } \, \frac{ \mQ_{2|2}^- }{\mQ_{2|2}^+ } \right|_{u_{\tilde3,k}}  ,& &\text{with }\mQ_{12|1}(u_{\tilde3,k} ) = 0 ,\\
-1 =& \left. \frac{ \mQ_{2|2}^{--} }{\mQ_{2|2}^{++} } \, \frac{ \mQ_{12|2}^+ }{\mQ_{12|2}^- } \, \frac{ \bQ_{2}^+ }{\bQ_{2}^- } \right|_{u_{2,k}} , & &\text{with }\mQ_{2|2}(u_{2,k} ) = 0 ,\\
1 =& \left. \frac{ \mQ_{2|2}^- }{\mQ_{2|2}^+ } \right|_{u_{\tilde1,k}}, & &\text{with }\bQ_{2}(u_{\tilde1,k} ) = 0 .
\label{eq:f1eta-1}
\end{align}
The main difference with respect to the derivation in the $\eta=+1$ case concerns the equations for the momentum-carrying nodes: for instance, (\ref{eq:f4sl2}) is obtained by taking the ratio of equation (\ref{eq:f4a}) evaluated at $u_{4,k}+i/2$ and equation (\ref{eq:f4b}) at $u_{4,k}-i/2$.
 As shown in the next Section \ref{sec:ABA}, equations (\ref{eq:f4sl2})-(\ref{eq:f1eta-1}) reduce to the $\eta=-1$ version of the ABA of \cite{Gromov:2008qe} in the large-$L$ limit. 

We may also consider subsets of Q functions whose zeros satisfy exact Bethe equations related to the so-called ``distinguished'' grading of the Dynkin diagram. An example of such a chain is:
\beq
\bQ_2 , \;\;\; \mQ_{\emptyset|12}, \;\;\; \mQ_{2|12}, \;\;\;   Q_{1|1}, \;\;\; Q^4_{\,|1}  . 
\eeq
The Bethe equations associated to the momentum-carrying nodes are (\ref{eq:4eta1}), (\ref{eq:bar4eta1}). To constrain the remaining Q functions, we may use $B_{1\ast}$ (\ref{eq:B1*}), $F_{2\ast}$ (\ref{eq:F2*}) and $B_{3\ast}$ (\ref{eq:B3*}) with indices $A,I=1$; $B,J=2$. 
Employing standard arguments, we find the Bethe equations:
\begin{align}
-1 &=&&\left. \frac{\mQ_{\emptyset|12}^{+} }{\mQ_{\emptyset|12}^{-} } \, \frac{ \bQ_2^{--} }{\bQ_2^{++}  } \right|_{u_{\tilde1,k}}  ,&&\text{with }\bQ_2(u_{\tilde1,k} ) = 0 ,& \\
1 &=&& \left. \frac{\mQ_{2|12}^{+} }{\mQ_{2|12}^{-} } \, \frac{ \bQ_2^{-} }{\bQ_2^{+}  } \right|_{u_{2,k}^d}  ,&&\text{with }\mQ_{\emptyset|12}(u_{2,k}^d ) = 0,&\\
-1 &=&& \left. \frac{\mQ_{2|12}^{++}}{\mQ_{2|12}^{--}}\,\frac{(Q_{1|1}Q^4_{\,|1})^-}{(Q_{1|1}Q^4_{\,|1})^+}\,\frac{\mQ_{\emptyset|12}^-}{\mQ_{\emptyset|12}^+}\right|_{u_{3,k}},&&\text{with }\mQ_{2|12}(u_{3,k})=0.&
\end{align}
At the leading weak coupling order these equations reduce to one of the variants of the 2-loop Bethe Ansatz of \cite{Minahan:2008hf}. However, it is well known that this grading is impractical when considering the large-volume limit and does not lead to simple Asymptotic Bethe equations.

\subsection{The ABA limit}
\label{sec:ABA}
Let us now argue that in the large volume limit a subset of Q functions -- in particular, the ones appearing in the chains (\ref{eq:chain1}) and (\ref{eq:chain2}) --  reduces to a simple explicit form parametrized by a finite set of Bethe roots living on two sheets only. The exact BA equations (\ref{eq:4eta1})-(\ref{eq:2eta1}) and (\ref{eq:f4sl2})-(\ref{eq:f1eta-1}) will then be shown to reproduce the Asymptotic Bethe Ansatz of \cite{Gromov:2008qe}. The following argument is very similar to the one presented in \cite{Gromov:2014caa}. 
The main origin of the simplification occurring in the large volume limit is that some of the Q functions vanish  at an exponential rate at large $L$. To keep track of the scaling of different quantities with $L$, we can rely heuristically on the asymptotics discussed in Section \ref{sec:asyQtau}. From (\ref{eq:chargesexcitations1}),(\ref{eq:chargesexcitations2}), we see that the charges scale as $\Delta, J_1, J_2 \sim L$, while $S, J_3 \sim \mathcal{O}(1)$ at large $L$, from which we get for example that
\beq
\nu_a\sim (1,1/\al,1/\al,1/\al^2),\quad \nu^a\sim (1/\al^2,1/\al,1/\al,1), 
\eeq
where $\al \sim u^{-L}$ represents a quantity exponentially suppressed in $L$. Similarly, we have
\beqa\label{eq:scaleQ}
&&Q_{a|i}\sim\left(
\begin{array}{cccc}
1 & \al & \al & \al^2\\
1/\al & 1 & 1 & \al\\
1/\al & 1 & 1 & \al\\
1/\al^2 & 1/\al & 1/\al & 1
\end{array}
\right), \;\;\;\;\;\;\; Q^{a|i}\sim\left(
\begin{array}{cccc}
1 & 1/\al & 1/\al & 1/\al^2\\
\al & 1 & 1 & 1/\al\\
\al & 1 & 1 & 1/\al\\
\al^2 & \al & \al & 1
\end{array}
\right),\\
&& \bP_1 , \bP_2 \sim \al , \;\;\;\; \bP_3 , \bP_4 \sim 1/\al, \;\;\;\; \bP_5, \bP_6 \sim 1 , \\
&& \bQ_1 , \bQ_2 \sim 1/\al , \;\;\;\; \bQ_3 , \bQ_4 \sim \al, \;\;\;\; \bQ_5, \bQ_{\circ} \sim 1 .
\eeqa 
Moreover, since the functions $\tau_i(u)$ approach constants at large $u$, we deduce that they scale as $\mathcal{O}(1)$ in the large volume limit. 
 Using this information, we obtain some simplified relations. Let us list the ones most relevant  for the derivation of the ABA. First, from the scaling (\ref{eq:scaleQ}) we find that (\ref{eq:nutau}) reduces to:
\beq
\nu_a \simeq Q_{a|1}^- \, \tau^1 , \;\;\;\;\; \nu^a \simeq (Q^{a|4} )^- \, \tau_4  . \label{eq:scaleABANu} 
\eeq
Second, from (\ref{eq:Pnu1}) we find, for $\alpha=1, 2$, 
\beq
\widetilde{\bP}_{\alpha} \sim   ( \bar{\sigma}_{\alpha} )^{ab} \, \widetilde{\nu}_a \, \nu_b \sim ( \bar{\sigma}_{\alpha} )^{ab} \, ( Q_{a|1}^+ \, Q_{b|1}^- ) \, \tau^1 \, \tau_4 = \mathcal{Q}_{\alpha|12} \,  \omega^{12} ,
\label{eq:Palpha2}
\eeq
where we used also the identity (\ref{eq:idB4}) in the last step, and we recall that $\omega^{12} = \tau^1 \, \tau_4$.  Similarly, in the large volume limit we have
\beq\label{eq:}
\mu_{12} \simeq \mQ_{12|12}^- \, \omega^{12}.
\eeq
Finally, it will be useful to consider the relation between Q functions analytic in the upper/lower half plane, which simplifies in the large volume limit. In particular, we have
\beq
( Q_{a|i}^{\text{LHPA}} )^-\simeq Q_{a|1}^- \, \( \delta^1_i - \tau^1 \, \tau_i \), 
\label{eq:ULHPAaba} 
\eeq
from which we see that equation (\ref{eq:Palpha2}) can be rewritten as
\beq
\widetilde{\bP}_{\alpha} \sim ( \bar{\sigma}_{\alpha} )^{ab} \, ( Q_{a|4}^{\text{LHPA}} )^+ \, ( Q_{b|4}^{\text{LHPA}} )^-  \, \frac{1}{ \omega^{12} } = \frac{\mathcal{Q}^{\text{LHPA}}_{\alpha|34}}{\omega^{12}}.\label{eq:Palpha2b} 
\eeq
\subsubsection*{Computing $\mu_{12}$, $\omega^{12}$ and $\mathcal{Q}_{12|12}$}
The first part of the argument is essentialy the same as in \cite{Gromov:2014caa}. We shall assume that $\nu_1$ and $\nu^4$ have each a finite number of zeros on the first sheet in physical kinematics, which we denote as $\left\{ u_{4, j} \right\}_{j=1}^{K_4}$, $\left\{ u_{\bar{4}, j} \right\}_{j=1}^{K_{\bar{4}}}$ respectively.
 We start by defining
\beqa\label{eq:defF}
F^2 &\equiv& \frac{\mu_{12} }{\widetilde \mu_{12}} \, \prod_{\balpha=4, \bar{4}} \, \frac{\mathbb{Q}_{\balpha}^+}{ \mathbb{Q}_{\balpha}^-} ,
\eeqa
where we remind the reader that $\mu_{12} = \nu_1 \, \nu^4$ and 
\beq
\mathbb{Q}_4 = \prod_{j=1}^{K_4} \( u - u_{4, j} \),  \;\;\;\;\; \mathbb{Q}_{\bar{4}}= \prod_{j=1}^{K_{\bar4}} \( u - u_{\bar4, j} \) .
\eeq
We will be concentrating on the case of real charges, so that we can take these Baxter polynomials to be real. The function $F(u)$ defined above is manifestly free of poles on the first sheet. Further, we can show that it has only a single branch cut. Indeeed, using (\ref{eq:scaleABANu}), we can rewrite this quantity as 
\beq
F^2
 = \frac{\mathcal{Q}_{12|12}^-}{\mathcal{Q}_{12|12}^+} \, \prod_{\balpha =4, \bar{4}} \, \frac{\mathbb{Q}_{\balpha}^+}{ \mathbb{Q}_{\balpha}^-}  ,
\label{eq:F^2}
\eeq
where the contribution of $\omega^{12}$ cancels due to its $i$-periodicity.  
The expression (\ref{eq:F^2}) shows that, within this approximation, $F^2$ is built out of quantities that have manifestly no cuts in the upper half plane. On the other hand, using (\ref{eq:ULHPAaba}) we see that $F^2$ could equivalently be rewritten in terms of LHPA Q functions only. We therefore conclude that it must have no singularities apart from a short branch cut running on the real axis. The discontinuity across the latter can be determined   from equation (\ref{eq:defF}), and reads
\beq\label{eq:FtildeF}
F\widetilde F=\prod_{\balpha=4, \bar{4}} \, \frac{\mathbb{Q}_{\balpha}^+}{ \mathbb{Q}_{\balpha}^-}.
\eeq
Besides, from (\ref{eq:F^2}) we deduce that $F(u) \rightarrow 1$ as $u \rightarrow \infty$. Supplementing equation (\ref{eq:FtildeF}) with the large-$u$ behaviour $F(u) \sim u^0$ 
already fixes this function in terms of the Bethe roots (but for a sign):
 \beq
 \label{eq:Fsol}
 F= \pm \prod_{\balpha=4, \bar{4}} \frac{B_{\balpha(+)}}{B_{\balpha(-)}}
 ,
 \eeq
 where 
 \beqa
B_{\balpha (\pm) }(u) &=& \prod_{j=1}^{K_{\balpha}} \sqrt{\frac{h}{x_{\balpha, j}^{\mp} } } \( \frac{1}{x(u)} - x^{\mp}_{\balpha,j} \) %
,\;\;\;\;\; x^{\mp}_{\balpha, k}=x( u_{\balpha, k} \mp i/2 ),  \\
 R_{\balpha (\pm) }(u) &=& \tilde B_{\balpha  (\pm) }(u) =\prod_{j=1}^{K_{\balpha}} \sqrt{\frac{h}{x_{\balpha, j}^{\mp} } } \( x(u) - x^{\mp}_{\balpha,j} \).
\label{eq:defBR}
 \eeqa
Plugging (\ref{eq:Fsol}) into (\ref{eq:F^2}), we can now solve for $\mQ_{12|12}^-$. Imposing the correct analyticity in the upper half plane, we find 
\beq\label{eq:Q1212sym}
\mQ_{12|12}= Q_{1|1} \, Q^4_{\,|1} \propto \prod_{\balpha=4,\bar{4}} \mathbb{Q}_{\balpha} \, ( f_{\balpha}^{[+]} )^2 ,
\eeq
where the functions $f_{4}(u)$, $f_{\bar{4}}(u)$ are defined as solutions of the difference equations
 \beq
 \frac{f_{\balpha} }{f_{\balpha}^{[+2]} }=\frac{B_{\balpha (+)}}{B_{\balpha  (-)} } ,
 \eeq
 analytic in the upper half plane and with power-like asymptotics. Apart for an overall factor, they are uniquely fixed by the following integral representation:
\beq
f_s(u) \propto \text{exp}\left( -\int_{-2h}^{2h} \frac{dz}{2 \pi i} \,\log\frac{ B_{s(+)}(z) R_{s(-)}(z) }{ R_{s(+)}(z) B_{s(-)}(z) }\,\partial_z \log\Gamma(i(z-u)  ) \right).
\eeq 
 Finally, one can determine $\mu_{12}$ imposing that it has the right discontinuity given by (\ref{eq:defF}) and that it satisfies $\mu_{12} \simeq \mQ_{12|12}^- \, \omega^{12}$, where $\omega_{12}$ should be an $i$-periodic function. The result is
\beq
\mu_{12}=\nu_1\nu^4\propto \prod_{\balpha=4, \bar{4} } f_{\balpha} \,  \bar f_{\balpha}^{[-2]} \, \,  \mathbb{Q}_{\balpha}^- , \quad \quad \omega^{12} =  \tau^1 \, \tau_4 \propto \prod_{\balpha=4,\bar{4}} \frac{ \bar{f}^{[-2]}_{\balpha}  }{ f_{\balpha}  } .
\label{eq:nunutautau}
\eeq
Indeed, one can easily verify that $\omega^{12}$ in (\ref{eq:nunutautau}) is $i$-periodic due to the reality of the set of Bethe roots.
\subsubsection*{Zero momentum condition and anomalous dimension}
Already at this stage, we can prove that the zero momentum condition (\ref{eq:ZMC0}) is contained in the QSC equations. Indeed, from (\ref{eq:nunutautau}) we have:
\beqa
\frac{\widetilde \mu_{12}}{\mu_{12} } = \prod_{\bu=4, \bar{4}} \frac{R_{\bu (+)} B_{\bu (-)}}{B_{\bu (+)} R_{\bu (-)}}, \label{eq:ratiomuABA}
\eeqa
in the ABA limit. Due to the mirror $i$-periodicity of $\mu_{12}$, the lhs of (\ref{eq:ratiomuABA}) should approach $1$ at large $u$. Expanding the rhs, we find 
\beq
\left( \prod_{j=1}^{K_{4} }\frac{x_{4, j}^+ }{x_{4, j}^-} \right) \, \left( \prod_{j=1}^{K_{\bar{4}} }\frac{x_{\bar{4}, j}^+ }{x_{\bar{4}, j}^-} \right) = 1 ,\label{eq:ZMC}
\eeq
which coincides with the zero-momentum condition (\ref{eq:ZMC0}) taking into account the relation between rapidity and momentum $p_{4, j} = -i \log( x_{4, j}^+/x_{4, j}^- )$, $p_{\bar{4}, j} = -i \log( x_{\bar{4}, j}^+/x_{\bar{4}, j}^- )$. 
The next order in the large-$u$ expansion can be compared with the  asymptotics (\ref{eq:NaNi})-(\ref{eq:Qaiasym}), and fixes the ABA limit of the anomalous dimension:
\beq
\gamma = 2 h i \sum_{j=1}^{K_4}\left(\frac{1}{x_{4, j}^+ }-\frac{1}{x_{4, j}^-} \right)  + 2 h i \sum_{j=1}^{K_{\bar{4}}}\left(\frac{1}{x_{\bar{4}, j}^+ }-\frac{1}{x_{\bar{4}, j}^-} \right).
\eeq
\subsubsection*{Computing $\nu_1$, $\nu^4$}
We now notice that the ratio between $Q_{1|1}$ and $Q^4_{\,|1} $ must be, in the large-$L$ limit, a meromorphic function without branch cuts. Indeed, equation (\ref{eq:ULHPAaba}) shows that
\beq
{ Q_{1|1} }/{Q^4_{\,|1} } \simeq {  Q_{1|1}^{\text{LHPA}} }/{ ( Q^4_{\,|1} )^{\text{LHPA}}  }.
\label{eq:ratio}
\eeq
Taken together, the regions of analyticity  of the two sides of (\ref{eq:ratio}) cover all the complex plane, showing that this ratio  indeed has no cuts. Therefore, ${ Q_{1|1} }/{Q^4_{\,|1} }$ must be a rational function of $u$. Combined with (\ref{eq:Q1212sym}), this shows that
\beq
Q_{1|1} \propto \mathbb{Q}_{4} \,  f^+_4 \, f_{\bar{4}}^+\,,\quad Q^{4}_{\,|1} \propto  \mathbb{Q}_{\bar{4}} \,  f^+_4 \, f_{\bar{4}}^+  .
\label{eq:Q11Q41}
\eeq
Let us now introduce the following parametrization
\beq
\nu_1 \propto \mathbb{Q}^-_4 \, \left( \prod_{\balpha=4, \bar{4} } f_{\balpha} \,  \bar f_{\balpha}^{[-2]} \right)^{\frac{1}{2} } \, \mathcal{F} \, e^{-i \sigtw/2 }, \;\;\;\;\; \nu^4 \propto \mathbb{Q}^-_{\bar{4}} \,  \left(\prod_{\balpha=4, \bar{4} } f_{\balpha} \,  \bar f_{\balpha}^{[-2]} \right)^{\frac{1}{2} } \, \mathcal{F}^{-1} \,  e^{+i \sigtw/2 } ,\label{eq:nuABA2}
\eeq
for some function $\mathcal{F}$ which should be free of zeros on the first sheet. The factors $e^{\pm i \sigtw/2}$ , with $\mathcal{P}$ defined in (\ref{eq:perioanti}), have been introduced for future convenience. To fix the form of the splitting factor $\mathcal{F}$ we should enforce the properties $\widetilde \nu_1 = e^{i \sigtw} \, \nu_1^{[+2]}$, $(\tau^1 )^{[+2]} = -e^{-i \sigtw} \, \tau_4$, and using (\ref{eq:scaleABANu}),(\ref{eq:Q11Q41}) we obtain the conditions
\beq
\mathcal{F}^{[+2]}=\mathcal{F}^{-1},\;\;\;\;\;
\mathcal{F} \widetilde{\mathcal{F} } =  \left(\frac{\mathbb{Q}_4^+}{\mathbb{Q}_4^-}\frac{\mathbb{Q}_{\bar4}^-}{\mathbb{Q}_{\bar4}^+}\right)^{\frac{1}{2} } \, e^{i \sigtw}.\label{eq:Ftil}
\eeq
The solution of the constraints (\ref{eq:Ftil}) may be found in terms of an integral representation\footnote{ A detailed derivation of essentially the same formula is given in another context in \cite{Gromov:2014eha}.}:
\beq
\log\mathcal{F}(u) =\sqrt{e^{2 \pi u} - e^{4 \pi h} }\sqrt{e^{2 \pi u} - e^{-4 \pi h} }\, \int_{-2 h}^{2 h} \frac{\log(\frac{\mathbb{Q}_4^+(z)}{\mathbb{Q}_4^-(z)}\frac{\mathbb{Q}_{\bar4}^-(z)}{\mathbb{Q}_{\bar4}^+(z)} \, e^{2 i \sigtw} ) \, e^{\pi (u+z)}}{\sqrt{(e^{2 \pi z} -e^{4 \pi h}) \, (e^{2 \pi z} - e^{-4 \pi h} )} \, (e^{2 \pi z} -e^{2 \pi u} )} \, \frac{dz}{2 i}.\label{eq:logFABA}
\eeq
We should also impose that $\log\mathcal{F}(u)$ has the correct bounded asymptotic behaviour as $u \rightarrow +\infty$, which leads to the condition
\beq
\sigtw=-\frac{1}{4 \pi \, \mathbb{E}(h) } \, \int_{-2 h}^{2 h}\frac{ \log(\frac{\mathbb{Q}_4^+(z)}{\mathbb{Q}_4^-(z)}\frac{\mathbb{Q}_{\bar4}^-(z)}{\mathbb{Q}_{\bar4}^+(z)} ) \, e^{ \pi z} }{\sqrt{ (e^{2 \pi z} -e^{4 \pi h}) \, (e^{2 \pi z} - e^{-4 \pi h} )} } \, dz , \label{eq:PABA} %
\eeq
where 
\beq\label{eq:Edef}
\mathbb{E}(h) \equiv -\frac{1}{2 \pi i } \int_{-2h}^{2h} \frac{dz \, e^{\pi z}  }{\sqrt{(e^{2 \pi z}- e^{4 \pi h} ) \, (e^{ 2 \pi z}- e^{- 4 \pi h} ) } } .
\eeq
Expanding (\ref{eq:PABA}) for small $h$, we see that it confirms the identification (\ref{eq:idePmom}) up to order $\mathcal{O}(h^2)$. Indeed, notice that the ABA expression for the total momentum of a single excitation species is given by: 
\beqa
P^{({4})}_{\text{ABA}} &=& \frac{1}{2}( P^{({4})}_{\text{ABA}}-P^{(\bar{ {4}})}_{\text{ABA}} ) =\frac{1}{2} \left( \sum_{i=1}^{K_4} p^{\text{ABA}}_{4, i} - \sum_{i=1}^{K_{\bar{4}} } p^{\text{ABA}}_{\bar{4}, i} \right) \nn \\
&=& -\frac{i}{2} \left(\sum_{i=1}^{K_4} \log\frac{x_{4, i}^+}{x_{4, i}^-} - \sum_{i=1}^{K_{\bar{4}} } \log\frac{x_{\bar{4}, i}^+}{x_{\bar{4}, i}^-} \right) = \frac{1}{2 \pi i} \, \int_{-2 h}^{2 h}\frac{ \log(\frac{\mathbb{Q}_4^+(z)}{\mathbb{Q}_4^-(z)}\frac{\mathbb{Q}_{\bar4}^-(z)}{\mathbb{Q}_{\bar4}^+(z)} ) }{\sqrt{ 4 h^2 - z^2 } } \, dz ,\nn
\eeqa
which agrees with the rhs of (\ref{eq:PABA}) at the first two orders at weak coupling. Further, one can verify that  the lowest transcendentality part of (\ref{eq:PABA}), seen as  a function of the positions of the Bethe roots, exactly agrees with $P^{({4})}_{\text{ABA}}$.  

As already discussed in Section \ref{subsec:P}, (\ref{eq:PABA}) is expected to hold only in the large-$L$ limit, or at the first $\sim L$ orders at weak coupling. A general exact integral formula for $\sigtw$, expressed in terms of quantities computable form the numerical solution of the QSC, can be found in Appendix \ref{app:Prec}. 

\subsubsection*{Computing $\bP_{\alpha}$, $\mQ_{\alpha|12}$ and $\mQ_{\alpha|\beta}$}
Let us now derive the ABA limit of $\bP_{\alpha}$, with $\alpha=1,2$ (again, we follow \cite{Gromov:2014caa} closely). 
Apart for an irrelevant factor, we define two functions $\sigmaBES_4$, $\sigmaBES_{\bar{4}}$ through
\beq\label{eq:crossing}
\sigmaBES_{\balpha} \, \widetilde\sigmaBES_{\balpha} \propto  \bar{f}_{\balpha}^{[-2]} \, f_{\balpha}^{[+2]} , \;\;\;\; \balpha=4, \bar{4} ,
\eeq
with the requirement that they have a single short cut connecting $\pm 2h$ and no other singularities on their defining sheet, and are bounded at infinity. Notice that equation (\ref{eq:crossing}) can be recognized as one of the crossing equations, and in particular $\sigmaBES_4$, $\sigma_{\bar{4}}$ are simply related to the Beisert-Eden-Staudacher dressing factor \cite{Beisert:2006ez} as in:
\beq
\sigma^+_{\balpha}(u)/\sigma^-_{\balpha}(u)= \prod_{j=1}^{K_{\balpha} } \sigma_{\text{BES}}(u , u_{\balpha, j} ), \;\;\;\; \balpha=4, \bar{4} .
\eeq
 Let us consider the quantity $\bP_{\alpha} /( \sigmaBES_4 \, \sigmaBES_{\bar{4}} )$, which by construction has a Laurent series expansion in $1/x(u)$. Using (\ref{eq:Palpha2}),(\ref{eq:Q1212sym}), we see that, on the second sheet, it may be written as
\beq
\widetilde\bP_{\alpha}/(\widetilde\sigmaBES_4 \, \widetilde\sigmaBES_{\bar{4}}) \sim \mQ_{\alpha|12} \, \omega^{12} /(\widetilde\sigmaBES_4 \, \widetilde\sigmaBES_{\bar{4}}) \propto  \mQ_{\alpha|12} \prod_{\balpha=4, \bar{4}} \, \sigmaBES_{\balpha}/( f_{\balpha} \, f_{\balpha}^{[+2]} ) ,
\eeq
which has no cuts in the upper half plane, or alternatively from (\ref{eq:Palpha2b}) as
\beq
\widetilde\bP_{\alpha}/(\widetilde\sigmaBES_4 \, \widetilde\sigmaBES_{\bar{4}}) \propto \mathcal{Q}^{\text{LHPA}}_{\alpha|34} \,\prod_{\balpha=4, \bar{4}} \, \sigmaBES_{\balpha}/(\bar{f}_{\balpha} \, \bar{f}_{\balpha}^{[-2]}),
\eeq
which has no cuts in the lower half plane. Hence, we find that $\bP_{\alpha}/(\sigmaBES_4 \,\sigmaBES_{\bar{4}}) $ must have a single cut also on the second sheet. Therefore, it is a two-sheeted function with power-like asymptotics everywhere, which implies that it can be written as a rational function in the Zhukovsky variable $x(u)$. Moreover, this function cannot have any poles at finite $u$. All these constraints fix
\beq
\bP_{\alpha} \propto x^{-L} \, B_{\alpha|12}\, R_{\alpha|\emptyset} \; \sigmaBES_4 \, \sigmaBES_{\bar{4}} , \quad \alpha=1,2 ,
\label{eq:Palpha1}
\eeq
where the  $x^{-L}$ prefactor is fixed by the large-$u$ asymptotics (\ref{eq:Pasy}), and the factors $R_{\alpha|\emptyset}(u)$ and $B_{\alpha|12}(u)$  denote generic polynomials in $x(u)$  and $1/x(u)$, respectively. 
 By consistency with (\ref{eq:Palpha2}), we then find:
\beq
\mathcal{Q}_{\alpha|12} \propto x^{+L} \,  R_{\alpha|12}\, B_{\alpha|\emptyset} \, \prod_{\balpha=4,\bar{4}} f_{\balpha}\, f_{\balpha}^{++} /\sigmaBES_{\balpha} ,\quad \alpha=1,2 \,,
 \label{eq:Qalpha12}
\eeq
where $R_{\alpha|12}(u)=\widetilde B_{\alpha|12}(u)$ and $B_{\alpha|\emptyset}(u) = \widetilde R_{\alpha|\emptyset}(u)$ are obtained through analytic continuation, which sends $x(u) \rightarrow 1/x(u)$. 
At this stage, we have computed four of the functions entering the chain (\ref{eq:chain1}); to complete the picture we still need to compute the Q functions corresponding to the second node. We start from relation 
\beq
Q_{1b|1j}^- = (Q^{\text{LHPA}}_{1b|1j} )^- \, \( 1  - \tau^1 \tau_1 \), \;\;\;\; \forall b, j ,
\eeq
which is a consequence of (\ref{eq:ULHPAaba}), and implies that ratios of the form\footnote{Notice the restriction of the indices to the set $ \left\{1,2\right\}$. This ensures that the ratios in (\ref{eq:Qratios}) are of order $\mathcal{O}(1)$ for large $L$, which is a prerequisite condition for obtaining nontrivial information in the asymptotic limit. }
\beq\label{eq:Qratios}
{ \mQ_{\alpha|\beta} }/{\mQ_{\alpha'|\beta'} } = { \mQ_{\alpha|\beta}^{\text{LHPA}} }/{\mQ_{\alpha'|\beta'}^{\text{LHPA}} }, \quad \quad \alpha, \beta, \alpha', \beta' \in \left\{1,2\right\},
\eeq
have no cuts and are therefore ratios of polynomials. 
We have therefore a parametrization
\beq
\mathcal{Q}_{\alpha|\beta}=\mathbb{Q}_{\alpha|\beta}\, f_4^+ \,f_{\bar4}^+ ,\ \alpha,\beta \in \left\{1,2 \right\} ,
\label{eq:Qalphabeta}
\eeq
where $\mathbb{Q}_{\alpha|\beta}$ is a polynomial function of  $u$,  and the $f_{4} \, f_{\bar{4}}$ factor was fixed by comparison with (\ref{eq:Q1212sym}). 

\subsubsection*{Asymptotic Bethe Ansatz in $\eta=+1$ grading}
Generalizing the arguments of Section \ref{sec:exactBA}, we see that the Q functions
\beqa\label{eq:QchainABA}
\bP_{\alpha}, \quad \mQ_{\alpha|\beta} , \quad \mQ_{\alpha|12}, \quad Q_{1|1}, \quad Q^4_{\;|1}, 
\eeqa
for any choice of $\alpha, \beta \in \left\{1,2\right\}$, satisfy exact Bethe equations of the form (\ref{eq:4eta1})-(\ref{eq:f3su2}). Using (\ref{eq:Q11Q41}), (\ref{eq:Palpha1}), (\ref{eq:Qalpha12}), (\ref{eq:Qalphabeta}), it is straightforward to verify that, in the large volume limit, these Bethe equations reduce precisely to the ABA of \cite{Gromov:2008qe} in $\eta=+1$ grading (see Appendix \ref{app:dictionary}). 
In each of these four equivalent sets of ABA equations, the role of roots of types $1$,$2$,$3$, is played by the zeros of the following polynomials in $u$: $\mathbb{Q}_{\alpha|\emptyset}(u) = R_{\alpha|\emptyset}(u) \, B_{\alpha|\emptyset}(u)$,  $\mathbb{Q}_{\alpha|\beta}(u)$, $\mathbb{Q}_{\alpha|12}(u) = R_{\alpha|12}(u) \, B_{\alpha|12}(u)$, respectively. 
 With our conventions for the ordering of asymptotics, the chain of Q functions with $\alpha=\beta=2$ in (\ref{eq:QchainABA}) gives the   simplest representation of the state, since it contains the least  number of Bethe roots on each node. 

\subsubsection*{Computing $\bQ_1$ and $\bQ_2$}
The large volume limit of $\bQ_{\beta}$  with $\beta=1,2$, may be computed from the Q-system relation $F_1$, namely:
\beq
\bP_{\alpha} \, \bQ_{\beta} = \mQ_{\alpha|\beta}^+ - \mQ_{\alpha|\beta}^- , 
\eeq
for $\alpha, \beta \in \left\{1,2\right\}$. Similarly, $\mQ_{12|\beta}$ may be determined from the $F_3$ equation: 
\beq
\mQ_{\alpha|12} \, \mQ_{12|\beta} = (Q_{1|1} \, Q^4_{\,|1} )^+ \, \mQ_{\alpha|\beta}^- -  (Q_{1|1} \, Q^4_{\,|1} )^- \, \mQ_{\alpha|\beta}^+ .
\eeq
Using the large-$L$ expressions (\ref{eq:Palpha1}), (\ref{eq:Qalpha12}) and (\ref{eq:Qalphabeta}), these relations yield
\beq
\bQ_{\alpha} \propto x^{L} \, R_{\emptyset|\alpha}\, B_{12|\alpha} \, \prod_{\balpha=4,\bar{4}} \frac{ f_{\balpha}^{++}  }{  B_{\balpha (-)} \, \sigmaBES_{\balpha} },\quad \mathcal{Q}_{12|\alpha} \propto x^{-L}\,  B_{\emptyset|\alpha}\, R_{12|\alpha} \, \prod_{\balpha=4, \bar{4}} \, \sigmaBES_{\balpha} \,  f_{\balpha}^{++}  \,  B_{\balpha (+)}, \label{eq:Q2}
\eeq
where the functions $R_{\emptyset|\alpha}$ and $R_{12|\alpha}$ ($B_{\emptyset|\alpha}$ and $B_{12|\alpha}$, respectively) are polynomials in $x(u)$ ($1/x(u)$)  defined by
\beqa
\label{eq:fermdual1}
&&R_{\alpha|\emptyset} \, R_{\emptyset|\beta} \, B_{12|\beta} \, B_{\alpha|12} \propto 
\left(\mathbb{Q}_{\alpha|\beta}^+ \, B_{4 (-)} \, B_{\bar{4} (-)}- \mathbb{Q}_{\alpha|\beta}^- \, B_{4 (+)} \,  B_{\bar{4} (+)} \right), \\
&& B_{\alpha|\emptyset} \, B_{\emptyset|\beta} \, R_{12|\beta} \, R_{\alpha|12} \propto 
\left(\mathbb{Q}_{\alpha|\beta}^+ \, R_{4 (-)} \,  R_{\bar{4} (-)}- \mathbb{Q}_{\alpha|\beta}^- \, R_{4 (+)} \,  R_{\bar{4} (+)} \right) .
\label{eq:fermdual3}
\eeqa
Notice that the fact that the newly defined $R$ and $B$ functions have no poles is a consequence of the ABA.
Equations (\ref{eq:fermdual3})-(\ref{eq:fermdual3}) are the well-known fermionic duality relations, which allow to switch between the $\eta=\pm 1$  versions of the ABA, see Section \ref{app:Fermionic}. 
 Using (\ref{eq:Q11Q41}), (\ref{eq:Qalphabeta}), (\ref{eq:Palpha1}), (\ref{eq:Q2}),  we may indeed check that the exact Bethe Ansatz satisfied by the chains of Q functions
\beqa
\bQ_{\beta}, \quad \mQ_{\alpha|\beta} , \quad \mQ_{12|\beta}, \quad Q_{1|1}, \quad Q^4_{\;|1},  
\eeqa
which in particular involves the fermionic form of the massive node equations, (\ref{eq:f4sl2}),(\ref{eq:fbar4sl2}), reduce precisely to the $\eta=-1$ ABA equations. 

\subsubsection*{Classical limit from the large-volume solution}
 Let us briefly discuss how the large volume solution can be used to obtain the semi-classical approximation (\ref{eq:classid1})-(\ref{eq:classid23}) (here, we follow closely Section 6 of \cite{Gromov:2014caa}). 
To this end, we exploit the important fact that the classical spectral curve emerges from a continuum limit of solutions of the ABA, where the length and the number of Bethe roots scale with the coupling as $L \sim K_s \sim h \rightarrow \infty$, and the roots condense on a finite number of contours in the $x$-plane; the quasi-momenta are explicitly parametrized in terms of the root densities \cite{Kazakov:2004qf,BKSZ,AFS}. The ABJM case is discussed in detail in \cite{Gromov:2008qe}.
 
Following \cite{Gromov:2014caa}, we will study this limit starting from the large-volume identity
\beq\label{eq:P2pP2m}
\frac{\bP_2^+(u)}{\bP_2^-(u)} \simeq \left(\frac{x^-(u)}{x^+(u)}\right)^{L}\frac{R_{2| \emptyset}^+(u)\,B_{2|12}^+(u)}{R_{2|\emptyset}^-(u)\,B_{2|12}^-(u)}\,\frac{\sigma^+_4(u) \sigma_{\bar{4}}^+(u)}{\sigma^-_4(u) \sigma_{\bar{4}}^-(u)}.
\eeq
At strong coupling, the natural variable is $z = u/h$. Sending $u \sim  h\rightarrow \infty$ with $z$ finite, the lhs of (\ref{eq:P2pP2m}) can be approximated as $i \partial_z \log \bP_2/h$. To treat the rhs we use $x^{\pm} \sim x \pm i x^2/(x^2-1)/h$, and the strong coupling limit of the dressing factor described by the AFS phase \cite{AFS}. Introducing the resolvents $H_s(x) = \sum_{j=1}^{K_s} \frac{ x^2 }{h (x-x_{s, j}) \, (x^2-1) }$, which remain finite in the classical limit, we find
\beq\label{eq:resolvents}
i\partial_z \log \bP_2/h \sim -i \frac{L/h \, x - 2 \pi m }{(x^2 -1)} + i \, ( H_1(x) + H_3(1/x) - H_4(1/x) - H_{\bar{4}}(1/x) ) + \mathcal{O}(1/h), 
\eeq
where $m \in \mathbb{Z}$ is same integer appearing in (\ref{eq:invsym}), related to the total momentum $
 2 \pi m \equiv P^{(4)}_{\text{ABA}} + P^{(\bar{4})}_{\text{ABA}} $, and we are labelling the roots appearing in (\ref{eq:P2pP2m}) as $u_{(2|\emptyset), j} \equiv  u_{1, j}$, $u_{(2|12), j} \equiv u_{3, j}$ according to their role as solutions of the ABA in $\eta=1$ grading. Comparing with \cite{Gromov:2008qe}, we see that   the rhs of (\ref{eq:resolvents}) is one of the quasi-momenta. Therefore we have
\beq
i\partial_z \log \bP_2 / h \sim -i q_3(x) + \mathcal{O}(1/h) ,
\eeq 
 which establishes one of the relations in (\ref{eq:classid23}). 
 
  The above reasoning can be repeated for the other Q functions that we have determined in the large-volume limit, such as $\bP_1$, $\bQ_1$ and $\bQ_2$, and confirms the corresponding semi-classical approximations in (\ref{eq:classid23}). As a technical comment, we point out that each of these functions is parametrized in terms of a different equivalent set of auxiliary Bethe roots; however, using duality equations such as (\ref{eq:fermdual1}),(\ref{eq:fermdual3}), it is straightforward to convert all classical expressions in terms of the same solution of the ABA, so that they can be compared with the resolvent expressions e.g. in \cite{Gromov:2008qe}. Indeed, it is well-known that duality transformations in the ABA simply amount to a relabelling of the sheets of the classical curve (see \cite{BKSZ,CompleteOneLoop}). 

Finally, let us briefly discuss the proof of the semi-classical limit of $\bP_5$ (the case of $\bP_6$ is analogous). 
We start from one of the equations in (\ref{eq:Pnu1}):
\beq\label{eq:rhsP5}
\bP_5 = \widetilde \nu_1 /\nu^4  - \bP_1 \, \nu^2/\nu^4 - \bP_2 \, \nu^3/\nu^4.
\eeq 
 In the classical limit, the last two terms on the rhs of this identity are suppressed, since $ \bP_1$ and $\bP_2$ are exponentially small and the ratios $\nu^a/\nu^b$, being $i$-periodic on the mirror section, are constants by the argument described in Section \ref{sec:classical}. Therefore, in this limit we have\footnote{By a slightly more refined analysis, one can argue that (\ref{eq:P5class}) is not only valid classically, but also  in the large-volume limit. In fact, all terms on the rhs of (\ref{eq:rhsP5}) scale with the same power of $u$ at large value of the spectral parameter, multiplied by  a coefficient which can be determined in terms of the charges. Inspecting these coefficients one finds that the last two terms on the rhs are suppressed by a power of $1/L$ at large volume. }
\beq\label{eq:P5class}
\bP_5 \sim \widetilde \nu_1/\nu^4 .
\eeq
 We can now evaluate the classical scaling of the rhs of (\ref{eq:P5class}) starting from its large-volume approximation.  This finally yields:
\beqa
 i/h \, \partial_z \log \bP_5  &\sim & i/h \, \partial_z \left( \widetilde \nu_1/\nu^4 \right)_{\text{classical}} = i/h \, \partial_z  \log \left( \frac{R_{4(+)} \, B_{4 (-) }}{R_{\bar{4} (-) } \, B_{\bar{4} (+) } } \right)_{\text{classical}} \\
 &=& i \left( H_{4}(x) + H_4(1/x) - H_{\bar{4}}(x) - H_{\bar{4}}(1/x) \right) = i q_5(x) .
\eeqa

\section{Conclusions}
\label{sec:conclusions}

In this paper, besides a detailed derivation of the equations proposed in \cite{Cavaglia:2014exa}, we presented several new results on the Quantum Spectral Curve associated to the $AdS_4/CFT_3$ duality, deepening our understanding of the basic integrable structures underlying this theory. 

There are many directions for future work. First of all, the results of this paper make it possible to develop a high-precision numerical algorithm for the computation of anomalous dimensions at finite coupling, inspired by \cite{Gromov:2015wca}. We already have partial results \cite{IGSTPoster,ContiNumerical} confirming the TBA data of \cite{LevkovichMaslyuk:2011ty}. The QSC method however allows us to move deeper in the strong coupling region, and therefore to test more accurately the AdS/CFT predictions. 

Secondly, we expect from the example of $AdS_5/CFT_4$ \cite{Drukker:2012de,Correa:2012hh,QSCCusp} that the QSC may be used, with minimal modifications, to describe also various open string configurations. In particular, it would be very interesting to find an integrable description of some kind of generalized cusp anomalous dimension, such as the one described in \cite{Griguolo:2012iq}. This would give a direct way to test the proposals of  \cite{Gromov:2014eha,Cavaglia:2014exa} for the ABJM/ABJ interpolating functions, by comparison with localization results for the Brehmsstrahlung function \cite{Aguilera-Damia:2014bqa,Forini:2012bb,Lewkowycz:2013laa, Bianchi:2014laa}. 

Third, these results should allow to extend the weak coupling algorithm of \cite{Anselmetti:2015mda} to a generic operator. 

It would be very interesting to gain a complete understanding of the algebraic structures underlying our results. Especially, it would be desirable to understand the interpretation of the Q-system described in Section \ref{sec:Qsystem} in terms of representation theory of the full supergroup $OSp(4|6)$. 

 We hope that the results presented in this paper, which exhibit some interesting differences from the  $AdS_5/CFT_4$ case, will also help to extend the QSC method to the integrable examples of $AdS_3/CFT_2$ and $AdS_2/CFT_1$, see e.g. \cite{Zarembo:2010sg,Wulff:2014kja, SfondriniReview,Hoare:2014kma}. These cases are less supersymmetric, and the construction may be expected to be even more complicated. It is important to stress that, since a TBA formulation for these models is at present still missing (and even the structure of the Asymptotic Bethe Ansatz is quite intricate and fully known only in one case, see \cite{Borsato:2016xns}), there is presently no way to rigorously derive the QSC for these theories. However, the two examples at hand, $AdS_5/CFT_4$ and $AdS_4/CFT_3$, show that the structure of the QSC is, in the end, quite  universal and rigidly constrained by the symmetry. It would be very nice if these examples could help to develop a classification of several types of QSC corresponding to different gauge and string theories.

\section*{Acknowledgements}
We thank Mikhail Alfimov, Lorenzo Anselmetti, Lorenzo Bianchi, Riccardo Conti, Martina Cornagliotto,  Vladimir Kazakov, Fedor Levkovich-Maslyuk, Christian Marboe, Carlo Meneghelli, Stefano Negro, Georgios Papathanasiou, Grigory Sizov, Alessandro Torrielli, Cristian Vergu and Dmytro Volin for interesting discussions and suggestions. 

In particular, we thank Riccardo Conti for collaboration on the project \cite{ContiNumerical}, during which many aspects of the present work were elucidated. 

This project was partially supported by the  INFN (I.S. {FTECP} and {GAST}),  UniTo-SanPaolo  Nr TO-Call3-2012-0088 {\it ``Modern Applications of String Theory'' ({MAST})}, ESF Network {HoloGrav} (09-RNP-092 (PESC)), MPNS--COST Action MP1210 {\it ``The String Theory Universe''}, and the EU network GATIS. 
AC thanks King's College London for kind hospitality during two visits in 2016, where part of this work was done. %

\appendix

\section{Derivation of the QSC from the analytic properties of T functions}
\label{app:derivations}
In this Appendix we present in detail the derivation of the QSC equations from the TBA/T-system framework, which was already outlined in \cite{Cavaglia:2014exa}. In particular, we will obtain the QSC equations in the ``$\bP\mu$'' vector form presented in Section \ref{sec:PmuForm}. 

\subsection{Summary on the properties of T functions}
\begin{figure}[t!]
\centering
\includegraphics[scale=0.4]{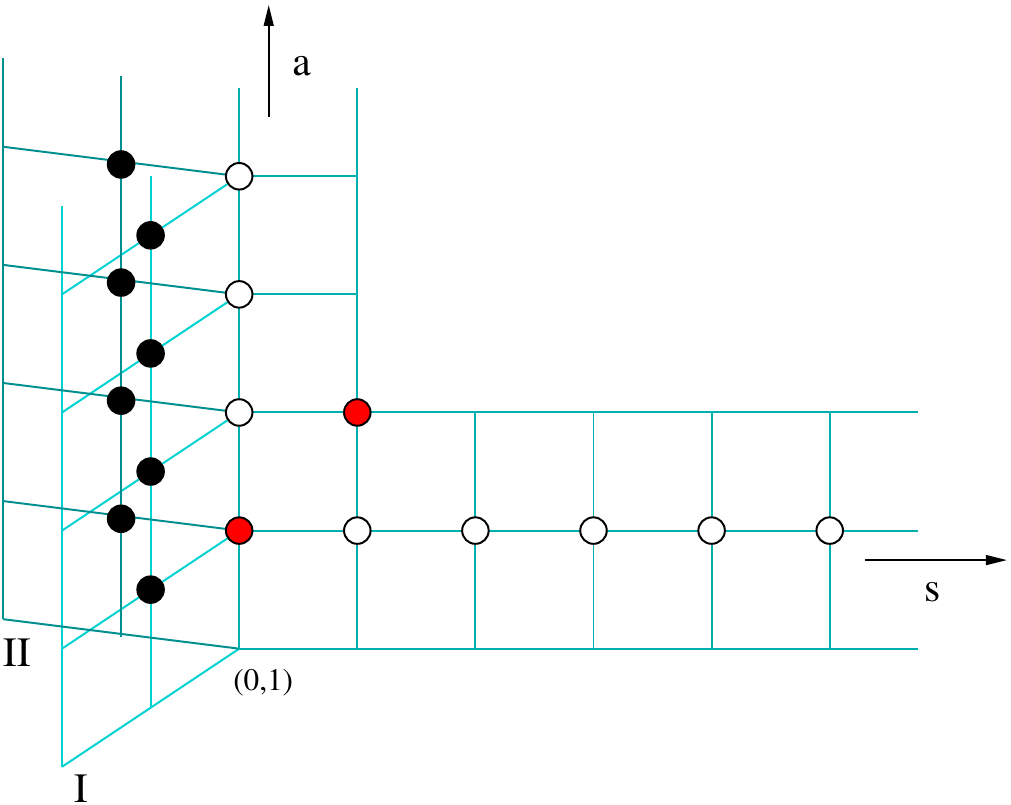}
\caption{Domain of definition of the T-system (\ref{eq:Hirota})-(\ref{eq:Hirotalast}). In our notations, T functions belonging to the two wings of the diagram are distinguished by the superscript $\alpha\in \left\{I, II\right\}$. }
\label{fig:Thook}
\end{figure}

Let us briefly summarize the starting point of the derivation (see \cite{ABJMdisco} for more details). 
 The discrete Hirota equation, or T-system, is the following difference equation for a set of T functions defined on the nodes of the ``T-hook'' diagram shown in 
 Figure \ref{fig:Thook}:
\beqa\label{eq:TT}\label{eq:Hirota}
T^{(+1)}_{a, s} T^{(-1)}_{a, s} &=& \prod_{(a' \sim a )_{\updownarrow} } T_{a', s} + \prod_{(s' \sim s )_{\leftrightarrow} } T_{a, s'}, \;\;\;\;\text{  for   } s > 0 ,\\
(T^{\alpha} )^{(+1)}_{a, 0} (T^{\beta} )^{(-1)}_{a, 0} &=& T_{a+1, 0}^{\alpha} \, T_{a-1, 0}^{\beta} + T_{a, 1} \,  T^{\beta}_{a, -1}, \;\;\;\; \alpha, \beta \in \left\{I, II\right\},\;\; \alpha \neq \beta  ,\\
(T^{\alpha} )^{(+1)}_{a, -1} (T^{\beta} )^{(-1)}_{a, -1} &=& T_{a+1, 0}^{\alpha} \, T_{a-1, 0}^{\beta} + T_{a, 1} \,  T^{\beta}_{a, -1},\;\;\;\; \alpha, \beta \in \left\{I, II\right\},\;\; \alpha \neq \beta ,\label{eq:Hirotalast}
\eeqa
where T functions with indices outside the diagram are taken to be zero and the products in (\ref{eq:Hirota}) are over horizontal ({\scriptsize$\leftrightarrow$}) and vertical ({\scriptsize$\updownarrow$}) neighbouring nodes,  with the subtlety that, for $s=0, -1$, the two wings of the diagrams need to be crossed\footnote{This subtlety was not reported in \cite{Cavaglia:2014exa} but was fully explained in \cite{ABJMdisco}.}.
 Notice that $T^{(n)} = T( u + \frac{i}{2}n )$ denotes shifts on a specific section of the $u$ domain where all cuts are long, connecting $\pm 2 h + i \mathbb{Z}$ to infinity.  
This is called the \emph{mirror} section and is the one where the Y-system and T-system are naturally defined \cite{Extended}. Throughout this Appendix we will use the special notation $f^{(n)}(u) \equiv f(u+i n/2 )$ to denote a function shifted on this particular sheet. \\
 T functions are related to Y functions, the objects appearing in the TBA formulation, by 
\beqa\label{eq:TtoY}
\hspace{-2cm}Y_{a,s} &=& \frac{\prod_{(s' \sim s )_{\leftrightarrow} } T_{a , \, s'}}{ \prod_{(a' \sim a )_{\updownarrow} } T_{a' , s}}, \;\;\;\; s > 0 ,
 \;\;\;\;\;\;\; Y^{\alpha}_{a, 0} = \frac{ T_{a , 1} \, T_{a, -1}^{\beta}}{ T_{a+1 ,  0}^{\alpha} \, T_{a-1, 0}^{\beta}}, \;\;\;\; \alpha, \beta \in \left\{I, II\right\},\;\; \alpha \neq \beta .
\eeqa 
This parametrization is not unique: there is a vast ``gauge'' freedom (which we will exploit) in choosing a set of T functions corresponding to a given solution of the TBA. 
 In order to furnish a complete formulation of the spectral problem, the T-system must be supplemented by some information on its analytic dependence on the spectral parameter. As learnt in the $AdS_5/CFT_4$ case, this extra input can be expressed in terms of discontinuity relations for the $Y(u)$ functions across their branch cuts in the $u$-plane \cite{Extended}, but can be simplified and much better understood in the T-system framework \cite{FiNLIE}. In the case of $AdS_4/CFT_3$, similar analytic constraints on the T functions were identified in \cite{ABJMdisco}. 
 They are expressed in terms of two special gauges, denoted as $\bT$ and $\wT$. The properties of the $\bf{T}$ gauge needed in the following derivation are:
\begin{description}
\item [(i)] \emph{Analyticity strips}: denoting as $\mathcal{A}_n$ the class of functions free of branch point singularities in the strip $|\text{Im}(u)|<\frac{n}{2}$, we have
\beq\label{eq:bbTbasic}
 ({ \bf T }_{n, 0}^{\alpha} ) \in \mathcal{ A }_{n+1} , \hspace{0.5cm} ({ \bf T }_{n, 1} )
\in \mathcal{ A }_{n}, \hspace{0.5cm} ({ \bf T }_{n, 2} ) \in \mathcal{ A }_{n-1} ,\hspace{0.5cm}n\in \mathbb{ N }, \;\;\; \alpha \in \left\{ I, II \right\} .
\eeq
Besides, on the leftmost edges of the diagram:  $\bT_{n, -1}^{\alpha} = 1$.  

\item [(ii)]
The two functions $\bfT_{0, 0}^I$, $\bfT_{0, 0}^{II}$ are equal, and  periodic on the mirror section:
\beqa\label{eq:perio}
({\bf T}_{0 , 0}^{I} )^{(+1)}(u) &=& ({\bf T}_{0 , 0}^{II} )^{(+1)}(u) \equiv \check{\mu}_{12} ,\\ 
\check{\mu}_{12}^{(+1)} &=& \check{\mu}_{12}^{(-1)}.
\eeqa 
The function $\check{\mu}_{12}$ defined above will eventually be identified with an element of the $\mu_{AB}$ matrix appearing in the QSC equations.  The notation $\check{\mu}_{12}$ signals that, throughout this Appendix, we will consider $\check{\mu}_{12}(u)$ as a function defined on the mirror Riemann section with long cuts, where it is $i$-periodic. This function agrees with $\mu_{12}(u)$ used in the rest of the paper in the strip $0 <\text{Im}(u) < 1/2$, and elsewhere is obtained by analytic continuation keeping all cuts long. Notice that the mirror $i$-periodicity of $\check{\mu}(u)$ is equivalent to the property (\ref{eq:periomu}). 
 
\item [(iii)] Finally, the $\bf{T}$ functions enjoy the following \emph{group-theoretical properties}:
\beq
{\bf T }_{0 , n } = ( \check{\mu}_{12}^{(n)} )^2,\hspace{1cm}{\bf T}_{n+1 , 2} = {\bf T}_{2 , n+1}, \hspace{2cm} n\in\mathbb{N}^+ \label{reflection}.
\eeq
\end{description}
We expect that the $\bT$ gauge defined by these properties is essentially unique (apart from rescalings by constants independent of $u$). 
 The $\bbT$ gauge may be defined by a transformation:
\begin{align}\label{eq:bbtrans}
\bT_{n , s}(u) &= 
\wT_{n , s }(u) \left(  \check{\mu}_{12}^{ (n + s - 1 )}(u) \right)^{ 2-n } , &s\in \mathbb{N}^+ , \; n \in \mathbb{N} , \\
\hspace{-1cm}\bT_{n , 0}^{\alpha}(u) &= 
\wT_{n , 0 }^{\alpha}(u) \, \left( \sqrt{ \check{\mu}_{12}^{ (n - 1) }(u) } \right)^{ 2-n } \; ( d^{(n)}(u) )^{ s_{\alpha} \, n}, \;\; &\alpha\in \left\{I, II\right\}, \; n \in \mathbb{N} \\
\hspace{-1cm}  \bT_{n , -1}^{\alpha}(u) &= \wT_{n , -1 }^{\alpha}(u)  = 1 , &\alpha\in \left\{I, II\right\}, \; n \in \mathbb{N},\label{eq:bbtrans3}
\end{align}
where $s_I = -s_{II} = + 1$, and $d(u) = d^{(+2)}(u)$ is a mirror $i$-periodic function, representing an additional degree of freedom in the definition which will be practically irrelevant for our derivation\footnote{In \cite{ABJMdisco},\cite{Cavaglia:2014exa}, a different convention was taken with a specific constant choice for $d(u)$. Here, we keep this degree of freedom explicit  since it is relevant for discussing the regularity properties of the $\wT$ gauge (see the explanation at the end of this Section).}. 
It is simple to check that  (\ref{eq:bbtrans})-(\ref{eq:bbtrans3}) leave invariant the form of the T-system due to the mirror periodicity of $\check{\mu}_{12}$ and $d(u)$. 
 
In general, we expect both the $\bT$ and $\wT$ functions to exhibit an infinite ladder of branch points for $u \in \pm 2 h + i \mathbb{Z}/2$. From the TBA analysis, we know that these singularities are all of square-root type and that analytic continuation around branch points symmetric with respect to the imaginary axis leads to the same sheets. 
This structure is further specified by the property $\bf{ (i)}$ above: some of the potential branch points in the $\bT$ functions fall inside the analyticity strips and therefore they must have trivial monodromy. 

Besides, the $\wT$ functions enjoy some special properties when continued to the short-cut section of the Riemann surface (also known as the $\emph{physical}$ sheet). We will denote their values on this section as $\hat{\mathbb{T}}$: in analogy with the case of $\check{\mu}$ and $\mu$, the convention is that $\wT$ and $\hat{\wT}$ are the same in the analyticity strip immediately above the real axis, while in the rest of the complex plane, they are defined by analytic continuation keeping long cuts for $\wT$ and short cuts for $\hat{\wT}$. The $\hat{\wT}_{a, s}$ functions have the following nontrivial properties:
\begin{description}
\item[(a)]
the functions $\hat{\mathbb{T}}_{1,n}$ with $n \geq 1$ have only two short branch cuts: $(-2h,2h)\pm in/2$,
\item[(b)]
the functions $\hat{\mathbb{T}}_{2,m}$ with $m \geq 2$ have only four short branch cuts, lying at $(-2h,2h)\pm i(m-1)/2$, $(-2h,2h)\pm i(m+1)/2$. 
\end{description}   
 The goal of the following derivation is to obtain the Riemann-Hilbert type equations characterizing the QSC. We will see that the whole structure can be derived by imposing the consistency of the conditions {\bf(i)},  {\bf(ii)} ,  {\bf(iii)}  and  {\bf(a)}, {\bf(b)}. 
 
 Let us make an additional comment. Here, we do not aim to derive the \emph{regularity} properties of the QSC, namely the statement that $\bP(u)$ and $\nu(u)$ functions are entire on the Riemann surface defined by the branch points at $u \in \pm 2 h \pm i \mathbb{Z}$. 
 However, it is natural to expect that this condition is equivalent to the requirement that the T functions are regular in appropriate gauges, and indeed one can verify a posteriori that, picking appropriately the function $d(u)$ in (\ref{eq:bbtrans}) and assuming the regularity of the QSC, all the $\bT$ and $\wT$ functions can be chosen to be regular. 
 For instance, it is possible to identify\footnote{These expressions for $\bT_{1, 0}^{\alpha}$ follow from the comparison between equation (\ref{eq:P5P6}) below and the $\bP \nu$-system.} $\bT_{1, 0}^I(u) = \nu_1 (u) \, \widetilde \nu_1(u)$, $\bT_{1, 0}^{II}(u) =  \nu^4 (u) \, \widetilde \nu^4(u)$. Therefore,  choosing $d(u) \propto (\nu_1(u)/ \nu^4(u) )^{\frac{1}{2} }$, one can set $\wT_{1, 0}^I(u) \propto \widetilde \nu_1(u)$, $\wT_{1, 0}^{II}(u) \propto \widetilde \nu^4(u)$, from which we have a clear indication that the regularity properties of the $\nu$ and T functions are equivalent. 
 This example also illustrates the fact that a requirement of regularity for the $\wT$ gauge specifies the function $d(u)$ uniquely, apart for an overall constant\footnote{In fact, from (\ref{eq:bbtrans}) it is evident that this function must be chosen in such a way that it cancels the extra singularities in $\wT^{\alpha}_{a, 0}$ introduced by the square root factors $\sqrt{\check{\mu}}_{12}$ in (\ref{eq:bbtrans}). For states with $4 \leftrightarrow \bar{4}$-symmetry, we can simply set $d(u)=1$, since in that case $\mu_{12}$ has only double zeros.}.  
However, we remark that, for the purposes of the following derivation, the precise form of $d(u)$ is irrelevant: this function  cancels out of all the equations reported below.

\subsection{Strategy of the derivation}\label{sec:Tsyst2}
The main tactic of the derivation is to choose a parametrization of the $\wT$ functions that makes {\bf(a)}, {\bf (b)} explicit; we will then reconstruct the $\bT$ functions through (\ref{eq:bbtrans}) and impose the validity of {\bf(i)}, {\bf (ii)}, {\bf(iii)}. 

To start, we notice that the properties {\bf(a)}, {\bf (b)} presented above can be encapsulated by the following parametrization (see Section 4.2 in \cite{FiNLIE}):
\beqa\label{eq:whitegauge}
\hat{\wT}_{1 , s} &=& \bP_1^{ \left[ +s \right] } \bP_2^{ \left[ -s \right] } - \bP_2^{ \left[ +s \right] } \bP_1^{ \left[ -s \right] } , \;\; \;,\;\;\;\; \hat{\wT}_{2 , s+1} = \hat{\wT}_{ 1 , 1 }^{ \left[ +s+1 \right] }  \hat{\wT}_{1 , 1 }^{ \left[ -s-1 \right]}, \;\; s \in \mathbb{N}^+ , \\
\hat{\wT}_{0, 0}^{\alpha} &=& 1 , \;\; \;, \;\;\;\; \hat{\wT}_{0 , s} = 1 , \;\; s \in \mathbb{N}^+, \;\; \alpha \in \left\{I, II\right\} ,
\eeqa
where ${\bf P}_1$, ${\bf P}_2$ are functions with a single short cut. 
Notice that this parametrization covers only the right tail of the T-hook diagram. To reach the rest of the diagram using the T-system relation  (\ref{eq:Hirota}), we need one more constraint involving at least one node outside this domain. For this purpose we may use
\beq\label{eq:T32}
\mathbb{T}_{3, 2} / \mathbb{T}_{2,3} = \check{\mu}_{12} ,
\eeq
which follows from the transformation (\ref{eq:bbtrans}) combined with the property {\bf (iii)}. 
We then see that, applying Hirota equation starting from any point in the right band, we may parametrize any of the $\wT$ functions in terms of only three building blocks, the functions $\bP_1$, $\bP_2$, $\mu_{12}$, which as we will see will be evaluated on various Riemann sheets. 
 The $\bf{ T }$ functions, defined through (\ref{eq:bbtrans}), can be expressed in terms of the same data, and one can check that they satisfy the constraints { \bf (ii)}, {\bf(iii)} by construction. However, 
 it is not obvious that they have the correct analyticity strips described by condition {\bf (i)}; we still need to impose an infinite ladder of relations:
\beq\label{eq:condn}
\DD \( (\bfT_{n+1, 0}^{ \alpha})^{(+n)} \) = \DD \, \(
\bfT_{n+2, 1}^{(+n)} \) = 0 ,
\eeq
where we use the symbol $\DD$ for the discontinuity $\DD f \equiv f - \tilde{f}$ expressing the monodromy around any of the branch points at $\pm 2 h$ on the real axis. The conditions (\ref{eq:condn}) place further constraints on $\bP_1$, $\bP_2$ and $\check{\mu}_{12}$ and will lead us to the QSC equations.

As a convenient notation, we will introduce a splitting function $g(u)$, defined through
\beq
g^2 \equiv \frac{\bT_{1,0}^I }{\bT_{1, 0}^{II} } =  \frac{\wT_{1,0}^I }{\wT_{1, 0}^{II} } \, d^2 .
\eeq  
In particular, in the $4\leftrightarrow \bar{4}$-symmetric subsector in which $\bT_{n, 0}^I=\bT_{n, 0}^{II}$, one has simply $g(u) = 1$. 
\subsection{Details}
Before discussing the derivation in detail, let us mention a technical point. In the following paragraphs, we will find relations between functions which are defined, by default, on different sections of the  Riemann surface covering the $u$ plane. To remove possible ambiguities, we specify that all the equations below are valid for $u$ in a strip slightly above the real axis. With this understanding, we will use interchangeably $\check{\mu}_{12} $ and $\mu_{12}$ in the following equations. 
\subsubsection*{First level $n=0$}
The first constraint coming from (\ref{eq:condn}) is that $\bT_{2,1} = \wT_{2,1}$ has no cut on the real axis. The consequences of this requirement were  already discussed in \cite{Cavaglia:2014exa}. Using Hirota equation and  carefully continuing the expressions (\ref{eq:whitegauge}) to the mirror sheet, we find
\beqa
\bbT_{2, 1} &=& \frac{ {\bbT}_{2,2}^{(+1)} \, {\bbT}_{2,2}^{(-1)}  - \bbT_{1 , 2} \bbT_{32} }{ \bbT_{23} } \\
&=&
( \bP_1^{ [ + 2  ]  } \bP_2 - \bP_2^{[ + 2] } \bP_1 )( { \widetilde \bP }_1 \bP_2^{ [ - 2 ] }  - { \widetilde \bP }_2  \bP_1^{ [- 2 ] } ) - {\mu}_{12} \, \bbT_{1 , 2}.
\eeqa
Imposing the absence of  a cut on the real axis, we obtain
\beq
\DD \( \bbT_{2, 1} \) = \wT_{1,2} \, \left( { \widetilde \mu }_{12} - \mu_{12} - \bP_1 {\widetilde \bP}_2 +   \bP_2 {\widetilde \bP}_1  \right)=0,
\eeq
and, since $\bbT_{1, 2}$ cannot be zero everywhere, we get a first relation of the $\bP\mu$-system (\ref{eq:Pmut}):
\beq\label{eq:appmut}
 \mu_{12} +  \bP_1 \, {\widetilde \bP}_2  -  \bP_2 {\widetilde \bP}_1 = { \widetilde \mu}_{12}.
\eeq
 Using the Hirota equation centered at the node $(1,1)$, we can also compute
\beq\label{eq:TI_II}
\bfT_{1, 0}^I \, \bT_{1,0}^{II} = \mu_{12} \, \frac{ {\wT}_{1,1}^{(+1)} \, {\wT}_{1,1}^{(-1)} - \wT_{2,1} }{\wT_{1,2} } =\mu_{12} \, ( \mu_{12} +  \bP_1 {\widetilde \bP}_2  -  \bP_2 {\widetilde \bP}_1 ) = \mu_{12} \, \widetilde{\mu}_{12} ,
\eeq
 which means we can parametrize 
\beq
\bfT_{1, 0}^I(u) = \sqrt{\mu_{12}(u) \widetilde{\mu}_{12}(u) } \, g(u) , \;\;\;\;\; 
\bfT_{1, 0}^{II}(u)=\frac{\sqrt{\mu_{12}(u) \widetilde{\mu}_{12}(u) } }{g(u) }.
\eeq
The requirement that $\bT_{1,0}^{\alpha}$ have no cuts on the real axis then imposes $\DD( g(u) )= 0 $. 

\subsubsection*{General level}
As illustrated in the previous example, the functions $\wT_{a, s}$ with $a > s$,  computed using the T-system relations, will depend not only on the values of $\bP_1$, $\bP_2$ and $\mu_{12}$ on their defining sheet, but also on their shifted values on the second sheet: this is due to the fact that Hirota equation is defined on the mirror section, while the cut in the definition of $\bP_1$ and $\bP_2$ is short. In general, the constraints (\ref{eq:condn}) can be translated as conditions on the monodromies of $\widetilde{\bP}_1$, $\widetilde{\bP}_2$ and $g$ around the branch points lying further and further from the real axis. 
 Remarkably, the content of (\ref{eq:condn}) can be recast in a very simple form: the constraints on the cuts in the upper half plane yield\footnote{ We verified the form of these equations for the first few values of $n$, and conjecture that the pattern is general. }
\beq
\hspace{-1cm}\mu_{12} \, \widetilde{(\widetilde \bP_A)^{(2n)}}=
+\bP_1^{(2n)} \, \DD( \bP_2 \, \bP_A) - \bP_2^{(2n)} \, \DD( \bP_1 \, \bP_A 
)
+(\widetilde \bP_A )^{(2n)} \,\widetilde{\mu}_{12}
+ 2 \, \DD(\bP_A) \, \eta_n , \label{eq:discnon_sym} 
\eeq
for $n \in \mathbb{N}^+$, $A \in \left\{1,2\right\}$, with
\beq
\label{eta_n}
\eta_n\equiv
\frac{1}{2} \( \frac{g^{(+2n)}}{g} +  \frac{g}{g^{(+2 n)}}\) \, \sqrt{ \widetilde \mu_{12}  \; (\widetilde \mu_{12} )^{(2n)}} +\widetilde \bP_1 \bP_2^{(2n)}-
\widetilde \bP_2 \bP_1^{(2n)} ,
\eeq
 together with the condition that
\beq
\DD { \Big ( } \frac{g^{(+2n)} }{ g} \; \sqrt{\frac{(\widetilde \mu_{12} )^{(2n)}}{ \mu_{12}} }  { \Big ) } = \DD { \Big ( } \frac{g }{ g^{(+2n)}} \; \sqrt{\frac{(\widetilde \mu_{12} )^{(2n)}}{ \mu_{12}} }  { \Big ) } .  
\eeq
One obtains very similar but not identical equations describing the discontinuities in the lower half plane. For conciseness, we will only refer to (\ref{eq:discnon_sym}) in the following arguments. 
 Remarkably, the form of these relations contains already the full structure of the QSC. 
 \subsubsection*{Constructing the $\bP\mu$-system}
The equations in (\ref{eq:discnon_sym}) can be rewritten as
\beq
-  {\tilde \mu}_{12} \mu_{12}\;\DD { \Big ( }\frac{  \widetilde \bP_A^{(2n)}}{\mu_{12} }  { \Big ) } =
+\bP_1^{(2n)} \, \DD( \bP_2 \, \bP_A )
-
\bP_2^{(2n)} \, \DD( \bP_1 \, \bP_A )
+
2 \, \DD(\bP_A) \, \eta_n ,  \label{eq:impeq}  
\eeq
for $A \in \left\{1,2\right\}$. 
Considering the discontinuity of these relations on the real axis, we see that $\DD( \eta_n ) = 0$. Inspecting expression (\ref{eta_n}), we then see that
\beq
 \frac{ { \Big ( } - \bP_1^{ (+2 n) } \DD(\bP_2)  
+ \bP_2^{ (+2n)}\, \DD(\bP_1)   { \Big ) } }{  \sqrt{ \widetilde \mu_{12} \,  \mu_{12}} } = \DD{ \Big ( } \frac{g_{\alpha}^{(+2n)} }{ g_{\alpha}} \; \sqrt{\frac{(\widetilde \mu_{12} )^{(2n)}}{ \mu_{12}} } { \Big ) } , \;\;\; \alpha \in \left\{I, II\right\}, \label{eq:defP56ssss}
\eeq
with  $g_I=g$, $g_{II}=1/g$. We will exploit (\ref{eq:defP56ssss}) to construct two new functions with a single short cut, which we denote as $\bP_5$ and $\bP_6$ in anticipation of their role in the QSC equations. They are defined through
\beqa\label{eq:P5P6}
\hspace{-2cm}\bP_5 \equiv \frac{\sqrt{\widetilde \mu_{12}}}{\sqrt{\mu_{12}}} \;  g - \bP_2 \, \om_{1, I}
 + \bP_1 \, \om_{2, I} , \;\;\;\;\;
\bP_6 \equiv \frac{\sqrt{\widetilde \mu_{12}}}{\sqrt{\mu_{12}} \; g}  - \bP_2 \,\om_{1, II} + \bP_1 \, \om_{2, II} ,
\eeqa 
where the functions $\om_{A, \alpha}$, with indices $A\in \left\{1,2\right\}$, $\alpha \in \left\{I, II\right\}$, are defined from the requirement that they are periodic on the mirror section, with power-like asymptotics and with discontinuities \footnote{These requirements specify $\om_{A, \alpha}$ uniquely apart from an additive constant independent of $u$.}
\beq\label{eq:defdiscom}
\sqrt{\widetilde \mu_{12}\mu_{12}} \; \DD(\om_{A, \alpha} ) = 
\DD(\bP_A) \, g_{\alpha},\;\;\;\; A\in \left\{1,2\right\} , \;\; \alpha \in \left\{I, II\right\}.
\eeq
Combining (\ref{eq:defdiscom}) and (\ref{eq:P5P6}), we can indeed verify that the newly constructed functions have vanishing discontinuities in the upper half plane $\DD(\bP_5^{(+2n)} )=\DD(\bP_6^{(+2n)} )=0$, $\forall n \in \mathbb{N}^+$. A simple extension of this analysis shows that (\ref{eq:P5P6}) defines a function with only a single short cut on the real axis.  

Let us point out that, when $g=1$ (which is appropriate for $4 \leftrightarrow \bar{4}$-symmetric states), by definition we have $\bP_5 = \bP_6$, in agreement with the rules described in Section \ref{sec:LR}.
 As another side remark, notice that the definitions (\ref{eq:P5P6}) can be recognized as two equations of the $\bP \nu$-system (\ref{eq:Pnu1}) provided the mirror-periodic functions $\om_{A, \alpha}$ are identified as ratios of $\nu$ functions, and $g(u) $ is identified as
\beq
g^2 = \frac{\nu_1 \, \widetilde{\nu}_1}{\nu^4 \, \widetilde{\nu}^4}. \label{eq:gNuNu}
\eeq 
 In the rest of this Appendix, for simplicity we will concentrate solely on obtaining  the ``$\bP\mu$'' vector form of the equations. 

Using (\ref{eq:P5P6}),(\ref{eta_n}), equations (\ref{eq:impeq}) 
becomes
\beqa\label{eq:impeq2}
-  {\widetilde \mu}_{12} \mu_{12}\;\DD  { \Big ( } \frac{  \widetilde \bP_{{A}}^{(2n)}}{\mu_{12} }  { \Big ) } &=&
+\bP_{1}^{(2n)}\left[ \DD( \bP_{{A}} \, \bP_2 ) - 2 \, \DD( \bP_{{A}} ) \, \(  { \widetilde \bP}_2  +\sqrt{  {\widetilde \mu}_{12}\mu_{12}} \sum_{\alpha=I, II}\frac{\om_{2, \alpha} }{2 g_{\alpha}}\) \right] \nn\\
&&-
\bP_2^{(2n)} \left[ \DD ( \bP_1 \, \bP_{{A}} ) - 2 \, \DD( \bP_{{A}} ) \, \(  { \widetilde \bP}_1  +\sqrt{  {\widetilde \mu}_{12}\mu_{12}} \sum_{\alpha=I, II}\frac{\om_{1, \alpha} }{2 g_{\alpha}} \) \right]  \nn\\
&&+
\DD (\bP_{{A}} ) \, \left( {\bP_5 }^{(2n)} \, \frac{ \sqrt{\widetilde \mu_{12} \mu_{12}}}{g} +  \bP_6^{(2n)} \, g \;\sqrt{\widetilde \mu_{12} \mu_{12}} \right),
\eeqa
with $A \in \left\{1, 2\right\}$. Let us now introduce four functions $\Om_{AB}$, periodic on the mirror sheet, $\Om_{AB}=\Om_{AB}^{(+2n)}$, for  $A, B\in \left\{1,2\right\}$, whose (periodically repeated) discontinuities are 
\beq\label{eq:defOOme}
{\mu_{12}\widetilde \mu_{12}} \, \DD(\Om_{AB})=
\DD(\bP_A\bP_B)-2\, \DD(\bP_A) \, \left(  { \widetilde \bP}_B  +\sqrt{  {\widetilde \mu}_{12}\mu_{12}} \sum_{\alpha=I, II}\frac{\om_{B, \alpha} }{2 g_{\alpha}} \right) .
\eeq
Then, defining the functions $\bP_3$ and $\bP_4$ as 
\beqa
-\bP_3&\equiv&
\frac{\widetilde \bP_1}{\mu_{12}}+\Om_{12} \, \bP_1 - \Om_{11} \,\bP_2+ {\om}_{1, II} \, \bP_5 + \om_{1, I} \, \bP_6 \label{eq:defP3} ,\\
-\bP_4&\equiv&
\frac{\widetilde \bP_2}{\mu_{12}}+\Om_{22} \,\bP_1 - \Om_{21} \,\bP_2+ {\om}_{2, II} \, \bP_5 + \om_{2, I} \, \bP_6 \label{eq:defP4} ,
\eeqa
we see that, due to (\ref{eq:impeq2}),
\beqa
\DD (  \bP_3^{(2n)} ) = \DD ( \bP_4^{(2n)} ) = 0 ,\;\;\;\;\; n \in \mathbb{N}^+ ,
\eeqa
 therefore $\bP_3$ and $\bP_4$ are free of branch points in the upper half plane and, by a small additional effort, we can show that they have just a single short cut on the real axis.

Let us summarize the situation: by a scrutiny of the equations, we have so far found six functions with a single short cut, and eight mirror-periodic functions $\Om_{AB}$ , $\om_{A, \alpha}$. It remains only to check that the relations between their monodromies can be written in a closed form. 

The fifteen components of the antisymmetric matrix $\mu_{AB}$ can be defined in terms of the periodic functions introduced above. Indeed, setting
\beqa\label{eq:defOmMu}
&&\mu_{14}=-\Om_{12} \, \mu_{12} - 1 , \quad\mu_{13}= -\Om_{11} \, \mu_{12} , \quad  
\mu_{15}=-\om_{1, I}\, \mu_{12} , \quad \mu_{16}= - {\om}_{1, II } \, \mu_{12} ,\\ 
&&\mu_{24}=-\Om_{22} \, \mu_{12} , \quad
\mu_{23}=-\Om_{21} \, \mu_{12} + 1 , \quad
\mu_{25}=-\om_{2, I} \, \mu_{12} , \quad
\mu_{26}= -{\om}_{2, II} \, \mu_{12} ,
\eeqa
we immediately recognize that (\ref{eq:defP3}),(\ref{eq:defP4}) are two equations of the $\bP\mu$-system. Besides, the form of these relations implies the existence of three quadratic constraints among the matrix elements defined in (\ref{eq:defOmMu}); let us discuss in detail how these conditions emerge. Consider the following equation:
\beq\label{eq:appOthMut}
 \mu_{13} - \widetilde \mu_{13} + \bP_1 \widetilde \bP_3 - \bP_3 \widetilde \bP_1 = -2 \, ( \mu_{12}- \widetilde \mu_{12} ) \, ( \Om_{11 }- \om_{1, I} \, \om_{1, II} ) ,
\eeq
which can be derived from (\ref{eq:defP3}) and its analytic continuation to the second sheet using the monodromy rules (\ref{eq:appmut}),(\ref{eq:defdiscom}),(\ref{eq:defOOme}). The form of (\ref{eq:appOthMut}) implies that the combination of mirror-periodic functions $ \Om_{11 }- \om_{1, I} \, \om_{1, II} $ must be free of cuts. Due to its power-like asymptotics, it must be a constant independent of $u$ and, using the freedom to redefine the $\Om$'s by a constant shift, we will assume that $ \Om_{11 }- \om_{1, I} \, \om_{1, II} = 0$. Therefore, (\ref{eq:appOthMut}) can be recognized as another equation of the $\bP \mu$-system. Moreover, the quadratic constraint we have just found can be rewritten as $\mu_{12} \mu_{13}- \mu_{15} \mu_{16} = 0$, which is one of the components of the matrix equation $(\mu \eta)^2 = 0$. By similar reasoning, we can impose two more constraints and all in all we can set
\beq
\Om_{11 }- \om_{1, I} \, \om_{1, II} =0 , \;\;\;\;\;\Om_{22 }- \om_{2, I} \, \om_{2, II} =0, \;\;\;\;\; \Om_{12} + \Om_{21} + \om_{1, I} \om_{2, II} + \om_{2, I} \om_{1, II} = 0 .
\eeq
The rest of the derivation goes along the same lines. The remaining independent entries of $\mu_{AB}$ are defined as:
\beqa\label{eq:defOmMu2}
&&\mu_{35} = \om_{1, I} \,( \mu_{23} -\mu_{12} \,{\om}_{1, II} \, \om_{2, I} ), \quad \mu_{36} = {\om}_{1, II}\,( \mu_{23} -\mu_{12} \, {\om}_{1, I} \, \om_{2, II} ) , \\
&&\mu_{45}= \om_{2, I} \, ( \mu_{23} -\mu_{12} \, {\om}_{1, II} \, \om_{2, I} ) , \quad \mu_{46} = {\om}_{2, II} \, ( \mu_{23} -\mu_{12} \, {\om}_{1, I} \, \om_{2, II} )  , \\
&& \quad
\mu_{34}= \frac{ \mu_{35} \, \mu_{36} }{ \mu_{12} \, \om_{1, I} \, {\om}_{1, II} }  , \quad
\mu_{5 6}= -\mu_{12} ( {\om}_{1, II}\, \om_{2, I} - \om_{1, I} \,{\om}_{2, II} ) , \label{eq:defOmMuLa}
\eeqa
and it is possible to verify that all equations of the $\bP \mu$ system, including the quadratic constraints on the $\bP$ and $\mu$ functions, follow from the relations listed above (and their analytic continuation through the branch cut on the real axis).   

 The specific form of the matrix $\eta_{AB}$ entering the $\bP \mu$-system equations does depend on the normalization of our definitions (\ref{eq:P5P6}),(\ref{eq:defP3}),(\ref{eq:defP4}),(\ref{eq:defOmMu})-(\ref{eq:defOmMuLa}), and could be changed by rescaling some of the $\mu$ or $\bP$ functions, or by a more general linear change of basis, $\bP_A(u) \rightarrow H_A^B \, \bP_B(u)$, $\mu_{AB}(u) \rightarrow H_A^C \, H_B^D \, \mu_{CD}(u)$, which would transform $\eta_{AB} \rightarrow H_A^C \, H_B^D \, \eta_{CD}$. However, $\eta_{AB}$ is clearly always a symmetric tensor, and besides its signature $(+++---)$ is invariant under all linear transformations with 
$H \in \mathbb{R}^{6 \times 6}$. This reality restriction is meaningful since it preserves the following property: for real values of the coupling, all the functions $\bP_A(u)$ can be chosen to be real\footnote{Correspondingly, one can choose all functions $\mu_{AB}^+(u)$ to be purely imaginary on the real axis.} on the Riemann section with short cuts. This property is verified with our choice of conventions, and follows from the reality of the solutions of the TBA. 
 
\section{Algebraic identities}
\label{app:gamma}
\subsection{Identities for gamma matrices}\label{app:identities}
In this Appendix we collect some useful algebraic identities, descending from the properties of gamma and sigma matrices for $SO(3,3)$ and $SO(3,2)$. 
The defining relation for the $SO(3,3)$ sigma matrices is
\beq
(\sigma_{ A } )_{ac} \; (\bar{\sigma}_{B} )^{cb} + (\sigma_{ B } )_{ac} \; (\bar{\sigma}_{A} )^{cb}  = \delta_a^b \; \eta_{AB},
\eeq
and we recall that $(\sigma_{AB} )_a^b$ is defined through
\beq
(\sigma_{ A } )_{ac} \; (\bar{\sigma}_{B } )^{cb} - (\sigma_{B} )_{ac} \; (\bar{\sigma}_{A} )^{cb}= -2 \; (\sigma_{AB} )_a^b ,
\eeq
so that we have
\beq
(\sigma_{ A } )_{ac} \; (\bar{\sigma}_{B } )^{cb} = \frac{1}{2} \, \delta_a^b \; \eta_{AB} -  (\sigma_{AB} )_a^b .
\eeq
A useful property, specific to orthogonal groups in six and five dimensions, is the fact that gamma matrices are anti-symmetric: $(\sigma_A )_{ab}=-(\sigma_A )_{ba}$. 
This allows us to prove the following very useful relation:
\beq\label{eq:sigss}
\( \bar{\sigma}_C \, \sigma_A \, \bar{\sigma}_B -  \bar{\sigma}_C \, \sigma_B \, \bar{\sigma}_A \)^{ab} = \eta_{AC} \, (\bar{\sigma}_B)^{ab} - \eta_{BC} \, (\bar{\sigma}_A)^{ab} ,
\eeq
and its consequence
\beqa\label{eq:trss}
\text{Tr} \( \sigma_{AB} \, \sigma^{CD} \) = \delta_A^D \, \delta_B^C - \delta_A^C \, \delta_B^D. 
\eeqa
Another identity that is specific to this dimension is
\beq
\bar{\sigma}^{ab} = -\frac{1}{2} \, \epsilon^{abcd} \, \sigma_{cd},
\eeq
which implies in particular that $(\sigma_{AB})$ is traceless: $(\sigma_{AB})_a^a = 0$, and moreover that, for any anti-symmetric matrix $4 \times 4$ matrix $G_{ab}$:
\beq
2 \,\text{Pf}(G_{ab}) = G_A \, \eta^{AB} \, G_B ,
\eeq
where the corresponding vector $\left\{G_A\right\}_{A=1}^6$ is defined by $G_{ab} = G_A \, (\sigma^A )_{ab}$. 
Another useful formula is:
\beq\label{eq:commu}
(\sigma_A )_{ab} (\bar{\sigma}_B )^{cd} - (\sigma_B )_{ab} (\bar{\sigma}_A
)^{cd}
= (\sigma_{AB} )^c_a \, \delta^d_b - (\sigma_{AB} )^c_b \, \delta^d_a - (\sigma_{AB} )^d_a \,
\delta^c_b + (\sigma_{AB} )^d_b \,\delta^c_a .
\eeq
All the properties listed above are independent on for any choice of  chiral representation of the gamma matrices. The situation is analogous for the representations of $SO(3,2)$. In that case we recall that we use the symbols $(\Sigma_I )_{ij}$, $(\Sigma_{IJ} )_i^j $, and denote the metric as $\rho_{IJ} \equiv \frac{1}{2} \, \text{Tr} ( \Sigma_{ I } \, \bar{\Sigma}_J )$. In particular the defining relation for the matrices $\Sigma_I$ and $\Sigma_{IJ}$ is:
\beq\label{eq:SigSig}
(\Sigma_{ I } )_{ki} \; (\bar{\Sigma}_{J } )^{il} = \frac{1}{2} \, \delta_k^l \; \rho_{IJ} -  (\Sigma_{IJ} )_k^l ,
\eeq
with $\Sigma_{IJ} = - \Sigma_{JI}$. 
On top of these properties, in the $SO(3,2)$ case the matrices $\Sigma$ and $\bar{\Sigma}$ are related by a similarity transformation:
\beq\label{eq:sigId}
(\Sigma_I )_{ij} = \(\ch_{ik} \, (\bar{\Sigma}_I )^{kl} \, \ch_{lj} \),
\eeq
where $\ch_{ij}$ is an anti-symmetric $4\times 4$ matrix. Equation (\ref{eq:sigId}) can be used to prove the additional symmetry property $(\Sigma_{IJ} )_{ij} = + (\Sigma_{IJ} )_{ji}$. 
Finally, the analogue of (\ref{eq:trss}), (\ref{eq:sigss}) are
\beqa\label{eq:propTr}
\text{Tr} \( \Sigma_{IJ} \, \Sigma^{KL} \) &=& \delta_I^L \, \delta_J^K - \delta_I^K \, \delta_J^L , \\
\( \bar{\Sigma}_K \, \Sigma_I \, \bar{\Sigma}_J -  \bar{\Sigma}_K \, \Sigma_J \, \bar{\Sigma}_I \)^{ij} &=& \rho_{IK} \, (\bar{\Sigma}_J)^{ij} - \rho_{J K} \, (\bar{\Sigma}_I)^{ij} .  
\eeqa

Finally, we report below some useful identities for a generic antisymmetric $4\times 4$ matrix $G_{ab}=-G_{ba}$:
\beqa
G_{ab} \, G_{cd} - G_{cb} \, G_{ad} - G_{ac} \, G_{bd} &=& \epsilon_{abcd} \, \text{Pf}( G ) ,\label{eq:trick1}\\
-\frac{1}{2} \, \epsilon^{ijkl} \, G_{kl} \, G_{jm} &=& \delta_{m}^i \; \text{Pf}( G ) ,\\
G_{ik} \; G_{jl} \; \epsilon^{klmn} &=& -\text{Pf}( G ) \; \( G_{ij} \; G^{mn} + \delta_i^m \, \delta_j^n - \delta_i^n \, \delta_j^m \), \\
G_{ij} = -\frac{1}{2} \, \epsilon_{ijkl} \, G^{kl} \, \text{Pf}( G ) ,\label{eq:Ginv}
\eeqa
where we recall that the Pfaffian is defined as
\beq
\text{Pf}( G) = \frac{1}{8} \, \epsilon^{abcd} \, G_{ab} \, G_{cd} = G_{12} G_{34}+ G_{14} G_{23} - G_{13} G_{24} . 
\eeq
In particular:
\beq
\ch_{ik} \; \ch_{jl} \; \epsilon^{klmn} = \( \ch_{ij} \; \ch^{mn} + \delta_i^m \, \delta_j^n - \delta_i^n \, \delta_j^m \).\label{eq:ide}
\eeq
\subsection{Relation between $Q_{ab|ij}$ and $Q^{ab}_{\,|ij}$ }\label{app:QabQabup}
In Section \ref{sec:constrQsyst}, we have defined the objects $Q_{ab|ij}$ as subdeterminants of the $4\times 4$ matrix $\left\{Q_{a|i}\right\}$. 
Notice that, in principle, one can also define
\beq
Q^{ab}_{\,|ij} = Q^a_{\,|i} \, Q^{b}_{\,|j} - Q^{a}_{\,|j} \, Q^{b}_{\,|i}.
\eeq
However, a simple linear algebra identity relates the minors of a matrix and its inverse, and shows that the two definitions are algebraically related:
\beq\label{eq:Qupdown}
Q_{ab|ij} =\frac{1}{2} \, \( \text{det}( Q_{\ast|\ast} ) \)\,  \epsilon_{abcd} \; \epsilon_{ijkl} \; Q^{c|k} \, Q^{d|l} = - \frac{1}{2} \, \epsilon_{abcd} \; \epsilon_{ijkl} \; Q^{c|k} \, Q^{d|l}.
\eeq
From (\ref{eq:Qupdown}), we see that
\beq
Q_{ab|ij} = -\frac{1}{2} \, \epsilon_{abcd} \, Q^c_{\,|j_1} \, Q^d_{\,|j_2} \,  \epsilon_{ijkl} \, \ch^{k j_1}\, \ch^{l j_2} , 
\eeq
and using (\ref{eq:ide}) we find
\beq\label{eq:lowup}
Q_{ab|ij} = -\frac{1}{2} \, \epsilon_{abcd} \; \( Q^{cd}_{\,|ij} + \ch_{ij} \; Q^{cd}_{\circ} \).
\eeq
Let us define  the projections:
\beq
Q_{ab|\circ} \equiv \frac{1}{2} Q_{ab|ij} \, \ch^{ij} , \;\;\;\;\;\;
 Q_{ab|\overline{(ij)}} \equiv Q_{ab|ij}  +  \frac{1}{2} \ch_{ij} \, Q_{ab|\circ}, 
\eeq
where $ Q_{ab|\overline{(ij)}} $ denotes the traceless part and satisfies $ Q_{ab|\overline{(ij)}}  \, \ch^{ij} = 0$. Identity (\ref{eq:lowup}) then splits as
\beq
Q_{ab|\circ} = \frac{1}{2} \, \epsilon_{abcd} \, Q^{cd}_{\,|\circ} , \;\;\;\;\;
Q_{ab|\overline{(ij)}}= - \frac{1}{2} \, \epsilon_{abcd} \, Q^{cd}_{\,|\overline{(ij)}}.\label{eq:lowup3APP}
\eeq
\subsection{Relation between $\bQ_{ij}$ and its inverse}\label{app:QQinv}
From (\ref{eq:Ginv}), we have
\beq
\bQ_{ij} = \frac{1}{2} \epsilon_{ijkl} \;\bQ^{kl} , \label{eq:QQinv}
\eeq
 and, using (\ref{eq:trick1}), we immediately find
\beq\label{eq:QQhat}
{\bQ}_{ij} = \ch_{i i_1} \, \ch_{j j_1} \, \bQ^{i_1 j_1} - \frac{1}{2} \, \ch_{ij} \, \widehat{\bQ}_{\circ} ,
\eeq
where
\beq
\widehat{\bQ}_{\circ} = {\bQ}^{mn} \, \ch_{mn} .
\eeq 
Contracting (\ref{eq:QQhat}) with $\ch^{ij}$, we find that in fact $ \widehat{\bQ}_{\circ} = {\bQ}_{\circ} =  \bQ_{ij} \, \ch^{ij}$, so that (\ref{eq:QQhat}) reduces to equation (\ref{eq:QQhatmain}) presented in the main text. 

\section{Derivation of constraints on large-$u$ asymptotics}\label{app:ABderive}
Here we derive the constraints (\ref{eq:AAasy}), (\ref{eq:BIBI}) on the asymptotics of $\bP$ and $\bQ$ functions 
using the QQ-relations derived in Section \ref{sec:Qsystem}. 
In order to find (\ref{eq:AAasy}), we start from relation  (\ref{eq:F1}). At large $u$, its rhs is given by
\beq
\bP_A(u) \, \bQ_{I}(u)\simeq \mathcal{A}_A \, \mathcal{B}_{I}u^{\hat M_{I}-M_A-1}\,,
\eeq
which constrains the asymptotic behaviour of $\mQ_{A|I}$ to be
\beq
\mQ_{A|I}(u) \simeq -i\frac{\mathcal{A}_A\mathcal{B}_{I} u^{\hat M_{I}-M_A}}{\hat M_{I}-M_A}. \label{asympQ_AI}
\eeq
We may now use the following relation, which is a consequence of the Q-system:
\beq
\bQ_{I} = \pm \bP^A \, Q_{A|I}^{\pm}, \label{eq:QPQ_AI}
\eeq
and gives, using the aymptotics (\ref{asympQ_AI}), the constraint
\beq
\sum_A\frac{\mathcal{A}_A\mathcal{A}^A}{\hat M_{I}-M_A}=0 , \quad I=1, \dots, 5 .
\eeq
These relations, together with the constraint $ \text{Pf}\left( \bP_{ij} \right) = 1$, may be solved for the terms $\mathcal{A}_A \, \mathcal{A}^A$, giving precisely (\ref{eq:AAasy}).
To derive (\ref{eq:BIBI}), it will be convenient to use the following equaton, which can be obtained with simple manipulations from the Q-system relations: 
\beq
\bP_A=\bQ^I \, Q_{A|I}^-+\frac{\bQ_{\circ}Q_{A|\circ}^-}{4} .
\label{eq:PQQAI}
\eeq 
The large-$u$ asymptotics of $\bQ_{\circ}$ can be fixed using the first constraint in (\ref{eq:constrQvec}), which yields
\beq
\bQ_{\circ}(u) = 4 + \frac{2 \, \mathcal{C} }{u^2} + \mathcal{O}\(\frac{1}{u^3}\) , \quad \mathcal{C}= \mathcal{B}_1\mathcal{B}_4-\mathcal{B}_2\mathcal{B}_3+\mathcal{B}_5^2.\label{eq:subleadQo}
\eeq
We will also need
\beq
\bP_A(u) \simeq  u^{-M_A} \, \[ \mathcal{A}_A  +\frac{ \mathcal{A}_A^{sub}}{u} + \mathcal{O}\(\frac{1}{u^2}\) \], \label{eq:subleadPA} 
\eeq
and, from (\ref{eq:Fcirc}),
\beq
Q_{A|\circ}(u) = u^{-M_A}\[ \mathcal{A}_A   +  \frac{\mathcal{A}_A^{sub}}{u} 
+ \mathcal{O}\(\frac{1}{u^2}\)\].\label{eq:subleadQAo}
\eeq 
Expanding (\ref{eq:PQQAI}) at NLO, we find, using (\ref{eq:subleadQo}), (\ref{eq:subleadPA}), (\ref{eq:subleadQAo}), 
\beq
\sum_{I=1}^5 \frac{\mathcal{B}^I\mathcal{B}_I}{\hat M_I-M_A}=\frac{M_A}{2}
,\quad A=1,\dots,6 .
\eeq
The solution of these equations finally yields (\ref{eq:BIBI}) and fixes the coefficient $\mathcal{C}$ as in (\ref{eq:Casimir}).

Finally, for the reader's convenience, we report more explicitly the constraints (\ref{eq:AAasy}),(\ref{eq:BIBI}):
{\small 
\beqa\label{eq:AsBsapp}
&&\mathcal{A}_1 \mathcal{A}_4 = \frac{(M_1^2 - \hat{M}_1^2 )(M_1^2 - \hat{M}_2^2 )}{(M_1^2- M_2^2) (M_1^2 - M_5^2 ) }, \;\;\;\; \mathcal{A}_2 \mathcal{A}_3 = \frac{(M_2^2 - \hat{M}_1^2 )(M_2^2 - \hat{M}_2^2 )}{(M_1^2- M_2^2) (M_2^2 - M_5^2 ) }, \;\;\;\; \mathcal{A}_5 \mathcal{A}_6 = \frac{(M_5^2 - \hat{M}_1^2 )(M_5^2 - \hat{M}_2^2 )}{(M_5^2- M_1^2) (M_5^2 - M_2^2 ) } ,\nn\\
&&\mathcal{B}_1 \mathcal{B}_4 = \frac{(\hat{M}_1^2 - M_1^2 )(\hat{M}_1^2 - M_2^2 )(\hat{M}_1^2 - M_5^2)}{4 \, \hat{M}_1^2\, (\hat{M}_1^2- \hat{M}_2^2)}, \;\;\;\; \mathcal{B}_2 \mathcal{B}_3 =  \frac{(\hat{M}_2^2 - M_1^2 )(\hat{M}_2^2 - M_2^2 )(\hat{M}_2^2 - M_5^2)}{4 \, \hat{M}_2^2\, (\hat{M}_1^2- \hat{M}_2^2)}, \nn \\
&&
\mathcal{B}_5^2 = -\frac{M_1^2 \, M_2^2 \, M_5^2}{4 \, \hat{M}_1^2\, \hat{M}_1^2}.
\eeqa
}

\section{Asymptotics and charges: consistency checks at weak coupling}\label{sec:appWC}
In this Section we show the emergence of a polynomial Bethe Ansatz in the weak coupling limit and use it to match the parameters entering the asymptotics of the QSC with the quantum numbers of the state, proving (\ref{MAdef}),(\ref{MIdef}). 
The discussion presented below may also be useful for developing an analytic weak coupling solution algorithm valid in every sector. 

We shall start from the following large-$u$ asymptotics
\beqa\label{eq:powerlikePQ}
\bP_A \sim \left( u^{-M_1}, u^{-M_2}, u^{M_2}, u^{M_1}, u^{-M_5}, u^{M_5} \right) , \\
\bQ_I \sim \left( u^{\hat{M}_1-1}, u^{\hat{M}_2-1}, u^{-\hat{M}_2-1}, u^{-\hat{M}_1-1}, u^{-1} \right),
\eeqa
where, for the moment, we view the five charges $\left\{M_1, \, M_2, \, M_5\right\} \in \mathbb{Z}^3$, $\left\{\hat{M}_1, \hat{M}_2 \right\} \in \mathbb{R}^2$  as generic parameters.  
 Choosing a conventional ordering, we assume that, for small real values of the coupling constant, they are ordered as
\beq\label{eq:orderMs}
\hat{M}_1 > \hat{M}_2 > M_2 > M_1 > | M_5| .
\eeq 
Since they are not quantized in integers, the charges $\hat{M}_i$ will depend on the coupling. We make the further assumption that, as $h \rightarrow 0$, $\hat{M}_2$ and $\hat{M}_1$ have integer limiting values\footnote{While we do not have a rigorous proof, we expect that this is true for all solutions of the QSC equations with power-like asymptotics (\ref{eq:powerlikePQ}). Notice that it is enough to impose this condition on only one of the two charges $\hat{M}_i$, since we proved in Section \ref{sec:gluing} that their difference $\hat{M}_1 - \hat{M}_2$ is integer.}:
 $\lim_{h \rightarrow 0^+} \hat{M}_i \in \mathbb{Z}$, with deviations of order $\mathcal{O}(h^2)$. This property will play an important role since it implies that the powers in the asymptotics  (\ref{eq:powerlikePQ}) are integer for $h \sim 0$. Therefore, at leading order in $h$ any Q functions which turns out to be free of singularities must reduce to a polynomial function of $u$. 

\subsection{Generic features of the weak coupling expansion}\label{sec:appWC2}
We now discuss some general features of the weak coupling limit (see also \cite{Marboe:2014gma}). 

\paragraph{1) Properties of $\bP$ functions: }
As $h \rightarrow 0^+$, the branch cuts of the QSC shrink to zero size; each of these cuts is in general replaced by a pole. 
    
 For a generic Q function analytic in the upper half plane, we may expect a string of poles for $-u \in i \mathbb{N}$ at weak coupling. However, since the $\bP$ functions originally had only a single cut on the first sheet, at each order in the weak coupling expansion they are rational functions of $u$, with no singularities apart for a multiple pole at $u=0$. 
 
 Consistently with relations (\ref{eq:AsBsapp}), we will choose a normalization so that the $\bP$'s scale like $\mathcal{O}(h^0)$ at weak coupling: $\bP_A(u) = \bP_A^{(0)}(u) + \mathcal{O}(h^2) $. Let us now introduce an important parameter: we denote as $\ell$ the order of the strongest pole  occurring among all the functions $\bP_A^{(0)}(u)$ at $u=0$, and we will write
 \beq\label{eq:poleprop}
 \bP_A^{(0)}(u) = u^{- \ell} \, p^{\text{pol} }_A(u) ,
 \eeq
where $p^{\text{pol}}_A(u)$ are polynomials in $u$. 
Notice that we must have $\ell  \geq M_1 > 0$, since otherwise $\bP_1^{(0)}(u)$ would not have decreasing asymptotics at infinity, contradicting (\ref{eq:orderMs}). We will see eventually that $\ell$ can be identified with the spin chain length entering the Bethe Ansatz equations at weak coupling. 

 \paragraph{2) Properties of $\nu$ functions: }
 In the leading approximation at weak coupling all functions $\nu_a(u)$, $\nu^a(u)$ are necessarily polynomials in $u$. 
 Let us review the argument leading to this conclusion, following \cite{Marboe:2014gma}. We start noticing that, at finite $h$, the functions 
 \beq\label{eq:nucombs}
 \nu_a(u) + \widetilde \nu_a(u), \;\;\;\;\; \frac{ \widetilde \nu_a(u) - \nu_a(u) }{\sqrt{u^2-4 h^2 } }
 \eeq 
 do not have cuts on the real axis. Therefore, when expanded at weak coupling these particular combinations of $\nu$ functions should have no pole at $u=0$, at every loop order. Combining this observation with the mirror-periodicity $\widetilde \nu_a(u) = e^{i \sigtw} \nu_a(u+i)$ shows that, at the leading order at weak coupling, $\nu_a(u)$ cannot have a pole neither at $u=0$ nor at $u=i$; further, the difference equation (\ref{eq:4thdiff}) shows that these functions cannot have singularities anywhere else. Therefore, at leading order in the weak coupling expansion they must reduce to polynomial functions of $u$.

Studying the large-$u$ asymptotics, we can also deduce that the components of $\nu_a(u)$ must have the same scaling at weak coupling. Furthermore, by an appropriate normalization, we can impose that $\nu_a$ and $\nu^a$ have the same scaling behaviour. Thus, we can write:
\beq\label{eq:scalenu}
\nu_a(u) = h^{- \ell_{\ast}} \, ( \nu_a^{(0)}(u) + \mathcal{O}(h^2) ), \;\;\;\; \nu^a(u) = h^{- \ell_{\ast}} \, ( \nu^{a(0)}(u) + \mathcal{O}(h^2) ) ,
\eeq
where $\nu_a^{(0)}(u)$, $\nu^{a(0)}(u)$ are polynomials of $u$. We will prove below that $\ell_{\ast} \geq \ell$. We will see that the   zeros of the polynomials $\nu_1^{(0)}(u+i/2)$, $\nu^{4(0)}(u+i/2)$ can be identified with $4$ and $\bar{4}$-type Bethe roots entering the 2-loop Bethe Ansatz: 
\beq
 \nu_1^{(0)}(u) \propto \prod_{j=1}^{K_4}( u - u^{(0)}_{4, j} -i/2), \;\;\;\;  \nu^{4(0)}(u) \propto \prod_{j=1}^{K_{\bar{4}}}( u - u^{(0)}_{\bar{4}, j} -i/2).
\eeq
Finally, let us study the regularity of the combinations (\ref{eq:nucombs}) at leading order at weak coupling: imposing the absence of a pole at $u=0$ we find the useful equations
\beq\label{eq:P0ideapp}
( \nu_a^{(0)}(0) - e^{i \sigtw^{(0)}} \nu_a^{(0)}(+i) )= ( \nu^{b(0)}(0) - e^{-i \sigtw^{(0)}} \nu^{b(0)}(+i) ) = 0, \;\;\;\; \forall a, b ,
\eeq
where $\sigtw^{(0)} = \lim_{h \rightarrow 0} \sigtw$.  From (\ref{eq:P0ideapp}), we see that 
\beq\label{eq:proofP0}
e^{i \sigtw} + \mathcal{O}(h^2) = \frac{ \nu_1^{(0)}(0) }{\nu_1^{(0)}(i)} = \prod_{j=1}^{K_4}\left(\frac{u_{4, j}^{(0)}+ i/2}{u_{4, j}^{(0)}-i/2 }\right) = \text{exp}(i \sum_{j=1}^{K_4} p_{4, j} ) + \mathcal{O}(h^2) ,
\eeq
which proves the statement anticipated in Section \ref{subsec:P}: at leading order, $\sigtw$ can be identified with the total momentum of a single species of spin chain magnons. 

\paragraph{3) Properties of $\widetilde \bP$: }
Important information on the weak coupling behaviour of  $\widetilde \bP_A$ can be obtained studying the properties of the $1/x(u)$ expansion (\ref{eq:expP}), which is valid at finite $h$.

Consider one of the $\bP_A$ functions which exhibits a pole of order $\ell$ in $u=0$. In order to be more explicit, we will take this function to be $\bP_1(u)$. Write its $1/x$ expansion as:
\beq\label{eq:expP1}
\bP_1(u) = \sum_{n=M_1 }^{ \infty} \frac{ b_{1, n} }{(hx(u))^{n} } ,
\eeq
then, taking into account that $x(u) \sim u/h + \mathcal{O}(h)$, we deduce that the coefficients scale as $b_{1, n} \sim \mathcal{O}(h^0)$ for $n \leq \ell$ at weak coupling, and at the very least\footnote{  Actually, coefficients with large index $n$ must decrease much faster with the coupling, since the radius of convergence of the $1/x$ expansion scales like $1/h$ at $h \sim 0$ (see \cite{Marboe:2014gma}).  } $b_{1, n} \sim \mathcal{O}(h^2)$ for $n > \ell$.

Sending $x(u) \rightarrow \widetilde x(u) = 1/x(u)$ in (\ref{eq:expP1}), we obtain the expansion
\beq\label{eq:appserPtil}
\widetilde \bP_1(u) = \sum_{n=M_1 }^{ \infty} \frac{ b_{1, n} \, x^n(u) }{(h)^{n} },
\eeq
which converges in a finite region described by $|x(u) | < | x( 2h + i/2 ) |$ \cite{Marboe:2014gma}. For $u$ in this region, we can safely re-expand the series at weak coupling. Examining (\ref{eq:appserPtil}), we can make the following observations:
\begin{enumerate}
\item 
 The term with $n = \ell$ in (\ref{eq:expP1}), at leading weak coupling order, generates 
 \beq\label{eq:bterm}
 b_{1, \ell} \, h^{-2 \ell} \, ( u^{\ell} + \mathcal{O}(h^2) ) ,
 \eeq
 where, by our assumption on the pole of $\bP_1^{(0)}(u)$, we have $b_{1, \ell} \neq 0$ for $h=0$. The scaling of the coefficients $b_{1, n}$ discussed above reveals that there is no way that other terms in the expansion would precisely cancel the contribution (\ref{eq:bterm}) and produce a milder behaviour for $\widetilde \bP_1(u)$ as $h \sim 0$. The only still conceivable possibility is that some term with $n > \ell$ would produce an even more singular scaling at weak coupling. From this we learn that\footnote{ In all examples we have studied, for instance for all states in the SL(2)-like sector, we have precisely $\ell = \ell_{\ast}$. We do not know whether this is a general rule. However for the following argument it is sufficient to work on the assumption that $\ell_{\ast} \geq \ell$.} $\widetilde \bP_1(u) =\mathcal{O}(h^{- 2 \ell_{\ast} } )$ with $\ell_{\ast} \in \mathbb{N}$ and $\ell_{\ast} \geq \ell$.
\item The scaling $h^{-\ell_{\ast}}$ is the same as the one introduced in (\ref{eq:scalenu}) for the $\nu_a(u)$ functions. Indeed, the two are related by equation (\ref{eq:Pnu1}), which at leading order becomes
\beq\label{eq:Ptilscale0}
\widetilde \bP_{A}^{(0)}(u) = - e^{i \sigtw^{(0)} }\left( \nu_a^{(0)}(u) \, \nu_b^{(0)}(u+i)\right) \, \bar \sigma_A^{ab},
\eeq
with\footnote{Notice that the function $\widetilde \bP_A^{(0)}$ is defined by (\ref{eq:Ptildefapp}) and does not imply any analytic continuation of $\bP^{(0)}(u)$. The two branches cannot be related by analytic continuation, since at weak coupling the cut has disappeared.} 
\beq\label{eq:Ptildefapp}
\widetilde \bP_A(u) \sim h^{-2 \ell_{\ast} } \widetilde \bP_A^{(0)} .
\eeq 
Notice that we have dropped the term $\bP_{A}$ from the lhs  of (\ref{eq:Ptilscale0}) since it is subleading at weak coupling. Equation (\ref{eq:Ptilscale0}) also shows that all functions $\widetilde \bP_A$ must have the same scaling at weak coupling, since all components of $\nu_a$ scale in the same way. 

\item Equation (\ref{eq:Ptilscale0}) also shows that  $\widetilde \bP_A^{(0)}(u)$ must be a polynomial in $u$. Besides, this polynomial must have a multiple zero in $u=0$ of order exactly $\ell_{\ast}$:
\beq\label{eq:Ptilzero}
\widetilde \bP_A^{(0)}(u) \sim u^{ \ell_{\ast} } , \;\;\; u \sim 0 . 
\eeq
This follows, again, from considering the expansion (\ref{eq:expP1}) (and similar for the other $\widetilde \bP$'s). Indeed, the only terms  that can contribute to $\widetilde \bP_1^{(0)}(u)$ at the leading order $\mathcal{O}(h^{-2 \ell_{\ast} } )$ are the ones with $n \geq \ell_{\ast}$, and each of them produces a positive power of $u^n$ which is subleading in (\ref{eq:Ptilzero}).  
\end{enumerate}

\paragraph{4) Properties of $\tau$ and $Q_{a|1}$: }
At leading order at weak coupling, the functions $\tau_i$ must be constants independent of $u$. Indeed, they can be computed through the definition 
\beq\label{eq:taudefapp}
\tau_i = \nu^a \, Q_{a|i}^-.
\eeq
Since $\nu^{a(0)}(u)$ is a polynomial, and $Q_{a|i}(u)$ cannot have singularities in the upper half plane, the $2i$-periodic function $\tau_i(u)$ must be analytic everywhere at the leading weak coupling order, and it is then a constant.

We can compute the value of these constants studying the large $u$-behaviour: first, from (\ref{eq:tau2tau3Van}) we know that components $i=2,3$ must vanish. Further, $\nu^a$ is proportional to $(Q^{a}_{|1})^-$  at large $u$, and this implies that the limiting value of $\tau_4(u)$  must be nonzero for consistency with $Q_{a|4} \, Q^{a}_{|1} = \ch_{41} \neq 0$. Finally, (\ref{eq:t1t4}) shows that the components $\tau_1$ must be subleading at weak coupling since $\tau_1 \tau_4 \sim \mathcal{O}(h^2)$. Therefore, using a normalization where $Q_{a|i}(u) \sim \mathcal{O}(1)$ at weak coupling, we find
\beq\label{eq:tauwc}
\(\tau_1(u), \tau_2(u), \tau_3(u), \tau_4(u) \) \propto h^{- \ell_{\ast}} \( 0,0,0,1 \) + \mathcal{O}(h^{2 - \ell_{\ast} }). 
\eeq
Finally, from (\ref{eq:tauwc}) and (\ref{eq:taudefapp}) we discover that, at leading order, the first columns\footnote{The other elements of these matrices will in general be more complicated and have an infinite string of poles even at the leading weak coupling order. } of the matrices $Q_{a|i}$, $Q^{a}_{|i}$ are polynomial in $u$ and proportional to $\nu_a^{(0)}(u+i/2)$, $\nu^{a(0)}(u+i/2)$, respectively:
\beq
  Q_{a|1}(u) \propto \nu_a^{(0)}(u+i/2) + \mathcal{O}( h^2), \;\;\;\;  Q^{a}_1(u) \propto \nu^{a(0)}(u+i/2) + \mathcal{O}( h^2) . \label{eq:propQainua}
\eeq

\paragraph{5) Properties of $\bQ_1$ and $\bQ_2$:} 
Finally, in order to show that part of the Q-system reduces to polynomials at leading order at weak coupling, we need to prove the polynomiality of some of the $\bQ$ functions. Only two of these functions have nice properties at weak coupling, namely $\bQ_1$ and $\bQ_2$. We will show that they reduce to polynomials with a multiple zero of order $\ell$ at $u=0$. 

First, we need to make some conventional choice: we pick a normalization such that all components of $Q_{a|i}(u) = \mathcal{O}(h^0)$. Consequently, we will also have $\bQ_I(u) \sim \mathcal{O}(h^0)$ at weak coupling. 

Then, let us prove a preliminary result: all functions $\widetilde \bQ_I(u)$ behave as
\beq\label{eq:Qtilpol0}
\widetilde \bQ_I(u) \sim h^{- 2 \ell_{\ast}} (u^{\ell_{\ast}} \, \text{Pol}_I(u) + \mathcal{O}(h^2) ),
\eeq
where $\text{Pol}_I(u)$ are polynomials. 
 This follows from considering the definition (\ref{Qijdef}) at weak coupling:
\beq\label{eq:Qtil0wc}
\widetilde \bQ_{I}^{(0)}(u) \Sigma_{ij}^I =  ( Q^{a(0)}_{\,|i}(u) )^{+} \, \bP_{A}^{(0)}(u) \, (Q^{b(0)}_{\,|j}(u) )^{+} \, \sigma^A_{ab} ,
\eeq
where we denote $ \widetilde \bQ_I(u) = h^{- 2 \ell_{\ast}} \, \widetilde \bQ_I^{(0)}(u) +  \mathcal{O}(h^{2 - 2 \ell_{\ast} } )$. 
Due to the cut structure of $\widetilde \bQ$, the lhs could possibly have poles only for $u \in i \mathbb{N}$. However, $\bP_A^{(0)}(u)$ is a polynomial, while $Q_{a|i}^+(u)$ is always analytic in the upper half plane: therefore at the leading order the rhs of (\ref{eq:Qtil0wc}) is regular at all these dangerous positions. This shows that $\widetilde \bQ^{(0)}$ has no singularities at all, so that it is a polynomial, and moreover it must factorize the multiple zero (\ref{eq:Ptilzero}) as in (\ref{eq:Qtilpol0}). 

We now wish to use the gluing conditions of Section \ref{sec:gluing} to deduce some properties of $\bQ_I(u)$. First, however, we need to understand the scaling of the constants $\delta_1$, $\delta_2$ appearing in these equations. Choosing a convenient normalization  where $y_i=1$, and using (\ref{eq:tauwc}) and (\ref{eq:t1t4}), we find
\beq
\delta_2 = e^{-i \sigtw} \,  \lim_{u \rightarrow \infty} (\, \tau_4(u) \, )^2 = \mathcal{O}( h^{-2 \ell_{\ast}} )\;\;,\;\;\;\;\; \delta_1 = \frac{\tan^2( \pi \hat{M}_1 ) }{ \delta_2 } = \mathcal{O}( h^{2 \ell_{\ast} + 4 } ).
\eeq
 Considering the two gluing equations in (\ref{eq:gluing1}) and (\ref{eq:gluing4}) where $\delta_2$ appears and dropping subleading terms we find:
\beq
\bQ_2(u) \propto  h^{ 2 \ell_{\ast} }\widetilde{ \bar{ \bQ}}_4(u) + \mathcal{O}(h^2), \;\;\;\;
\bQ_1(u) \propto  h^{ 2 \ell_{\ast} }\widetilde{ \bar{\bQ}}_3(u) + \mathcal{O}(h^2),
\eeq
and recalling (\ref{eq:Qtil0wc}) we see that
\beq\label{eq:Q12prop}
\bQ_{\alpha}(u) = u^{\ell_{\ast} } \, q_{\alpha}^{\text{pol}}(u) + \mathcal{O}(h^2) ,\;\;\; \alpha=1,2,
\eeq
with $q_{\alpha}^{\text{pol}}(u)$ polynomials in $u$. 
 The rest of the gluing condition also contain some information\footnote{In particular, they imply the scaling $\widetilde \bQ_I(u) = \mathcal{O}(h^0)$ for $I=1,2,5$, so that for these $\bQ$ functions several cancellations must occur in (\ref{eq:Qtil0wc}).}, but we will not need to use them in the following. 

\subsection{Recovering the 2-loop Bethe Ansatz}\label{sec:appWC3}
Let us show that, at leading order, a subset of the Q functions can be parametrized in terms of polynomials of $u$ at leading order at weak coupling, namely at order  $\mathcal{O}(h^0)$. 

For the following argument, it is convenient to first restrict to the case where $\lim_{h \rightarrow 0} \hat{M}_2 - M_2 \in \mathbb{N}^+$. From the constraints (\ref{eq:AsBsapp}), we see that this condition ensures  that no $\bP$'s or $\bQ$'s are vanishing at leading order, and this will make it simpler to draw conclusions from the Q-system equations. 

The main observation is that, due to (\ref{eq:poleprop}) and (\ref{eq:Q12prop}), the products $\bP_{\alpha}^{(0)} \, \bQ_{\beta}^{(0)}$ with $\beta \in \left\{1, 2\right\}$ are polynomials in $u$. The QQ relation (\ref{eq:F2}) then implies that, at weak coupling, $\mathcal{Q}_{\alpha|\beta}$ is also a polynomial in $u$ for $\beta \in \left\{1, 2\right\}$. At the same time, we have shown the polynomiality for the Q functions $Q_{a|1}(u)$ at leading order. Using the QQ relations, we can then prove the following polynomial parametrization for a set of Q functions\footnote{These are precisely the chains of Q functions for which we computed the ABA limit, and indeed at weak coupling the large-volume expressions are consistent with (\ref{eq:polyparapp}).}:
\beqa
&&\bP_{\alpha}^{(0)}(u) = u^{-\ell} \, p^{\text{pol} }_{\alpha}(u), \;\;\;\; 
\bQ_{\alpha}^{(0)}(u) = u^{\ell} \, q^{\text{pol} }_{\alpha}(u) , \;\;\;\; \mathcal{Q}_{\alpha |\beta}^{(0)}(u) = q_{\alpha|\beta}^{\text{pol} }(u), \label{eq:polyparapp0}\\
&& \;\;\;\; \mathcal{Q}_{12|\alpha}^{(0)}(u) = u^{-\ell} \, q_{12| \alpha}^{\text{pol}}(u) , \;\;\;\; \mathcal{Q}_{\alpha | 12}^{(0)}(u) = u^{\ell} \, q_{\alpha | 12}^{\text{pol}}(u) , \label{eq:polyparapp1}\\
&&  Q_{1|1}^{(0)}(u) \propto  \nu_1^{(0)}(u) , \;\;\;\;\;   Q^{4 (0) }_{|1}(u) \propto \nu^{4(0)}(u) , \label{eq:polyparapp}
\eeqa
where $q_{\ast| \ast}^{\text{pol}}(u)$ denotes a polynomial of $u$ and indices are restricted to the set $\alpha$, $\beta \in  \left\{1,2\right\}$. We then see that exact Bethe equations such as (\ref{eq:4eta1})-(\ref{eq:2eta1}) reduce to  the 2-loop polynomial Bethe equations of \cite{MZ}. In particular, notice that $\ell$ plays the role of spin chain length parameter $L$, entering the equations as in (\ref{eq:ABAplus1})-(\ref{eq:ABAplus4b}). Indeed, the presence of the terms $u^{\pm \ell}$  in  (\ref{eq:polyparapp1}) produces momentum factors such as $\left(\frac{u_{4, j} + i/2 }{u_{4, j}-i/2 }\right)^{\ell} = e^{i p_{4, j} \ell }$ in the Bethe equations for the massive nodes.  

The degree of the polynomials in (\ref{eq:polyparapp0})-(\ref{eq:polyparapp}), hence the number of Bethe roots, is related to the large-$u$ asymptotics of the Q functions (\ref{eq:powerlikePQ}). For instance, picking the simplest\footnote{This is the simplest such sequence since, given the ordering of charges (\ref{eq:orderMs}), it involves the least number of roots for every node of the diagram. 
} sequence of Q functions leading to the Bethe equations in $\eta=1$ grading, we find, from the Q-system, the large-$u$ scaling:
\beqa
p_2^{\text{pol} }(u) \sim u^{- M_2 + \ell} , \;\;\;\; q_{2|2}^{\text{pol}}(u) \sim u^{-M_2 + \hat{M}_2 } , \;\;\;\; q_{2|12}^{\text{pol}}(u) \sim u^{M_2 + \hat{M}_2 + \hat{M}_1  - 1 - \ell }, \label{eq:appchain0}\\
\nu_1^{(0)}(u) \sim u^{\frac{1}{2}( - M_1 - M_2 - M_5 + \hat{M}_1 + \hat{M}_2 ) } , \;\;\;\;  \nu^{4(0)}(u) \sim u^{\frac{1}{2}( - M_1 - M_2 + M_5 + \hat{M}_1 + \hat{M}_2 ) } . \label{eq:appchain}
\eeqa 
 Identifying the degrees of these polynomials  with the excitation numbers $K_1$, $K_2$, $K_3$, $K_4$, $K_{\bar{4}}$ establishes the map between quantum numbers and the asymptotics of the QSC. More precisely, we have now obtained the limit of this map for $h=0$.  However, on the assumption that the powers in the asymptotics can be written as a linear combination of the quantum numbers, it is unambiguous how to extend this prescription to finite coupling including the anomalous dimension $\gamma$. Rigorous tests of this prescription can be obtained using the connection with TBA as explained in \cite{Gromov:2014caa} or comparing with the large volume limit. 

For completeness, let us make a final comment on the case where $\lim_{h \rightarrow 0} \hat{M}_2 - M_2 = 0$, which corresponds to states with $K_2=K_1=0$. In this subsector, $\mathcal{A}_2 \mathcal{A}_3 \sim \mathcal{B}_2 \mathcal{B}_3 \sim 0$ at weak coupling, and it is most natural to choose a normalization where $\bP_2^{(0)} = \bQ_2^{(0)} = 0$. Rigorously speaking, some of the very last steps of the proof presented above need to be modified since some of the polynomials in (\ref{eq:polyparapp0})-(\ref{eq:polyparapp}) now vanish. An alternative argument valid for this case is presented below, and shows that the map between QSC asymptotics and quantum numbers (\ref{MAdef}),(\ref{MIdef}) holds unchanged in this subsector as well. 

 The only peculiarity of this case is that, while the parametrization (\ref{eq:polyparapp0})-(\ref{eq:polyparapp}) is perfectly valid, the polynomial $q_{2|12}^{\text{pol}}(u)$ always factors a zero at $u=0$. This adds one unit of length in the $\eta=1$ Bethe Ansatz, and yields a more refined identification of the parameter $\ell$:
\beq
\ell = L - \delta_{K_2, 0} = \widetilde L ,
\eeq
where $L$, $\widetilde L$ are the natural length parameters appearing in the Bethe equations in $\eta=1$ and $\eta=-1$ gradings, respectively (see equations (\ref{eq:ABAplus1})-(\ref{eq:ABAminus4b}) and the following Appendix for more details). 
 To prove the above statements, consider the following equations:
  \beqa
  \bP_1^{(0)} \, \widetilde \bP_2^{(0)} &=& ( \nu_1^{(0)} \, \nu^{4(0)} )^{[+2]} - \nu_1^{(0)} \, \nu^{4(0)} 
\label{eq:lastappweak1}\\
  &\propto& ( Q_{1|1}^{(0)} \, Q^{4(0)}_{|1} )^{[+1]} - ( Q_{1|1}^{(0)} \, Q^{4(0)}_{|1} )^{[-1]}  \label{eq:lastappweak2} \\
  &=& (\mathcal{Q}_{12|12}^{(0)} )^+ - (\mathcal{Q}_{12|12}^{(0)} )^- \propto \mathcal{Q}_{2|12}^{(0)} \, \mathcal{Q}_{12|2}^{(0)} \label{eq:lastappweak3} ,
 \eeqa
where we have started from one of the $\bP \mu$-system equations setting $\bP_2^{(0)}=0$ and the subsequent lines follow from the Q-system and (\ref{eq:propQainua}). From (\ref{eq:Ptilscale0}), we see that $\widetilde\bP_2^{(0)} \propto \mathcal{Q}_{2|12}^{(0)} $, and from (\ref{eq:lastappweak1})-(\ref{eq:lastappweak3}) we then find that $\mathcal{Q}_{12|2}^{(0)} \propto \bP_1^{(0)}$, confirming the parametrization of this function in (\ref{eq:polyparapp1}). 
 Besides, from (\ref{eq:P0ideapp}) we see that the rhs of (\ref{eq:lastappweak1}) is a polynomial with a zero at $u=0$. This implies that the parametrization (\ref{eq:polyparapp1}) is correct, and that, as stated above, the polynomial $q_{2|12}^{\text{pol}}(u)$ always has a zero at $u=0$ in this subsector.  

\section{State/charges dictionary}\label{app:dictionary}
\label{app:charges}
The purpose of this Appendix is to review the different versions of the Asymptotic Bethe Ansatz existing in the literature, and provide a practical dictionary between excitation numbers and the parameters appearing in the QSC in these different conventions. 

\subsection{Asymptotic Bethe Ansatz equations}\label{app:ABA}
In \cite{Gromov:2008qe} two equivalent versions of the ABA were introduced, characterized by the gradings $\eta=\pm 1$. 
The ABA equations in $\eta=+1$ grading read
\begin{align}
1&=\left. \frac{\mathbb{Q}_2^+ \, B_{4 (-)} \, B_{\bfo (-)}}{\mathbb{Q}_2^- \,  B_{4 (+)} \, B_{\bfo (+)}} \right|_{u_{{1},j}}\;\;, &j=1, \dots, K_1 , \label{eq:ABAplus1}
 \\
-1&=\left. \frac{\mathbb{Q}_2^{--} \, \mathbb{Q}_{{1}}^+ \, \mathbb{Q}_{{3}}^+ }{\mathbb{Q}_2^{++} \, \mathbb{Q}_{{1}}^- \, \mathbb{Q}_{{3}}^- } \right|_{u_{2,j}}\;\;, &j=1, \dots, K_2 ,
\\
1&=\left. \frac{\mathbb{Q}_2^+ \, R_{4 (-)} \, R_{\bfo (-)}}{\mathbb{Q}_2^- \,  R_{4 (+)} \, R_{\bfo  (+)}} \right|_{u_{{3},j}}\;\;,&j=1, \dots, K_3, \label{eq:ABAplus3}
\\
- 1 &= \left( \frac{x_{4,j}^-}{x_{4,j}^+}\right)^{-{L}} \, \left. \frac{\mathbb{Q}_{4}^{[-2]}}{ \mathbb{Q}_{4}^{[+2]}} \, \frac{  B_{{1}}^+ \, R_{{3}}^+ }{ B_{{1}}^- \, R_{{3}}^- } \, \frac{\sigmaBES^-_4 \, \sigmaBES^-_{\bar{4}}  }{ \sigmaBES^+_4 \, \sigmaBES^+_{\bar{4}}} \right|_{u_{4,j}} ,\;\;\; & j=1, \dots, K_4, 
\\
- 1 &= \left(\frac{x_{\bfo,j}^-}{x_{\bfo,j}^+}\right)^{-{L}}  \, \left. \frac{ \mathbb{Q}_{\bfo}^{[-2]}}{ \mathbb{Q}_{\bfo}^{[+2]}} \, \frac{  B_1^+ \, R_{{3}}^+  }{ B_1^- \, R_{{3}}^-  } \, \frac{\sigmaBES^-_4 \, \sigmaBES^-_{\bar{4}}}{ \sigmaBES^+_4 \, \sigmaBES^+_{\bar{4}}}\right|_{u_{\bfo,j}} , \;\;\; &j=1, \dots, K_{\bar{4}} ,\label{eq:ABAplus4b}
\end{align}
while the $\eta=-1$ grading version is
\begin{align}
1&=\left. \frac{\mathbb{Q}_2^+ \, B_{4  (-)} \, B_{\bfo (-)}}{\mathbb{Q}_2^- \,  B_{4 (+)} \, B_{\bfo (+)}} \right|_{u_{\tilde 1,j}}, &j=1, \dots, \tilde K_1  , \label{eq:ABAminus1}\\
-1&=\left. \frac{\mathbb{Q}_2^{--} \, \mathbb{Q}_{\tilde 1}^+ \, \mathbb{Q}_{\tilde 3}^+ }{\mathbb{Q}_2^{++} \, \mathbb{Q}_{\tilde 1}^- \, \mathbb{Q}_{\tilde 3}^- } \right|_{u_{2,j}}, &j=1, \dots, K_2 , \\
1&=\left. \frac{\mathbb{Q}_2^+ \, R_{4 (-)} \, R_{\bfo (-)}}{\mathbb{Q}_2^- \,  R_{4 (+)} \, R_{\bfo (+)}} \right|_{u_{\tilde 3,j}},  &j=1, \dots, \widetilde K_3  , \\
1 &= \left(\frac{x_{4,j}^-}{x_{4,j}^+}\right)^{ \widetilde L} \, \left. \frac{ \mathbb{Q}_{\bar{4}}^{[-2]}}{ \mathbb{Q}_{\bar{4}}^{[+2]}} \, \frac{  B_{\tilde 1}^+ \, R_{ \tilde 3}^+ \, B_{4  (+) }^+ \, B_{\bar{4}  (+) }^+ }{ B_{\tilde 1}^- \, R_{\tilde 3}^- \,  B_{4 (-) }^- \, B_{\bar{4} (-) }^- } \, \frac{\sigmaBES^+_4 \, \sigmaBES^+_{\bar{4}}}{\sigmaBES^-_4 \, \sigmaBES^-_{\bar{4}}} \right|_{u_{4,j}} , &j=1, \dots, K_4, \\
1 &= \left(\frac{x_{\bfo,j}^-}{x_{\bfo,j}^+}\right)^{\widetilde L} \, \left. \frac{ \mathbb{Q}_{4}^{[-2]}}{ \mathbb{Q}_{4}^{[+2]}} \, \frac{  B_{\tilde 1}^+ \, R_{\tilde 3}^+ \, B_{4 (+) }^+ \, B_{\bar{4} (+) }^+ }{ B_{\tilde 1}^- \, R_{\tilde 3}^- \,  B_{4 (-) }^- \, B_{\bar{4} (-) }^- } \, \frac{\sigmaBES^+_4 \, \sigmaBES^+_{\bar{4}} }{ \sigmaBES^-_4 \, \sigmaBES^-_{\bar{4}} } \right|_{u_{\bfo,j}} , &j=1, \dots, K_{\bar{4}} , \label{eq:ABAminus4b}
\end{align}
for a different set of Bethe roots. The precise relation between the two sets of roots is reviewed in Section \ref{app:Fermionic} below. 
Above and in the main text, we have used the notations:
\beqa
&&\hspace{-2cm}\mathbb{Q}_{\bu}(u) = \prod_{j=1}^{K_{\bu}}( u - u_{\bu, j} ) , \\
&&\hspace{-2cm}R_{\bu}(u) = \prod_{j=1}^{K_{\bu}} \sqrt{\frac{h}{x_{\bu, j} }}\left( x(u) - x_{\bu, j} \right) , \;\;\;\;  B_{\bu}(u) = \prod_{j=1}^{K_{\bu}} \sqrt{\frac{h}{x_{\bu, j} }}\left( 1/x(u) - x_{\bu, j} \right),  \\
&&\hspace{-2cm}R_{\balpha (\pm)}(u) = \prod_{j=1}^{K_{\bu}} \sqrt{\frac{h}{x_{\balpha, j}^{\mp} }}\left( x(u) - x_{\balpha, j}^{\mp} \right) , \;\;\;\;  B_{\balpha (\pm)}(u) = \prod_{j=1}^{K_{\balpha}} \sqrt{\frac{h}{x_{\balpha, j} }}\left( 1/x(u) - x_{\balpha, j}^{\mp} \right), \\
&&\hspace{-2cm}\frac{\sigma^+_{\balpha}(u)}{\sigma^-_{\balpha}(u)} = \prod_{j=1}^{K_{\bu}}\sigma_{\text{BES}}(u, u_{\bu, j}) ,  \;\;\; \;\; x^{\pm}_{\bu, j}=x(u_{\bu} \pm i/2 ), \;\;\;\; x_{\bu, j}=x(u_{\bu, j} ) ,
\label{eq:dressing}
\eeqa
where $\sigma_{\text{BES}}(u, v)$ is the Beisert-Eden-Staudacher dressing factor \cite{Beisert:2006ez}. 

\subsection{Fermionic duality: from $\eta=+1$ to $\eta=-1$}\label{app:Fermionic}
It is expected that every state (or, more precisely, every multiplet) can be represented by a \emph{regular} solution of the Asymptotic Bethe Ansatz, where regular means that for every type of root $x_i$ we have $x_i\neq 0,\, x_i \neq  \infty$. Let us now review (see  Appendix A in \cite{Gromov:2008qe}) how to switch from a regular solution of the $\eta= + 1$ ABA, 
characterized by the roots
\beqa
\left\{ u_{1, j} \right\}_{j=1}^{K_1}, \;\;\; \left\{ u_{2, j} \right\}_{j=1}^{K_2}, \;\;\; \left\{ u_{3, j} \right\}_{j=1}^{K_3}, \;\;\; \left\{ u_{4, j} \right\}_{j=1}^{K_4}, \;\;\; \left\{ u_{\bar{4}, j} \right\}_{j=1}^{K_{\bar{4}}},\label{eq:rootseta1}
\eeqa
to a regular solution of the $\eta=-1$ ABA. This type of duality transformations is well known from the $\mathcal{N}$=4 SYM case \cite{Beisert:2005fw}.  
Following the standard argument, we consider the polynomial in $x(u)$:
\beqa
P(x) &=& \prod_{j=1}^{K_4} (x - x^+_{4, j}) \, \prod_{j=1}^{K_{\bar{4}}} (x - x^+_{\bar{4}, j}) \,  \prod_{j=1}^{K_{2}} (x - x^-_{2}) \, (x - 1/x^-_{2}) \\
&-& \prod_{j=1}^{K_4} (x - x^-_{4, j}) \, \prod_{j=1}^{K_{\bar{4}}} (x - x^-_{\bar{4}, j}) \,  \prod_{j=1}^{K_{2}} (x - x^+_{2}) \, (x - 1/x^+_{2}).
\eeqa
Due to the ABA equations (\ref{eq:ABAplus1}),(\ref{eq:ABAplus3}), 
 we see that this polynomial has zeros at all roots of type $x=x(u_{3,j})$ and $x=1/x(u_{1, j})$; besides, due to the zero momentum condition, it vanishes for $x=0$.  One may then write
\beq
P(x) = x \prod_{j=1}^{K_1} (x - 1/x_{1, j}) \, \prod_{j=1}^{\widetilde K_1}  (x - 1/ x_{\tilde1, j}) \, \prod_{j=1}^{K_3}  (x - x_{3, j}) \, \prod_{j=1}^{ \widetilde{K}_3}  (x -  x_{\tilde3, j}) ,
\eeq
where $\left\{ x_{\widetilde3, j} \right\}_{j=1}^{\widetilde K_{ 3} }$ and $\left\{ 1/x_{\widetilde1, j} \right\}_{j=1}^{\widetilde K_{1} }$ label the  extra zeros of $P(x)$ outside/inside the unite circle, respectively. 
By considering the weak coupling limit of $P(x)$, and considering that $x_{\bu, j} \sim h^{-1}$, one may count the two new types of roots:
\beqa\label{eq:duality}
K_4 + K_{\bar{4}} +  K_2 - 1 - \delta_{K_2, 0}  = K_3 +  \widetilde{ K}_{3} , \quad K_2 -1 + \delta_{K_2, 0}  = K_1 + \widetilde{K}_{1} .
\eeqa
We have then found the fermionic duality equation\footnote{Notice that the prefactor $x^{\delta_{K_2, 0}}$ appears here due to the fact that we insisted  on enumerating only regular Bethe roots in both gradings. }:
\beq
R_{4 (-)} \, R_{\bar{4} (-)} \, \mathbb{Q}_2^+ - R_{4 (+)} \, R_{\bar{4} (+)} \, \mathbb{Q}_2^- \propto x^{\delta_{K_2, 0} } \, R_3 \, R_{\tilde{3}} \, B_1 \, B_{\tilde{1}} , \label{eq:app_fermionic}
\eeq
with an inessential proportionality factor independent of $u$. It is now standard to verify that the set of roots
\beq
\left\{ u_{\tilde 1, j} \right\}_{j=1}^{\widetilde K_1}, \;\;\; \left\{ u_{2, j} \right\}_{j=1}^{K_2}, \;\;\; \left\{ u_{\tilde 3, j} \right\}_{j=1}^{\widetilde K_3}, \;\;\; \left\{ u_{4, j} \right\}_{j=1}^{K_4}, \;\;\; \left\{ u_{\bar{4}, j} \right\}_{j=1}^{K_{\bar{4}}},\label{eq:rootsetam1}
\eeq
satisfy the $\eta=-1$ ABA, where the spin chain length parameter is
\beq\label{eq:dualityL}
 \widetilde L := L_{\eta=-1}  = L_{\eta=+1} - \delta_{K_2, 0} .
\eeq
\subsection{Asymptotics of the QSC and excitation numbers}
\label{app:asymeta-1}
The charges entering the asymptotics of the QSC are, in terms of the number of Bethe roots in $\eta=+1$ grading: 
\beqa\label{eq:quant1}
&M_1= L + K_3 -K_4 - K_{\bar{4}} + 1 , \quad M_2 = L -K_1\quad M_5= K_{\bar{4}} - K_4, & \\
&\hat M_1 = \gamma +L + K_3 - K_2 +1 , \quad \hat M_2 = \gamma + L + K_2 -K_1 .&
\label{eq:quant2}
\eeqa
Using the rules (\ref{eq:duality}) and (\ref{eq:dualityL}), (\ref{eq:quant1})-(\ref{eq:quant2}) can be rewritten as 
\beqa
&&M_1=\widetilde L-\widetilde K_3 + K_2 , \quad M_2 = \widetilde L +\widetilde K_1 -K_2 +1 , \quad M_5= K_{\bar{4}} - K_4 \\
&&\hat M_1 = \gamma + K_4 + K_{\bar{4}} + \widetilde L -\widetilde K_3 , \quad \hat M_2 = \gamma + \widetilde L + \widetilde K_1 +1,
\eeqa
where we have denoted $\widetilde{L} = L_{\eta=-1}$. 

\subsection{Important subsectors}

In what follows we list a set of special cases corresponding to different subsectors of the theory, described by different values of excitation numbers and subsets of BA equations in $\eta=\pm 1$ gradings. 
\paragraph{$\bf{SL(2|1)}$ sector: }
This  sector can be represented by operators made of scalars $Y^1Y_4^{\dagger}$, covariant derivatives and fermions $\psi_{4+}$, $\psi_+^{1\dagger}$. The corresponding large-volume spectrum is described by the solutions of the ABA equations (\ref{eq:ABAminus1})-(\ref{eq:ABAminus4b}) in $\eta=-1$ grading without any auxiliary root, namely $\widetilde{K}_3= \widetilde{K}_1=K_2=0$. The classical dimensions of these operators as realized in the $\eta=-1$ grading is $\Delta^{(0)}=\widetilde L+\frac{1}{2} ( K_4 + K_{\bar{4}} )$, and their spin is $S_{\eta=-1}=\frac{1}{2}(K_4 + K_{\bar{4}})$. 
The corresponding subset of ABA equations in $\eta=-1$ grading is
\beqa
1 &=& \left(\frac{x_{4,k}^-}{x_{4,k}^+}\right)^{\widetilde L} \, \left. \frac{ \mathbb{Q}_{\bar{4}}^{[-2]}}{ \mathbb{Q}_{\bar{4}}^{[+2]}} \, \frac{  B_{4 (+) }^+ \, B_{\bar{4} (+) }^+ }{ B_{4 (-) }^- \, B_{\bar{4} (-) }^- } \, \frac{\sigma_4^+ \, \sigma_{\bar{4}}^+ }{ \sigma_4^- \, \sigma_{\bar{4}}^- } \right|_{u_{4,k}} , \;\;\;\;\;\;\text{ with }  \mathbb{Q}_4(u_{4,k} ) = 0 ,\\
1 &=& \left(\frac{x_{\bfo,k}^-}{x_{\bfo,k}^+}\right)^{\widetilde L} \, \left. \frac{ \mathbb{Q}_{4}^{[-2]}}{ \mathbb{Q}_{4}^{[+2]}} \, \frac{ B_{4 (+) }^+ \, B_{\bar{4} (+) }^+ }{ B_{4 (-) }^- \, B_{\bar{4} (-) }^- } \, \frac{\sigma_4^+ \, \sigma_{\bar{4}}^+ }{ \sigma_4^- \, \sigma_{\bar{4}}^- } \right|_{u_{\bfo,k}} , \;\;\;\;\;\;\text{ with }  \mathbb{Q}_{\bfo}(u_{\bfo,k} ) = 0 ,
\eeqa
and the asymptotics of the corresponding QSC solution is parametrized by:
\beqa
&&M_1=\widetilde L, \;\;\;\;  M_2=\widetilde L+1, \;\;\;\; M_5= K_{\bar{4}} - K_4, \\
&&\hat M_1=\widetilde L+K_4+K_{\bar{4}}  + \gamma, \;\;\;\;\; \hat M_2=\widetilde L+\gamma+1.
\eeqa
In the grading $\eta=+1$, the description of this sector involves some of the auxiliary roots: $K_3=K_4 + K_{\bar{4}}-2$, while $\widetilde K_1= 0$. 

\paragraph{$\bf{SL(2)}$-like sector: }
Rather than a sector, this is a subset of states belonging to the $SL(2|1)$ sector, which satisfy the condition $K_4=K_{\bar 4}$ and $\left\{u_{4, j} \right\}=\left\{u_{\bar{4}, j} \right\}$ (see \cite{Papathanasiou:2009zm} and \cite{Klose:2010ki} for a detailed discussion). In this case $M_5=0$ and the ABA equations reduce to the following single equation:
\beq
1 = \left(\frac{x_{4,k}^-}{x_{4,k}^+}\right)^{\widetilde L} \, \left. \frac{ \mathbb{Q}_{4}^{[-2]}}{ \mathbb{Q}_{4}^{[+2]}} \,\left( \frac{ B_{4 (+) }^+ }{ B_{4 (-) }^- } \, \frac{\sigma_4^+}{ \sigma_4^-}\right)^2 \right|_{u_{4,k}} , \;\;\;\;\;\;\text{ with }  \mathbb{Q}_{4}(u_{4,k} ) = 0 .
\eeq
This set of states were studied at weak coupling using the QSC in \cite{Anselmetti:2015mda}. 

\paragraph{$\bf{SU(4)}$ sector: }
The operators belonging to this sector are made of all the complex scalars of the theory: $ Y^a$, $Y_b^{\dagger}$, $a, b=1, \dots, 4$. 
The corresponding scaling dimensions are described most conveniently by the ABA equations in $\eta=+1$ grading (\ref{eq:ABAplus1})-(\ref{eq:ABAplus4b}),  where only Bethe roots of type $4$, $\bar{4}$ and $3$ are excited:
\begin{align}
\label{eq:SU2SU2a}
- 1 &= \left( \frac{x_{4,k}^-}{x_{4,k}^+}\right)^{-L} \, \left. \frac{\mathbb{Q}_{4}^{[-2]}}{ \mathbb{Q}_{4}^{[+2]}} \,\frac{\sigma_4^- \, \sigma_{\bar{4}}^- }{ \sigma_4^+ \, \sigma_{\bar{4}}^+ }\, \frac{ R_{3}^+ }{ R_{3}^-} \right|_{u_{4,k}} , &\text{ with }  \mathbb{Q}_4(u_{4,k} ) = 0 , \\
- 1 &= \left(\frac{x_{\bfo,k}^-}{x_{\bfo,k}^+}\right)^{-L}  \, \left. \frac{ \mathbb{Q}_{\bfo}^{[-2]}}{ \mathbb{Q}_{\bfo}^{[+2]}} \, \frac{\sigma_4^- \, \sigma_{\bar{4}}^- }{ \sigma_4^+ \, \sigma_{\bar{4}}^+ } \,  \frac{ R_{3}^+ }{ R_{3}^-}\right|_{u_{\bfo,k}} , &\text{ with }  \mathbb{Q}_{\bfo}(u_{\bfo,k} ) = 0 , \\
1&=\left. \frac{R_{4 (-)} \, R_{\bfo(-)}}{R_{4(+)} \, R_{\bfo(+)}} \right|_{u_{3,k}} , &\text{ with }  \mathbb{Q}_{3}(u_{3,k} ) = 0 ,
\end{align}
and the excitation numbers are constrained by the conditions
\beqa
L + K_3 -2 K_4 \geq 0 , \;\;\;\;\; L + K_3 -2 K_{{4}}\geq 0 , \;\;\;\;\; K_4 + K_{\bar{4}} \geq 2 K_3 ,
\eeqa
(which are stricter than the general unitarity constraints). 
In this case the parameters entering the asymptotics of the QSC read 
\beqa
&&M_1=L+ K_3-K_4 - K_{\bar{4}} + 1, \quad M_2=L, \quad M_5= K_{\bar{4}} - K_4,\\ &&\hat M_1=L+K_3+1+\gamma, \;\;\;\;\;\hat M_2 = L+\gamma.
\eeqa
In the $\eta=-1$ grading, these states are represented with $\tilde K_3=K_4+K_{\bar4}-K_3-2$, $K_2=\widetilde K_1 = 0$. 

\paragraph{$\bf{SU(2)\times SU(2)}$ sector:}
This can be realized considering only scalars $Y^2$ and $Y_3^{\dagger}$ as excitations on top of the vacuum $\tr[(Y^1Y_4^{\dagger} )^L]$. The corresponding Bethe Ansatz solutions have only massive Bethe roots excited in $\eta=+1$ grading, with $K_3=0$.

\subsection{Distinguished grading}
Finally, a further very common form of the Bethe Ansatz equations is the one related to the distinguished Dynkin diagram. This is the form in which the 2-loop BA was originally written in \cite{Minahan:2008hf}; it is known that it does not admit an all-loop generalization in terms of explicit  functions of the Bethe roots. 
At two loops, one can relate the roots appearing in this version of the BA to the ones featuring in the other two versions by a chain of fermionic dualities  (see \cite{Papathanasiou:2009zm}, Appendix A). The relation between the excitation numbers in the distinguished-grading Bethe Ansatz, denoted as $K_{\bu}^{d}$ for $s=1,2,3,4,\bar{4}$, and the excitation numbers in the $\eta=-1$ grading, is
  \beqa\label{eq:dist2}
&&\hspace{-2cm}  K_{1}^d=\widetilde K_1, \;\;\;\; K_{2}^d = 
  K_4 + K_{\bar{4}} +\widetilde K_1-\widetilde K_3-2,\;\;\;\; K_{3}^d =
 K_4+K_{\bar{4}} + K_2-1-\widetilde K_3  , \label{eq:K3d}\\
 &&\hspace{-2cm}  K_4^d=K_4, \;\;\;\; K_{\bar{4}}^d=K_{\bar{4}} ,\nn
 \eeqa
 and the length entering this version of the BA is the same as in the $\eta=-1$ grading, $L^{d}=\widetilde L$. 
 The translation between excitation numbers of distinguished and $\eta=+1$ gradings can be obtained comparing equations (\ref{eq:dist2}) and (\ref{eq:duality}):
\beq
K_{1}^d = K_2 - K_1 - 1 + \delta_{K_2,0},\quad K_{2}^d = K_3 - K_1 - 2 + 2 \delta_{K_2,0},\quad K_{3}^d = K_3 + \delta_{K_2,0} ,
\eeq
while $L^d=L_{\eta=+1}-\delta_{K_2,0}$.
 
Finally, let us make contact with the Dynkin labels $[\Delta, j; p_1, q, p_2]$ defined in relation to the distinguished diagram, which are widely used in the literature, e.g. \cite{Papathanasiou:2009zm}. In terms of these charges, the parameters entering the asymptotics of the QSC are given by
\beqa
&M_1 = 1 + r_2 , \quad M_2 = 2 + r_1 , \quad M_5 = r_3,&\\
&\hat M_1 = \Delta + j + 2 , \quad \hat M_2 = \Delta - j + 1  , &
\eeqa
where
\beq
r_1=\frac{1}{2} \, (p_1+p_2 + 2 q ), \;\;\;\; r_2 = \frac{p_1+p_2}{2}, \;\;\;\;\; r_3=\frac{p_2-p_1}{2} .
\eeq

\section{An integral formula for $\sigtw$}\label{app:Prec}
In this Appendix we prove an exact integral formula for $\sigtw$, which could be useful for computing this quantity from the numerical solution of the QSC. 
The expression is 
\beqa\label{eq:intePapp}
\sigtw  &=& \frac{1}{2 \pi \, \mathbb{E}(h) } \int_{-2 h }^{2 h} \frac{dz \, e^{\pi z} \, \log\left( \frac{-\tau_4(z) }{\tau^1(z) } \right) }{\sqrt{(e^{2 \pi z}- e^{4 \pi h} ) \, (e^{ 2 \pi z}- e^{- 4 \pi h} ) } } \\&=&  \frac{1}{4 \pi \, \mathbb{E}(h) } \int_{-2h}^{2 h} \frac{dz \, e^{\pi z} \, \log\left( \frac{\tau_4(z) \, \widetilde \tau_4(z)  }{\tau^1(z) \, \widetilde \tau^1(z) } \right) }{\sqrt{(e^{2 \pi z}- e^{4 \pi h} ) \, (e^{ 2 \pi z}- e^{- 4 \pi h} ) } } ,\label{eq:intePapp2}
\eeqa
where $\mathbb{E}(h)$ is an elementary function of $h$ defined in (\ref{eq:Edef}). 
To prove (\ref{eq:intePapp}), we use (\ref{eq:tauUP}) to write
\beq
 \log\left( \frac{-\tau_4(z) }{\tau^1(z) } \right) = i \sigtw + \mathcal{A}(z), \quad \quad \quad \quad \mathcal{A}(z) =\log \frac{\tau_4(z) }{\tau_4(z+i) } ,
\eeq
where $\mathcal{A}(z+i)=-\mathcal{A}(z)$. Assuming that $\mathcal{A}(z)$ has no singularities on the first sheet, we can open up the integration contour circling the cut to a couple of infinite horizontal lines lying at $\text{Im}(z)=\pm i/2$. Thus we see that the integral over $\mathcal{A}(z)$ exactly cancels due to the periodicity of the integrand, leading to (\ref{eq:intePapp}). Notice that the ABA expression (\ref{eq:PABA}) for $\sigtw$ is just a an application of this formula where $\tau_4/\tau^1$ takes its large volume value, which can be read from (\ref{eq:scaleABANu}),(\ref{eq:Q11Q41}),(\ref{eq:nuABA2}),(\ref{eq:logFABA}).

\bibliography{Biblio3} 

\end{document}